\documentclass[12pt]{article}

\usepackage{bbm}
\usepackage{amssymb,amsmath,amsthm}

\usepackage[dvips,xdvi]{graphics}
\usepackage[dvips,xdvi]{graphicx}

\usepackage{epsfig}

\renewcommand{\arraystretch}{1.2}

\newcommand{\bone}{\mathbbm{1}}
\newcommand{\diag}{\mbox{diag}\,}
\newcommand{\zz}{\mathbbm{Z}}
\newcommand{\one}{\mathbf{1}}
\newcommand{\two}{\mathbf{2}}
\newcommand{\three}{\mathbf{3}}
\newcommand{\four}{\mathbf{4}}
\newcommand{\five}{\mathbf{5}}
\newcommand{\six}{\mathbf{6}}
\newcommand{\eight}{\mathbf{8}}

\newcommand{\oo}{\mbox{ord}\,}

\newcommand{\be}{\begin{equation}}
\newcommand{\ee}{\end{equation}}
\newcommand{\ba}{\begin{eqnarray}}
\newcommand{\ea}{\end{eqnarray}}

\newcommand{\bmat}{\begin{pmatrix}}
\newcommand{\emat}{\end{pmatrix}}

\newcommand{\s}{\hspace{0.5mm}}

\newcommand*{\bigtimes}{\mathop{\raisebox{-.5ex}{\hbox{\huge{$\times$}}}}}
\newcommand{\llhd}{\,\lhd\,}

\allowdisplaybreaks

\theoremstyle{definition} 
\newtheorem{define}{Definition}[section]

\theoremstyle{definition} 
\newtheorem{theorem}[define]{Theorem}

\theoremstyle{definition} 
\newtheorem{lemma}[define]{Lemma}

\theoremstyle{definition} 
\newtheorem{corollary}[define]{Corollary}

\begin{document}

\title{
\normalsize \hfill UWThPh-2011-30 \\[12mm]
\LARGE Finite flavour groups of fermions
}

\author{
Walter Grimus\thanks{E-mail: walter.grimus@univie.ac.at}
\setcounter{footnote}{6}\
and Patrick Otto Ludl\thanks{E-mail: patrick.ludl@univie.ac.at} 
\\*[3mm]
\small University of Vienna, Faculty of Physics \\
\small Boltzmanngasse 5, A--1090 Vienna, Austria
}

\date{25 April 2012}

\maketitle

\vspace{-1cm}
\begin{center}
\epsfig{scale=0.4,file=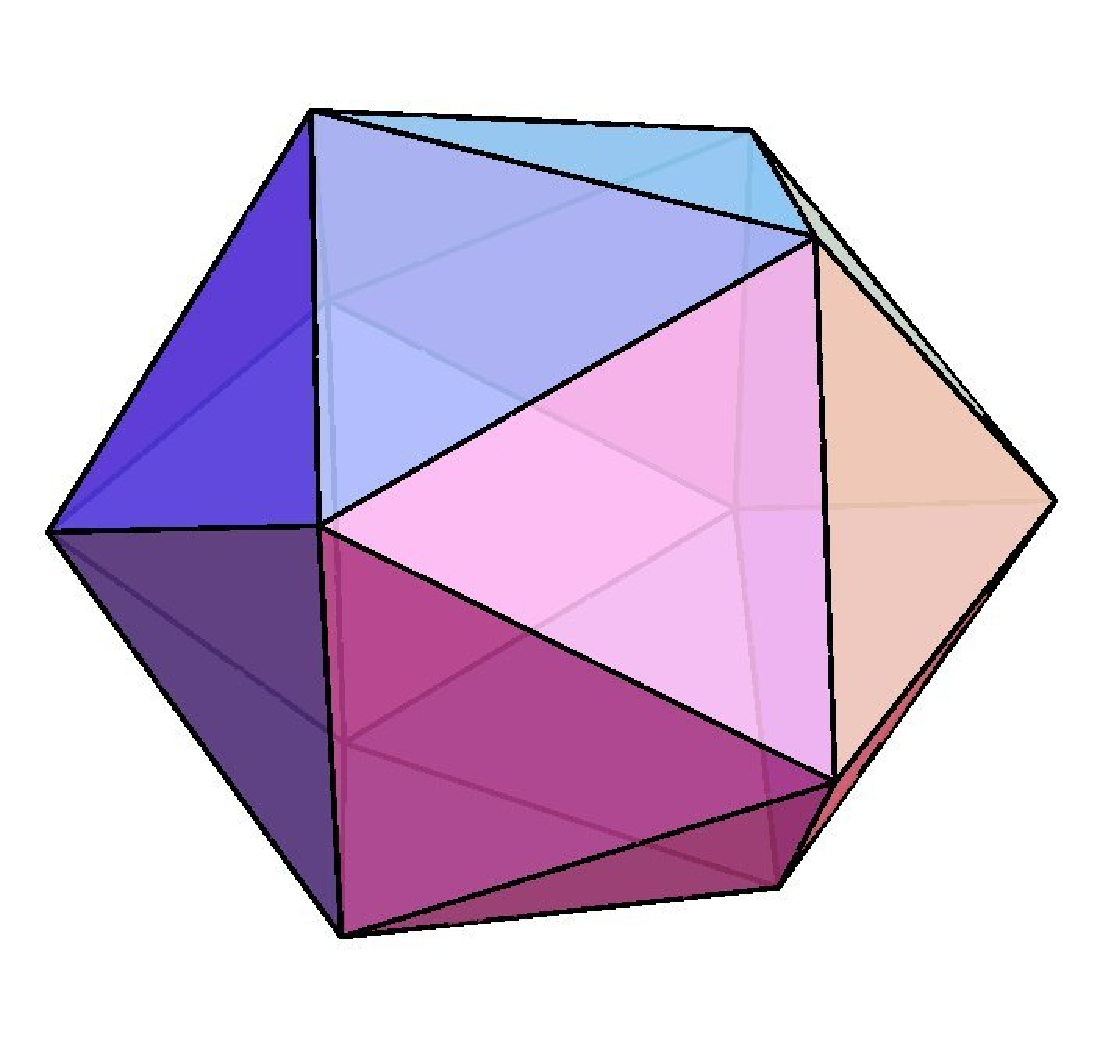}
\end{center}

\begin{abstract}
We present an overview of the theory of finite groups, with regard to their
application as flavour symmetries in particle physics. In a general part, we
discuss useful theorems concerning group structure, conjugacy classes,
representations and character tables. In a specialized part, we attempt to
give a fairly comprehensive review of finite subgroups of $SO(3)$ and
$SU(3)$, in which we apply and illustrate the general theory. Moreover, we also 
provide a concise description of the symmetric and alternating groups and
comment on the relationship between finite subgroups of $U(3)$ and finite
subgroups of $SU(3)$. Though in this review we give a detailed description of
a wide range of finite groups, the main focus is on the methods which allow 
the exploration of their different aspects.
\end{abstract}

\newpage

\tableofcontents

\newpage

\section{Introduction}

The gauge principle is one of the most successful principles in
particle physics. It determines the Lagrangians 
of electroweak interactions and QCD. However, the gauge symmetry
pertaining to the gauge group $SU(2)_L \times U(1)_Y$ must be
broken in order to generate gauge boson and fermion masses and the quark and
lepton mixing matrices. This is usually achieved by the Higgs mechanism, which
entails a new sector in the Lagrangian, the scalar sector. Again, the gauge
interactions of the scalars are fixed by the gauge principle, but this
principle has no bearing on the flavour structure of the Yukawa Lagrangian
$\mathcal{L}_Y$ and the scalar potential $V$.
While the gauge interactions are flavour-blind, 
the flavour dependence of the Yukawa couplings is crucial because this 
is, in conjunction with the vacuum expectation values determined by the minimum
of the scalar potential, the origin of fermion masses and the mixing
matrices.\footnote{Also in models without the Higgs mechanism 
  a flavour-dependent sector is indispensable in order to obtain the mass
  spectrum of the known fermions and the
  mixing matrices.} 
In the general case, the fermion masses and the parameters of the mixing
matrices are completely free. 
Up to now, we do not know of any fundamental principle,
comparable to the importance of the gauge principle, which would allow us to
constrain $\mathcal{L}_Y$ and $V$ such that the fermion mass spectra and the
entries of the mixing matrices find a satisfactory explanation. 

In view of this situation and bearing in mind that 
symmetry principles have proven very successful in physics, one
resorts to flavour symmetries, also called family or horizontal
symmetries, for obtaining restrictions on $\mathcal{L}_Y$ and $V$. 
At any rate, under such symmetries the experimentally tested gauge
couplings are automatically invariant. 
Moreover, flavour symmetries are, for instance, supported by 
the famous formula~\cite{gatto} 
\begin{equation} \label{c}
\sin\theta_c \simeq \sqrt{\frac{m_d}{m_s}},
\end{equation}
expressing the Cabibbo angle as a function of the ratio of the down
and strange quark mass, which has been derived in a number of flavour
models. More recently,  
the observation by Harrison, Perkins and Scott~\cite{HPS} that the lepton
mixing matrix $U$ is not far from \emph{tri-bimaximal}, i.e. 
\begin{equation}\label{HPS}
U \simeq \left( \begin{array}{rrr}
2/\sqrt{6} & 1/\sqrt{3} & 0\hphantom{xx} \\ 
-1/\sqrt{6} & 1/\sqrt{3} & -1/\sqrt{2} \\ 
-1/\sqrt{6} & 1/\sqrt{3} & 1/\sqrt{2}
\end{array} \right), 
\end{equation}
has strongly promoted the idea of family symmetries. 
There is a striking difference between the CKM matrix and the lepton
mixing matrix; while the CKM matrix is close to the unit
matrix~\cite{rpp}, the lepton mixing matrix has two large mixing
angles---see~\cite{results} for recent fits---and is, therefore, very
far from it. 

If family symmetries are broken
spontaneously together with the gauge symmetry, 
the simplest way to avoid Goldstone bosons is model building with
finite groups.\footnote{Spontaneous breaking of discrete symmetries
  possibly leads to the cosmological problem of domain
  walls~\cite{vilenkin}.}  
It is well-known that the usage of Abelian groups is
synonymous with texture zeros in fermion mass
matrices~\cite{GJLT}. Abelian groups are very simple but have a
limited capacity to enforce a mixing matrix; for instance, it is not
possible to enforce tri-bimaximal mixing~\cite{low}. Thus one is led
to non-Abelian finite groups as obvious candidates for model building.
For this purpose, $A_4$, the smallest group with an irreducible
three-dimensional representation, 
has become very popular for its capacity to
enforce tri-bimaximal mixing~\cite{ma}, provided one finds a
solution to the vacuum alignment problem.
There are models with extra dimensions without fundamental scalars;
in such models the family symmetries are broken by boundary
conditions---see for instance~\cite{serone}.
For recent reviews on model building in the lepton sector
see~\cite{review}. 

The aim of this review is to provide a short mathematical 
introduction to the theory of finite groups and their
representations and to discuss comprehensively most non-Abelian
groups which have been used in particle physics phenomenology.
We focus on aspects which are most expedient for applications, and
the groups we discuss serve as illustrations for the mathematical issues
presented in the review. We assume that the reader is familiar with
linear algebra and the basics of group theory. 
In particular, we will assume that the reader is familiar 
with the group axioms, definitions of group homomorphisms and
isomorphisms, notions like the order of a group, subgroups, 
invariant subgroups, conjugacy classes, and the concept of equivalent,
reducible and irreducible representations.
With these prerequisites our review can serve as an
introductory manual for the usage of finite groups in model
building. For additional groups, not considered in this review, we
refer the reader to~\cite{tanimoto}. The discussion of some ``exceptional''
subgroups of $SU(3)$ can be found in~\cite{he,GL10}. 
Since in this review we focus on the \emph{application} of mathematical theorems
on finite groups, we do not supply the corresponding proofs, but we do
quote adequate literature instead. 
In a few occasions, however, where we noticed that useful mathematical
issues are not well presented in the literature, we also provide the 
corresponding mathematical arguments.

The review is organized as follows.
In section~\ref{flavour symmetries} we present some general
considerations concerning flavour symmetries and Lagrangians. A 
mathematical introduction to the theory of finite groups and their
representations is given in section~\ref{properties}.
An excursion into permutation groups is made in
section~\ref{chapterpermutation}, while section~\ref{su2so3} is
devoted to finite subgroups of $SU(2)$ and $SO(3)$. 
We move on to finite subgroups of $SU(3)$ in section~\ref{finitesu3}.
After having made some remarks on finite subgroups of $U(3)$ in 
section~\ref{unitary-vs-su}, we conclude with 
miscellaneous remarks in section~\ref{conclusions}.
Some material of a more technical nature is found in the
appendices~\ref{example}--\ref{irrepsD}.

\section{Flavour symmetries and Lagrangians}
\label{flavour symmetries}
Flavour symmetries can be applied to extensions of the Standard Model (SM)
as well as to Grand Unified Theories. The symmetry structure is in
both cases given by the direct product 
$G_\mathrm{gauge} \times G_\mathrm{flavour}$
of the symmetry groups. For definiteness we assume that we have an
extension of the SM, i.e. 
$G_\mathrm{gauge} = SU(3)_c \times SU(2)_L \times U(1)_Y$ and the
fermionic gauge multiplets are left-handed doublets $q_L$, $\ell_L$
and right-handed singlets $d_R$, $u_R$, $e_R$, $\nu_R$ with obvious
notation for quark and lepton fields. In the following these chiral
fields will be denoted generically by $\psi$. 
Since there are three known families, for definiteness we will assume
that for each $\psi$ there will be three copies.
In this notation, the Lagrangian has the form
\begin{equation}
\mathcal{L} = \sum_\psi \sum_{j=1}^3 i\bar\psi_j \gamma^\mu \partial_\mu \psi_j
  + \cdots,
\end{equation}
where the dots indicate the terms beyond the kinetic terms of the SM
fermions. Let us assume that for some physical reason we postulate a
set of flavour symmetries. 
We enumerate flavour symmetries by the index 
$p = 1,\ldots,N$. 
For each $\psi$ and for every flavour symmetry 
there is a $3 \times 3$ matrix $A^{(\psi)}_p$ such that 
the symmetries of $\mathcal{L}$ are given by transformations
\begin{equation}\label{A}
\psi_j \to \left( A^{(\psi)}_p \right)_{jk} \psi_k .
\end{equation}
Because of the invariance of the kinetic term, the matrices
$A^{(\psi)}_p$ must be unitary. In addition, for the Yukawa couplings
we have to introduce scalar gauge multiplets. Since we have in mind
extensions of the SM, the scalar sector must at least
contain Higgs doublets. 
Assuming $n_H$ Higgs doublets $\phi_k$ we need $N$  
$n_H \times n_H$ unitary matrices $A^{(\phi)}_p$ for the symmetry
transformations of the Higgs doublets.\footnote{In multi-Higgs doublet
models there will in general be flavour-changing neutral
interactions. The experimental constraints on such interactions can
usually be taken care of by raising the mass of such scalars above the
electroweak scale or by imposing family symmetries which forbid such
interactions.} 
If there are further scalar
gauge multiplets like singlets or triplets, 
then further transformation matrices are needed, one
for each type of scalar gauge multiplet. 
The transformations of equation~(\ref{A}), together with the
corresponding transformations acting on the Higgs doublets, constrain the
Yukawa couplings. Using the Yukawa interactions of the right-handed
lepton singlets $e_R$ as an example, we find the following
invariance conditions on the $3 \times 3$ coupling matrices $\Delta_i$:
\begin{equation}\label{Delta}
\mathcal{L}_Y^{(e)} = -\sum_{i=1}^{n_H} \bar \ell_L \Delta_i \phi_i e_R 
+ \mbox{H.c.} \quad \Rightarrow \quad
\sum_{k=1}^{n_H}
{A^{(\ell)}_p}^\dagger \Delta_k A^{(e)}_p \left( A^{(\phi)}_p
\right)_{ki} = \Delta_i \quad \forall p.  
\end{equation}
In this approach, the matrices $A^{(\ell)}_p$, $A^{(e)}_p$ 
and $A^{(\phi)}_p$ can
be conceived as representations of group generators 
from which we can infer the group $G$. 

Conversely, one can directly postulate a family group $G$ and introduce
multiplets of fields which transform according to 
representations of $G$. In this way one determines $\mathcal{L}$ from the
symmetry group and the multiplets.

Given a family symmetry group $G$, one can solve equations
like equation~(\ref{Delta}) for the Yukawa coupling matrices $\Delta_i$ by
resorting to irreducible representations (irreps) of $G$. In this way,
as we will see shortly, Yukawa coupling matrices are related to 
Clebsch--Gordan coefficients. For this purpose, 
we consider generic Yukawa couplings in 
Majorana notation
\begin{equation}\label{Lgeneric}
\mathcal{L}_Y(\psi,\psi',S) = \psi_\alpha^T C^{-1} y_{i\alpha\beta} S_i\, 
\psi'_\beta + \mbox{H.c.},
\end{equation}
where $C$ is the charge-conjugation matrix. One can readily translate
equation~(\ref{Lgeneric}) 
to Dirac notation
by using 
$\psi_\alpha^T C^{-1} = -\overline{(\psi_\alpha)^c}$ with the
charge-conjugation operation 
$(\psi_\alpha)^c = C \gamma_0^T \psi_\alpha^*$. 
We  assume that $\psi$ and $\psi'$ transform according to the
irreps $D$ and $D'$, respectively, the irrep $D_S$ occurs in the
tensor product $D \otimes D'$ and $S$ transforms
according to the complex-conjugate irrep $D_S^*$. As discussed before, 
these irreps are given in the form of unitary matrices. We 
conceive the representation matrices as operators 
with respect to the orthonormal bases
$\{ e_\alpha \}$ of $D$, $\{ f_\beta \}$ of $D'$ and 
$\{ b_i \}$ of $D_S$, i.e. 
\begin{equation}
e_\alpha \to D_{\gamma\alpha} e_\gamma, \quad 
f_\beta \to D_{\delta\beta}' f_\delta, \quad 
b_i \to (D_S)_{ji} b_j.
\end{equation}
The sets $\{ e_\alpha \}$ and $\{ f_\beta \}$ can simply be considered
as Cartesian bases. 
Since $D_S$ is an irrep in the tensor product $D \otimes D'$,
its basis can be written as
\begin{equation}
\{ b_i = \Gamma_{i\alpha\beta}\, e_\alpha \otimes f_\beta \},
\end{equation}
where the $\Gamma_{i\alpha\beta}$ are the Clebsch--Gordan
coefficients. It is easy to check that these fulfill the conditions 
\begin{equation}\label{cond}
\Gamma_i = \left( D^\dagger \Gamma_j {D'}^* \right) (D_S)_{ji}.
\end{equation}
Then, comparing this equation with the condition obtained from the
requirement that
$\mathcal{L}_Y(\psi,\psi',S)$ of equation~(\ref{Lgeneric})
is invariant under the transformation 
\begin{equation}
\psi_\alpha \to D_{\alpha\gamma} \psi_\gamma, \quad 
\psi'_\beta \to D'_{\beta\delta} \psi'_\delta, \quad 
S_i \to \left(D_S \right)_{ij}^* S_j
\end{equation}
allows to deduce the result~\cite{grimus}
\begin{equation}
y_{i\alpha\beta} = y \left( \Gamma_{i\alpha\beta} \right)^*,
\end{equation}
where $y$ is a free parameter.
This means that the scalar fields transform with the irrep
complex-conjugate to $D_S$ and 
for every triplet $(\psi,\psi',S)$ the Yukawa coupling matrices are,
up to a common factor, identical with the complex-conjugate 
Clebsch--Gordan coefficient matrices.

For three fermion families the above-mentioned irreps $D$ and $D'$ can
only have the dimensions one, two or three.

One may wonder if, in the case of three families, one can confine the
discussion to finite subgroups of $U(3)$. Clearly, this class of
groups is very important in practice, consider e.g.\ the groups $A_4$,
$S_4$, $\Delta(27)$ etc. 
However, this is not the general case: There are models whose flavour
symmetry group cannot be conceived as a subgroup of $U(3)$. An
instance of such a case can be found in~\cite{GL07}. 
The essence of this model and its flavour group~$G$ is explained in
appendix~\ref{example}; the smallest
unitary group $U(n)$
of which $G$ can be
conceived as a subgroup is $U(6)$.

From the example in appendix~\ref{example} we infer the following
criterion that the symmetry group used for the construction of a model
can be conceived as a subgroup of $U(3)$. We distinguish two cases.
\begin{enumerate}
\renewcommand{\theenumi}{\alph{enumi}}
\item 
\textit{The symmetry group $G$ has a faithful three-dimensional
representation:} If we denote this representation by $D$, which can be
  reducible or irreducible, obviously $D$ has the 
  same information as the symmetry group $G$; thus we can replace $G$
  by $D$, with the matrices of $D$ being a subgroup of $U(3)$. 
\item 
\textit{The symmetry group $G$ has no faithful three-dimensional
representation:} 
In this case, we have to consider specifically the
representations $D_i$ of $G$ under which the field multiplets 
$q_L$, $\ell_L$, $d_R$, $u_R$, $e_R$, etc.\ 
of the model transform. Suppose one of the three-dimensional
representations, say $D_1$, has the property that its kernel
$K_1$ is contained in the kernels $K_i$ of all other representations $D_i$
($i>1$). With this requirement the $D_i$ with
$i > 1$ can be considered as representations of the group of unitary
$3 \times 3$ matrices $D_1$.
\end{enumerate}
If in a model with three fermion families the flavour symmetry group
$G$ does not fulfill the above criterion, then $G$ cannot be
conceived as a subgroup of $U(3)$. Obviously, the criterion could be
reformulated for any number of fermion families.

In general, every finite group $G$ can be considered as a subgroup
of a $U(n)$ with $n$ sufficiently large---see next section. 
Naturally, there will be a minimal $n$ where this is possible. 
This $n$ is simply the minimum of the dimensions of faithful
representations.  

In this review we have simply postulated the existence of finite
flavour symmetry groups. However, according to the discussion above, they
could originate from a $U(n)$ or $SU(n)$ by symmetry breaking. 
Recently, in~\cite{merle} this mechanism has been studied for the case
of $SU(3)$.

\section{Properties of finite groups}
\label{properties}
In this section we present a collection of properties of finite groups
that may satisfy the basic needs of model building in particle
physics---see also~\cite{Ramond,Hamermesh} for textbooks on group theory
from the physicist's point of view. Some theorems we have taken from
the book of Speiser~\cite{Speiser} which is a cornucopia of
information on finite groups. In general, since we focus on
applications, we do not present mathematical proofs but refer the
reader to the above-mentioned books and also to~\cite{Hall}.  

The striking feature of finite groups, which has no counterpart in
infinite groups, is that many of their 
properties are expressed in terms of the integers associated
with the group. Such integers are, for instance, the order of a group,
the number of conjugacy classes, the dimensions of its irreps, etc. 
Corresponding theorems will be emphasized in the following because, in
particular for small groups, these are quite useful.

Since in the present section we focus on a general discussion of finite
groups, we have deferred the discussion of classes of groups and specific
groups to the following sections. In this vein, also examples
illustrating definitions and theorems will be postponed.

\subsection{Generators and presentations}

A set of \emph{generators} or \emph{generating set} of a group $G$
is a subset $S$ of $G$ such that every 
element of $G$ can be written as a finite product of elements of $S$ and their
inverses. Note that in section~\ref{flavour symmetries} we have introduced
symmetries of the Lagrangian; these symmetries can be regarded as
representations of the set of group generators on the field multiplets. 
A group is called finitely generated if there is a finite set $S$ of
generators. Since we will be dealing with finite groups, all our groups will
be finitely generated.  

The precise definition of 
a \emph{presentation} of a group $G$ is complicated. Here it is sufficient to
have an intuitive understanding. A presentation consists of a set $S$ of
generators and a set $R$ of relations among the generators which completely
characterize the group. This means that writing strings of the generators and
using $R$ to shorten the strings one obtains all group elements.
We stress that a presentation of a group is by no means unique. It is
often useful to choose different presentations for different purposes.

The simplest example of a presentation is that of the cyclic group
$\zz_n$. It has one generator $a$ and one relation, $a^n = e$, which
completely characterizes the group.

We now want to prove a powerful theorem on the generating sets of a
group. For this proof we need two new notions, namely the normalizer
of a subset of a group and the conjugate subgroups of a group. In the
following we will use the notation $\oo G$ for the order of a finite
group $G$.

\begin{define}
Let~$M$ be a set of elements of a finite group~$G$. Then the set of
all elements~$a\in G$ for which 
	\be
	aMa^{-1}=M
	\ee
forms a subgroup of~$G$ which is called the \textit{normalizer} of~$M$
in~$G$. We will denote the normalizer of~$M$ in~$G$ by~$N_{G}(M)$. If
$M$ is a subgroup of~$G$ then it is also a subgroup of its
normalizer~$N_G(M)$. 
\end{define}

\begin{define}
Let~$S$ be a subgroup of a finite group~$G$. Then the groups
	\be
	gSg^{-1},\enspace g\in G
	\ee
are called the \textit{conjugate subgroups} of~$S$ in~$G$. All
conjugate subgroups $gSg^{-1}$ are isomorphic but not all of them need
to be equal.
\end{define}

\begin{theorem}\label{conjugatesubgroups}
The number of non-identical conjugate subgroups of a
subgroup~$S\subset G$ of a finite group~$G$ is given by 
	\be
	\frac{\mathrm{ord}\,G}{\mathrm{ord}\,N_{G}(S)}.
	\ee
\textit{Proof:} See e.g. \cite{Hall}, theorem 1.6.1 (p.14).
\end{theorem}
\noindent
Now we can prove the following theorem.
\begin{theorem}\label{generatortheorem}
Let~$G$ be a finite group with~$m$ conjugacy classes~$C_1,...,C_m$ and let
	\be
	M:=\{a_1,...,a_m\}
	\ee
be a subset of~$G$ such that~$a_i\in C_i$. Then~$M$ generates~$G$.
\medskip
\\
\textit{Proof:} Let $S$ denote the group generated by $M$. Since $S$
contains an element of every conjugacy class of $G$ we find 
	\be\label{unionofconjsubgroups}
	\bigcup_{g\in G} gSg^{-1}=G.
	\ee 
According to theorem~\ref{conjugatesubgroups} there are
	\be
	r:=\frac{\mathrm{ord}\,G}{\mathrm{ord}\,N_G(S)}
	\ee
different conjugate subgroups of $S$ in $G$. Since the unit element is
contained in every conjugate subgroup we find 
	\be\label{ordGinequality}
	\mathrm{ord}\,G=\mathrm{ord}\left( \bigcup_{g\in G} gSg^{-1}
        \right)\leq r\,\mathrm{ord}\,S-r+1. 
	\ee
Now we use that $S$ is a subgroup of its normalizer $N_G(S)$, thus
	\be
	\mathrm{ord}\,S\leq\oo N_G(S)\Rightarrow r\,\oo S\leq \mathrm{ord}\, G.
	\ee
Inserting this into (\ref{ordGinequality}) we find
	\be
	\mathrm{ord}\,G\leq r\,\mathrm{ord}\,S-r+1\leq
        \mathrm{ord}\,G-r+1\Rightarrow r\leq 1\Rightarrow r=1. 
	\ee
Thus $S$ itself is the only conjugate subgroup of $S$ in $G$, and, therefore,
from (\ref{unionofconjsubgroups}), we conclude $S=G$. \\ Q.E.D.
\end{theorem}

\begin{corollary}\label{generatorcorollary}
A subset~$X\subset G$ of a finite group $G$ is a generating set if and
only if in every conjugacy class of $G$ there exists an element which
can be expressed as a product of elements of $X$. 
\end{corollary}

\subsection{Subgroups and group structure}

\subsubsection{Subgroups}

\begin{define}
Let $G$ be a finite group and let $H\subset G$ be a subgroup of $G$. The sets
\begin{equation}
aH:=\{ah \vert h\in H\},\quad a\in G
\end{equation}
are called \textit{left cosets} of $H$ in $G$. 
Analogously, the sets 
\begin{equation}
Hb:=\{hb \vert h \in H\}, \quad b \in G
\end{equation}
are called \textit{right cosets}. 
\end{define}
\noindent
It is easy to see that two left or two right cosets are either identical or they
have no element in common. Since each coset has the same number of
elements which is identical with the order of $H$, we come to the
following conclusion.
\begin{theorem}
\textbf{(Lagrange)} The order of a subgroup of a finite group is a
divisor of the order $\oo G$ of the group. 
\end{theorem}
\noindent
If we choose any element $a$ of a finite group $G$, then the set 
$\{ e, a, a^2, a^3 , \ldots\}$ must also be finite. Therefore, there is a
smallest power $\nu > 0$ such that $a^\nu = e$. This power is called the
\emph{order of the group element} $a$. Every $a \in G$ is the generator of a
cyclic group $\zz_\nu$ which is a subgroup of $G$. Therefore, 
using the theorem of Lagrange we immediately draw the following conclusion.
\begin{theorem}
The order of a group element is a divisor of $\oo G$.
\end{theorem}
\noindent
All elements of a conjugacy class have the same order. 
Therefore, we can also speak of the \emph{order of a conjugacy class}.

Obviously, $H$ is an invariant (or normal) subgroup if and only if left and
right cosets are identical. 
If $H$ is a non-trivial normal subgroup of $G$, i.e. 
$H \neq \{e\}$ and $H \neq G$, we write $H \llhd G$. 
Another criterion for normal subgroups is the following.
\begin{theorem}
A subgroup $H$ of a finite group $G$ is invariant if and only
if $H$ consists of complete conjugacy classes of $G$.  
\\
\textit{Proof:} See e.g. \cite{Hamermesh} (p.28).
\end{theorem}
It is often tedious to compute the conjugacy classes. The following
theorem gives at least a clue how large conjugacy classes can be. 
\begin{theorem}
The number of elements residing in a conjugacy class of a finite group is
a divisor of the order of the group. 
\\
\textit{Proof:} See e.g. \cite{Hall}, corollary to theorem
1.6.1 (p.14); \cite{Speiser}, corollary~1 to theorem~62 (p.61). 
\end{theorem}

A factor group is, in a certain sense, the result of a division of a
group by a normal subgroup.
\begin{define}
Let $N$ be a normal subgroup of a finite group $G$. 
Then the cosets $aN = Na,\,a\in G$ together
with the multiplication law 
	\be
	(aN)(bN) = \{(an_1)(bn_2)\vert n_1,n_2\in N\}
= \{abn \vert n \in N \}
	\ee
form a group, the \textit{factor group} $G/N$.
\end{define}
\noindent 
Note that only a normal subgroup allows a definition of a
multiplication law on the cosets.

The following two theorems give a handle on how to obtain the conjugacy
classes of a group $G$ if the conjugacy classes of a proper normal
subgroup $N$ are known. For the proofs we refer the reader to~\cite{GL10}.
\begin{theorem}\label{cc1}
If $C_k$ is a conjugacy class of $N \llhd G$ and $b \in G$ 
but $b \not\in N$, then either $bC_kb^{-1} = C_k$ or 
the intersection between $bC_kb^{-1}$ and $C_k$ is empty.
\end{theorem}
\noindent
\begin{theorem}\label{cc2}
If $N \llhd G$ such that $G/N \cong \zz_n$ ($n \geq 2$) and 
$Nb$ is a generator of $G/N$, then every conjugacy class of $G$ 
can be written in the form $S b^\nu$
where $S$ is a subset of $N$ and $\nu \in \{0,\,1\, \ldots,n-1 \}$. 
The conjugacy classes of $G$ which are subsets of $N$ can be obtained
from the conjugacy classes of $N$ in the following way:
\begin{enumerate}
\renewcommand{\labelenumi}{\roman{enumi}.}
\item
$C_k$ is a conjugacy class of $N$ such that $bC_kb^{-1} = C_k$. In
  this case $C_k$ is also a conjugacy class of $G$.
\item
$C_k$ is a conjugacy class of $N$ with empty 
intersection between $bC_kb^{-1}$ and $C_k$. Then the 
conjugacy class of $G$ which contains $C_k$ is given by 
\begin{equation}
\overset{n-1}{\underset{\nu=0}{\bigcup}}\, b^\nu C_k b^{-\nu}.
\end{equation}
\end{enumerate}
\end{theorem}

\subsubsection{Direct and semidirect products of groups}
\label{GroupstructureI} 

\begin{define}\label{DirectProduct}
Let $G$ and $H$ be finite groups. Then the set $G\times H$ together
with the multiplication law 
	\be
	(g_1,h_1)(g_2,h_2):=(g_1g_2,h_1h_2)\quad g_1,g_2\in G,\,h_1,h_2\in H
	\ee
forms a group which is called the \textit{direct product} of $G$ and $H$. 
\end{define}
\noindent
Evidently, each factor in a direct product is a normal subgroup. 

Surprisingly, the definitions of direct products
and cyclic groups suffice to classify all finite Abelian groups due to
theorems~\ref{AbelianStructure1} and~\ref{AbelianStructure2}. 
\begin{theorem}\label{AbelianStructure1}\textbf{The structure of the
    finite Abelian groups.} A finite Abelian group $A$ of order
  $p_1^{a_1}p_2^{a_2}\cdots p_n^{a_n}$, where $p_1,\ldots,p_n$ are
  \textit{distinct} prime numbers, is a direct product of $n$ Abelian
  groups $A_i$  with $\mathrm{ord}\,A_i=p_{i}^{a_i}$, i.e. 
	\be
	A\cong\bigtimes_{i=1}^n A_i 
\quad \mbox{with} \quad
\mathrm{ord}\, A_i=p_i^{a_i}.
	\ee
\textit{Proof:} See e.g. \cite{Hall}, theorem 3.3.1 (p.40).
\end{theorem}
\begin{theorem}\label{AbelianStructure2}
Every Abelian group $A$ of prime power order $p^b$ is isomorphic to
a direct product of cyclic groups whose orders are powers of $p$:
	\be
	A\cong\bigtimes_{j} \zz_{p^{b_{j}}}\quad\mbox{where}\quad \sum_{j}b_{j}=b.
	\ee
\textit{Proof:} See e.g. \cite{Hall}, theorem 3.3.1 (p.40).
\end{theorem}
\noindent
While the decomposition of theorem~\ref{AbelianStructure1} only
depends on the decomposition of $\oo G$ into prime numbers, the
decomposition of the factors $A_i$ into cyclic groups depends on their 
group structure. 
For instance, if the order is $4 = 2^2$, there are
two possibilities: Klein's four-group $\zz_2 \times \zz_2$ and the
cyclic group $\zz_4$. Clearly, these two groups are different.

Direct products offer the simplest possibility for the construction of
larger groups from smaller ones. A more complicated but much more
versatile product is the \textit{semidirect product} of finite groups, 
which is a generalization of the direct product~\ref{DirectProduct}.
\begin{define}
Let $G$ and $H$ be finite groups and let 
$\phi: H \rightarrow \mathrm{Aut}(G)$ 
be a homomorphism from $H$ into the group of 
automorphisms on $G$, i.e. the isomorphisms $G \to G$. 
Then the set $G \times H$ together with
multiplication law 
	\be
	(g_1,h_1)(g_2,h_2):=(g_1\phi(h_1)g_2,h_1h_2)
	\ee
forms a group, the \textit{semidirect product} of $G$ and $H$,
denoted by $G \rtimes_\phi H$.  
\end{define}
\noindent
Note that, in this definition, $\phi(h_1) \in \mathrm{Aut}(G)$, i.e. 
$\phi(h_1)$ is an isomorphism acting on $G$, 
therefore $\phi(h_1)g_2 \in G$. One can also say that ``$H$ acts on $G$.''
The proof that $G \rtimes_\phi H$ fulfills the group axioms is tedious
but straightforward. Moreover, one can show the following property.
\begin{theorem}
$G \times \{ e' \}$ is a normal subgroup and 
$\{ e \} \times H$ a subgroup of $G \rtimes_\phi H$, where $e$ is the
unit element of $G$ and $e'$ the unit element of $H$.
\end{theorem}
\noindent
In the following we will usually drop $\phi$
in the symbol $\rtimes_{\phi}$, though there are instances where 
one can define more than one semidirect product $G \rtimes H$,
differing in the mapping $\phi$. 
Choosing $\phi(h)=\mathrm{id}_G$ $\forall h\in H$ one recovers the
direct product. 
We emphasize that a \emph{non-trivial} homomorphism 
$\phi: H \rightarrow \mathrm{Aut}(G)$ does not necessarily exist in
general. 

The following theorem summarizes the properties of a group $S$ 
that allows a decomposition into a semidirect product.
\begin{theorem}\label{semidp}
A finite group $S$ is isomorphic to the semidirect product
$G \rtimes_\phi H$ if and only if 
	\begin{enumerate}
	\item $G$ is a normal subgroup of $S$,
	\item $H$ is a subgroup of $S$,
	\item $\forall s\in S\,\,\exists g\in G,\,h\in H$ such that $s=gh$,
	\item $G\cap H=\{e\}$.
	\end{enumerate}
Then the decomposition $s=gh$ is unique, 
the isomorphism $S \to G \rtimes_\phi H$ is given by 
$gh \mapsto (g,h)$
and the homomorphism $\phi: H \rightarrow \mathrm{Aut}(G)$ is realized
as $\phi(h)g = hgh^{-1}$.\\ 
\textit{Proof:} 
See e.g.~\cite{GL10} and \cite{Hall}, theorems 6.5.2 and 6.5.3 (p.89).
\end{theorem}
\noindent
In essence the theorem follows from 
$(g_1h_1)(g_2h_2) = \left (g_1h_1 g_2 h_1^{-1} \right) (h_1h_2)$,
which is exactly the relation which motivates the definition of a semidirect
product. In theorem~\ref{semidp} the subgroup $H$ is actually
isomorphic to the factor group $S/G$ because with the requirements of
the theorem the mapping $h \leftrightarrow Gh$ is well-defined and
bijective. Thus the decomposition can also be written as 
$S \cong G \rtimes (S/G)$. In general, without the conditions~3 and~4 of
theorem~\ref{semidp}, one \emph{cannot} reconstruct the group $S$ from
$G$ and $S/G$---see also the remarks on composition series in
section~\ref{sub-normal}.

We want to finish the discussion of semidirect products with the illustrative
example of $\zz_n \rtimes \zz_m$---see~\cite{Ramond,bovier}. 
Suppose $\zz_n$ and $\zz_m$ are generated by $a$ and $b$, respectively.
Thus, $\phi(b)$ acts as an automorphism on $\zz_n$. Since this group is
cyclic, this action is necessarily of the form $a \mapsto a^r$ with some power
$r$. In the light of theorem~\ref{semidp} we consider both $a$ and $b$ as
an element of the same group, which allows to write
\begin{equation}
a^n = b^m = e, \quad \phi(b)a = b a b^{-1} = a^r.
\end{equation}
Moreover, by successive application of $\phi(b)$ we obtain
\begin{equation}
b^2 a b^{-2} = a^{r^2}, \quad b^3 a b^{-3} = a^{r^3}, \ldots,
\quad b^m a b^{-m} = a^{r^m} = a.
\end{equation}
Therefore, we find the consistency condition
\begin{equation}\label{r}
r^m = 1\,\,\mbox{mod}\, n.
\end{equation}
For a given pair of positive integers $n$ and $m$ there will, in general, be
several solutions for $r$ with $1 \leq r \leq n-1$, 
leading to distinct mappings $\phi$ and thus to
distinct semidirect products. 
Clearly, there is always the solution $r=1$ which corresponds to the direct
product. In the rest of the discussion we will skip this trivial case and look
for a solution in the range $2 \leq r \leq n-1$.
We consider first $\zz_3 \rtimes \zz_2$ leading to $r^2 = 1\,\,\mbox{mod}\,3$.
Here the only possibility is $r=2$, which is indeed a solution.
We will later see that $\zz_3 \rtimes \zz_2 \cong S_3$, where $S_3$ is
the
permutation group of three letters. A more complicated example is furnished
by $n=8$, $m=4$---see~\cite{Ramond}. Since $\phi$ is uniquely characterized by
the integer $r$, we write $\zz_8 \rtimes_r \zz_4$. In this case, the
semidirect product is indeed non-unique because $r = 3,5,7$ are all 
solutions of equation~(\ref{r}).

\subsubsection{Subgroups versus normal subgroups}
\label{sub-normal}
A group does not necessarily have non-trivial normal subgroups. 
This motivates the following definition.
\begin{define}
A finite group is called \textit{simple}, if it possesses no
non-trivial normal subgroup. 
\end{define}
\noindent
Important examples for simple groups are the 
cyclic groups $\zz_p$ of prime order $p$ 
and the alternating groups $A_n$ for $n>4$---for a proof see
e.g.~\cite{Speiser}, theorem~96 (p.110). 
The alternating groups furnish straightforward examples that 
theorems similar to~\ref{AbelianStructure1} 
and~\ref{AbelianStructure2}, with direct products replaced by 
semidirect products, will in general not hold for non-Abelian groups.
Consider $A_5$, which is simple though
its order $60=2^2\times 3\times 5$ is not a prime number; 
since it does not have any non-trivial normal subgroups, it 
can neither be a direct nor a semidirect product. 

Finite simple groups are often treated as the
``basic building blocks'' of all finite groups and their role in group
theory is sometimes compared with the role of the primes in number
theory. Unfortunately, this comparison is too farfetched. 
Though there are many non-simple groups which can be written as direct or
semidirect products of simple groups via theorem~\ref{semidp}, 
it is in general not possible to decompose a group which possesses a
normal subgroup---see discussion at the end of
section~\ref{GroupstructureI}. 

Consider for example the
cyclic group $\zz_4$. It is not simple, since it possess the normal
subgroup $\zz_2$, and it cannot be written as a direct or semidirect
product of any other groups. 

Though there is no analogon of theorem~\ref{AbelianStructure1} for
non-Abelian groups, there is, at least, a connection between the
decomposition of the group order into prime numbers and the existence
of subgroups with corresponding order.
\begin{theorem}\label{smallsylow}
A finite group $G$ of order
  $p_1^{a_1}p_2^{a_2}\cdots p_n^{a_n}$, where $p_1,\ldots,p_n$ are
  \textit{distinct} prime numbers, 
possesses subgroups of all orders $p_i^{s_i}$ with
$0 \leq s_i \leq a_i$ ($i = 1, \ldots, n$). 
The subgroups of order $p_i^{a_i}$ are called the
\textit{Sylow subgroups} of $G$ associated to the prime $p_i$. All Sylow
subgroups associated with the same prime number $p_i$ are equivalent,
i.e. for any two $p_i$-Sylow groups $S$, $S'$ exists an element 
$a \in G$ such that $S' = a S a^{-1}$. \\
\textit{Proof:} See e.g. \cite{Hall}, theorem 4.2.1 (p.44). 
\end{theorem}
\noindent
Theorem~\ref{smallsylow} contains a part of the famous Sylow theorems 
which can be found in many textbooks on group theory, 
e.g. in~\cite{Ramond} (p.27). From this theorem we learn that all
groups, except cyclic groups $\zz_p$ of prime order $p$, have
non-trivial subgroups. 
Therefore, all simple groups, except $\zz_p$, have subgroups
though no non-trivial normal ones. 
Moreover, groups with $\oo G$ as given in
theorem~\ref{smallsylow} have elements of order $p_1, \ldots, p_n$.
This statement is called Cauchy's theorem.

Though in general one cannot decompose a non-Abelian group into 
factors by direct or semidirect products, it is 
nevertheless helpful for the characterization and understanding of the
structure of a group if it possesses normal subgroups.
This idea leads to the concepts of composition and principal series.

A finite group $G$ is either simple or it has at least one non-trivial
maximal normal subgroup $N$. The group $N$ can itself be simple, or it
possesses a non-trivial normal subgroup. If we carry on searching for
maximal normal subgroups we will at some point end up with a simple
group which does not have any non-trivial normal subgroups. In this
way we obtain a series of maximal normal subgroups which is called a
\textit{composition series} of $G$. 
\begin{define}
A composition series of a finite group $G$ is a series of subgroups 
	\be
	\{e\} \llhd N_1 \llhd N_2 \llhd \cdots \llhd N_m=G
	\ee
such that each $N_i$ is a maximal normal subgroup of $N_{i+1}$. Due to
maximality every factor group $N_{i+1}/N_i$ is simple. The simple
factor groups $N_{i+1}/N_i$ are also called 
\textit{prime factor groups} of $G$. 
\end{define}
\noindent
In this way we arrive at a chain of simple
groups, namely the prime factor groups $N_{i+1}/N_i$. Of course this
concept only makes sense if for a given group the prime factor
groups are uniquely determined. The following theorem
ensures that this is the case. 
\begin{theorem}\label{Jordan-Hoelder}
\textbf{(Jordan--H\"older)} For any two compositions series
	\[
	\begin{split}
	& \{e\}\llhd A_1\llhd A_2\llhd \cdots \llhd A_m,\\
	& \{e\}\llhd B_1\llhd B_2\llhd \cdots \llhd B_n
	\end{split}
	\]
of a finite group, one necessarily has $n=m$ 
and for each prime factor group $A_{i+1}/A_{i}$
there is an isomorphic prime factor group $B_{j+1}/B_j$. 
In a nutshell, 
any two composition series of a finite group have the same
length and the prime factor groups are isomorphic up to ordering. 
\\
\textit{Proof:} See e.g. \cite{Hall}, theorem 8.4.4 (p.126).
\end{theorem}
\noindent
Note that the Jordan--H\"older theorem does not tell us that the
composition series uniquely defines the group. The simplest example
for two non-isomorphic groups with the same composition series is
provided by $\zz_4$ and $\zz_2\times\zz_2$. The two composition
series are 
	\be\label{zzzz}
	\{e\}\llhd \zz_2\llhd\zz_4\quad\mbox{ and }\quad
        \{e\}\llhd\zz_2\llhd \zz_2\times\zz_2,
	\ee
with the prime factor groups being two copies of $\zz_2$ for each case. 
This example also shows that in general the 
\emph{group extension problem} has no unique solution. 
This problem can be formulated in the following way: Given two
groups $N$ and $F$, what are the finite groups $G$ which possess $N$
as a normal subgroup such that $G/N\cong F$?
In the example above we have $N = F = \zz_2$, with two solutions: 
the trivial one, which is the direct product 
$G = N \times F = \zz_2 \times \zz_2$, 
and the non-trivial one $G = \zz_4$. 
For a treatment of group extensions we refer the reader to \cite{Hall}. 

If a group possesses many normal subgroups, then its so-called
principal series might be quite useful to find its irreps~\cite{GL10}.
It has the following definition.
\begin{define}\label{princ}
The principal series of a group is a maximal series of normal subgroups
	\be
	\{e\} \llhd N_1 \llhd N_2 \llhd \cdots \llhd N_m=G
	\ee
such that
	\be
	N_i\llhd N_j \quad\forall\,i<j.
	\ee
By ``maximal series'' it is meant that there is no group $N$ that fits
into the series such that the above conditions are still fulfilled. 
\end{define}
\noindent 
Note that in the case of the principal series, $N_{i+1}/N_i$ is not
necessarily simple, i.e. the normal subgroups do not need to be
maximal. Also for the principal series a Jordan--H\"older theorem
exists.  

\subsection{Representation theory}
\subsubsection{Fundamentals}
The most important group-theoretical application to physics is the
theory of group representations. If not stated otherwise, we always
have in mind representations on complex linear spaces. The following
theorem tells us that for finite groups, without loss of generality,
we can confine ourselves to unitary representation matrices.
\begin{theorem}\label{unitaryrep}
Every finite-dimensional 
representation $D$ of a finite group $G$ is equivalent to a
unitary representation, i.e. 
	\be
	\exists\,S:\, S^{-1}DS=D'\mbox{ with }
        D'(a)^{\dagger}=D'(a)^{-1}\enspace\forall\, a\in G. 
	\ee
\end{theorem}

As a mathematical tool for finite groups, the regular representation is of
eminent importance.
\begin{define}
The \textit{regular representation} $\mathcal{R}$ of a finite group 
$G=\{a_1,...,a_m\}$ with $\oo G = m$ is defined as
	\be
	\mathcal{R}(a_i) a_j = a_i a_j= \sum_{k=1}^m \mathcal{R}_{kj}(a_i)a_k.
	\ee
\end{define}
\noindent
By definition, $\mathcal{R}$ is a faithful $m$-dimensional 
representation of $G$. Therefore, we find the following existence 
theorem.
\begin{theorem}\label{repexistence}
Every finite group has faithful finite-dimensional representations.
\end{theorem}
\noindent
Moreover, the $m \times m$ matrices $(\mathcal{R}_{kj}(a))$ are
permutation matrices, i.e.\ orthogonal matrices such that every entry
is either zero or one. Therefore, every finite group $G$ of order $m$ can be
considered as a subgroup of the symmetric group $S_m$, 
though in practice this knowledge is often of little value.

The regular representation has the following very important property.
\begin{theorem}
The regular representation of a finite group $G$ contains each of 
its inequivalent irreducible representations $D^{(\alpha)}$ with the
multiplicity of its dimension, 
i.e. 
	\be
	\mathcal{R}=\bigoplus_{\alpha}n_{\alpha}D^{(\alpha)}
	\ee
with $\mathrm{dim}\s D^{(\alpha)}=n_{\alpha}$. \\
\textit{Proof:} See e.g. \cite{Ramond} (p.38). 
\end{theorem}
\noindent
An interesting question is whether one of the irreps of a finite group is 
necessarily faithful. The answer is no: there are finite groups which
\textit{do not possess any faithful irreducible representation}. In
section~\ref{flavour symmetries} we have already mentioned such a
group. However, the simplest example for a finite group of that kind 
is Klein's four-group $K \cong \zz_2\times \zz_2$. 
Denoting the two generators of the group by $a$ and $b$, 
the four irreducible representations of $K$ are given by  
	\be
	\one^{(p,q)}:\, a\mapsto (-1)^p,\,b\mapsto (-1)^q,\quad p,q=0,1.	
	\ee
None of these irreducible representations is faithful, but the
reducible representation 
	\be
	\one^{(0,1)} \oplus \one^{(1,0)}:\, a \mapsto
	\bmat
	1 & \hphantom{-}0\\
	0 & -1
	\emat,\,
	b\mapsto
	\bmat
	-1 & 0\\
	\hphantom{-}0 & 1
	\emat
	\ee
is a faithful one. 

Now we collect some fundamental properties of irreps.
Theorems on the number and dimensions of irreducible
representations are particularly helpful in applications of group
theory. In particular, the first of the theorems is fundamental. 
\begin{theorem}\label{conjclassesirreps}
The number of irreducible representations of a finite group is equal
to the number of its conjugacy classes. 
\\
\textit{Proof:} See e.g. \cite{Ramond} (p.41).
\end{theorem}
\begin{theorem}\label{go}
The dimension of an irreducible representation of a finite group is a
divisor of the order of the group. 
\\
\textit{Proof:} See e.g. \cite{Hall}, theorem 16.8.4 (p.288);
\cite{Speiser}, theorem 155 (p.177). 
\end{theorem}
\begin{theorem}\label{n2}
Let $D^{(\alpha)}$ with $\mathrm{dim}\,D^{(\alpha)}=n_{\alpha}$ denote
the inequivalent irreps of a finite group
$G$. Then 
	\be
	\sum_{\alpha}n_{\alpha}^2=\mathrm{ord}\,G.
	\ee
\\
\textit{Proof:} See e.g.~\cite{Ramond} (p.39).
\end{theorem}
\begin{lemma} \textbf{(Schur)} 
This lemma consists of two parts.
\begin{enumerate}
 \item
Let $D^{(1)}$ and $D^{(2)}$ be finite-dimensional irreps 
of a finite group $G$ on the linear spaces $\mathcal{V}_1$ and
$\mathcal{V}_2$, respectively, 
and let $S: \mathcal{V}_1 \to \mathcal{V}_2$ be a linear operator such that 
	\be
	D^{(2)}(a)S = SD^{(1)}(a) \quad \forall a \in G.
	\ee 
Then, $S$ is either zero or invertible; in the second case the
two irreps are equivalent.
 \item 
If $D$ is an irrep on $\mathcal{V}$ and $S \neq 0$ a linear operator on
$\mathcal{V}$ with $D(a) S = SD(a)$ $\forall a \in G$, then
$\exists\,\lambda\in\mathbbm{C}\backslash \{0\}$ such that 
$S=\lambda\,\mathrm{id}$; i.e. operators which
commute with all $D(a)$ must be
proportional to the identity. 
\end{enumerate}
\textit{Proof:} See e.g.~\cite{Sternberg} (p.55).
\end{lemma}
\noindent
Note that in the first part of Schur's lemma the vector spaces can be
real or complex, whereas in the second part a complex field is
essential. A consequence of the second part of Schur's lemma is that
all irreps of an Abelian group are one-dimensional.

\subsubsection{Characters}
\begin{define}
Let $D$ be a finite-dimensional representation of a group
$G$. The character $\chi_D: G\rightarrow \mathbbm{C}$ is defined by 
	\be
	\chi_D(a):=\mathrm{Tr}\,D(a),\quad a \in G.
	\ee
\end{define}
\noindent
Thus the characters of a group are the traces of its
representation operators. Using the group axioms and the properties of the
trace it is straightforward to show the following properties.
\begin{theorem}\label{characterproperties} 
\textbf{Properties of characters:} 
	\begin{enumerate}
	 \item Equivalent representations have the same characters,
           i.e. $D \cong D'\Rightarrow \chi_{D}=\chi_{D'}$, 
	 \item the value of a character $\chi_D$ is the same on
           conjugate group elements, i.e. $a = g^{-1} b g$
           $\Rightarrow$ $\chi_D(a)=\chi_D(b)$, 
	 \item $\chi_D(a^{-1})=\chi_{D}^{\ast}(a)\enspace\forall a\in G$,
	 \item $\chi_{D\oplus D'}(a)=\chi_D(a)+\chi_{D'}(a)\enspace
           \forall a\in G$, 
	 \item $\chi_{D\otimes
           D'}(a)=\chi_D(a)\s\chi_{D'}(a)\enspace\forall a\in G$. 
	\end{enumerate}
\end{theorem}
\noindent
Point~2 says that characters are class functions, i.e. they have the
same value for all elements in a class. Therefore, we completely know
the character $\chi_D$ if we know its value $\chi_{Di}$ on every class
$C_i$. 

The possible values of the characters
are quite restricted. On the one hand, for finite groups 
we have the orthogonality relations, 
which will be presented in section~\ref{orthogonality}, on
the other hand there are inequalities on the absolute values of the
characters. According to theorem~\ref{unitaryrep}, any operator or matrix
$D(a)$ in a representation can be assumed to be unitary. Therefore,
$D(a)$ is diagonalizable with eigenvalues on the unit circle in the
complex plane. Since the character $\chi_{D}(a)$ is the sum of these
eigenvalues, one obtains
	\be
	\vert \chi_D(a)\vert\leq\mathrm{dim}\,D\quad\forall a\in G.
	\ee
Moreover, the trace of an $n\times n$ unitary matrix $U$ is $n$ if and
only if $U=\bone_n$. Therefore, we find 
	\be\label{chidim}
	\chi_{D}(a) = \mathrm{dim}\,D 
        \quad \Leftrightarrow \quad D(a)=\bone_{\mathrm{dim}\,D}. 
	\ee
Another useful inequality takes advantage of the fact that every
$D(a)$ has a finite 
order $m$, and therefore the eigenvalues of $D(a)$ are $m$-th roots of
unity. The character $\chi_D(a)$ is thus a sum of $m$-th roots of
unity. Plotting the sums of the $m$-th roots of unity for
$m=2,3,4,6$ in the complex plane one arrives at the following statement.
\begin{theorem}
Let $D$ be a representation of a finite group $G$ and let $a\in G$
such that $\mathrm{ord}\,D(a)\in\{2,3,4,6\}$. Then 
	\be\label{Characterinequality}
	\chi_{D}(a)=0 \enspace\mbox{  or  }\enspace \vert\chi_D(a)\vert\geq 1.
	\ee
\end{theorem}
\noindent
Note that this result is independent of the dimension of the
representation and it holds for both irreducible and reducible
representations. Another interesting theorem on the values
of characters, $\chi^{(\alpha)}_i$, on the class $C_i$ 
is the following. 
\begin{theorem}
Let $G$ be a finite group with an irrep $D^{(\alpha)}$. If
the number of elements $c_i$ of the conjugacy class $C_i$ and 
$\dim D^{(\alpha)}$ have no common divisor then 
	\be
	\chi^{(\alpha)}_i = 0 \quad \mbox{or} \quad c_i = 1.
	\ee
\\
\textit{Proof:} See e.g. \cite{Speiser}, theorem 164 (p.190).
\end{theorem}

We finish this section with a consideration of real and complex
irreps. If the character of an irrep $D^{(\alpha)}$ has complex values
on some classes, then by complex conjugation of the irrep we obtain a
distinct irrep; in this case $D^{(\alpha)}$ is called \emph{complex}. 
If the characters are real, there are two
possibilities. Either there is a basis in which all representation
matrices are real, then the irrep is called \emph{real}. If such a
basis does not exist, it is called \emph{pseudoreal}.
With the character $\chi^{(\alpha)}$ of the irrep $D^{(\alpha)}$, one
can establish the following criterion~\cite{Ramond}.
\begin{theorem}
The character $\chi^{(\alpha)}$ of an irrep $D^{(\alpha)}$ of a finite
group $G$ has the following property:
\begin{equation}
\label{psrc}
\frac{1}{\oo G}\,
\sum_{a \in G} \chi^{(\alpha)} \left( a^2 \right) = 
\left\{ \begin{array}{rcl} 
+1 & \Rightarrow & \mbox{real}, \\
-1 & \Rightarrow & \mbox{pseudoreal}, \\
 0 & \Rightarrow & \mbox{complex}.
\end{array} \right.
\end{equation}
\end{theorem}
\noindent
For a real or pseudoreal irrep given by the unitary matrices $D^{(\alpha)}(a)$
there exists a matrix $S$ such that
$S^{-1} D^{(\alpha)}(a) S = \left( D^{(\alpha)}(a) \right)^*$. 
It is easy to show that the matrix $S$ fulfills 
$S^T = \pm S$~\cite{Ramond}, where the plus sign refers to a real
and the minus sign to a pseudoreal irrep. We conclude that pseudoreal irreps
have even dimension.

\subsubsection{Orthogonality relations and character tables}
\label{orthogonality}
For finite groups there are orthogonality
relations for group representations and characters. In order to
formulate these in a convenient way one defines a bilinear 
form on the space of functions $G\rightarrow \mathbbm{C}$. 
\begin{define}
Let $f$ and $g$ be functions $G\rightarrow\mathbbm{C}$ defined on a
finite group $G$. Then we define a bilinear form on these functions via 
	\be
	\langle f\vert g\rangle:=\frac{1}{\mathrm{ord}\,G}\sum_{a\in
          G}f(a^{-1})g(a). 
	\ee
\end{define}
\noindent
Evidently, this bilinear form is symmetric, i.e. 
$\langle f \vert g \rangle = \langle g \vert f \rangle$. 
Moreover, on the real vector space of functions which fulfill 
$f(a^{-1}) = f^*(a)$ $\forall a \in G$ the bilinear form 
$\langle f\vert g\rangle$ assumes only real values and is, therefore, a scalar
product. 
\begin{theorem}
\textbf{Orthogonality relations.} Let $D^{(\alpha)}$ with 
$\mathrm{dim}\,D^{(\alpha)}=n_{\alpha}$ denote the inequivalent
irreducible representations of a finite group $G$ and let
$\chi^{(\alpha)}:=\mathrm{Tr}\,D^{(\alpha)}$, where the index $\alpha$ labels
the inequivalent irreps. Then the following orthogonality relations hold: 
	\be\label{o1}
	\langle (D^{(\alpha)})_{ij} \vert (D^{(\beta)})_{kl} \rangle = 
        \frac{1}{n_{\alpha}}\delta_{\alpha\beta} \delta_{il}\delta_{jk}. 
	\ee
The corresponding relations for the characters are 
	\be\label{characterorthogonality1}
	\langle\chi^{(\alpha)}\vert\chi^{(\beta)}\rangle=\delta_{\alpha\beta}.
	\ee
\textit{Proof:} See e.g.~\cite{Ramond} (p.36).
\end{theorem}
\noindent
Some remarks are at order. 
In equation~(\ref{o1}) the objects on the left-hand side are
representation matrices. The switching between linear operators $D(a)$
($a \in G$) of the representation and representation matrices is carried
out in the usual way by choosing a basis $\{ b_k \}$ in the linear
space. Then the action of $D(a)$ on the basis gives the corresponding
matrix via 
\begin{equation}
D(a) b_k = \sum_l \left( D(a) \right)_{lk} b_l.
\end{equation}
Equation~(\ref{o1}) is true in any basis. In particular, the basis
is not required to be orthonormal. 
Concerning equation~(\ref{characterorthogonality1}), we can
reformulate it by using the values of the characters on the classes,
$\chi^{(\alpha)}_i$, and the numbers $c_i$ of elements in the class
$C_i$. Then this equation reads 
\begin{equation}\label{o2}
\sum_i c_i \left( \chi^{(\alpha)}_i \right)^* \chi^{(\beta)}_i =
\oo G \,\delta_{\alpha\beta}. 
\end{equation}

We can specify 
all characters of a finite group by listing the values of the
characters on the different conjugacy classes. Since the number of
conjugacy classes equals the number of inequivalent irreducible
representations---see theorem~\ref{conjclassesirreps}---this
leads to a quadratic scheme, the so-called \textit{character table} of
the group. A schematic description of a character table of a finite
group is depicted in table~\ref{ct}. It contains $n$ lines and $n$ columns
where $n$ is the number of classes. 
Note that to the character table usually two further
lines are added---see table~\ref{ct}---which provide information on the
number of elements in a class and the order of these elements. 
\begin{table}
\begin{center}
\begin{tabular}{|c|cccc|}\hline
$G$        & $C_1$   & $C_2$   & $\cdots$ & $C_n$ \\
(\# $C$) & $(c_1)$ & $(c_2)$ & $\cdots$ & $(c_n)$ \\ 
$\oo(C)$ & $\nu_1$ & $\nu_2$ & $\cdots$ & $\nu_n$ \\ \hline
\rule{0pt}{5mm}
$D^{(1)}$  & $\chi^{(1)}_1$ & $\chi^{(1)}_2$ & $\cdots$ & $\chi^{(1)}_n$ \\
$D^{(2)}$   & $\chi^{(2)}_1$ & $\chi^{(2)}_2$ & $\cdots$ & $\chi^{(2)}_n$ \\
$\vdots$ & $\vdots$ & $\vdots$ & $\vdots$ & $\vdots$ \\
$D^{(n)}$ & $\chi^{(n)}_1$ & $\chi^{(n)}_2$ & $\cdots$ & $\chi^{(n)}_n$ \\ 
\hline
\end{tabular}
\end{center}
\caption{Schematic description of a character table. In the first line, after
  the name of the group $G$, the classes are listed, below each class $C_k$
  its number of elements $c_k$ can be found, and in the second line
  below the class the 
  order $\nu_k$ of its elements is stated. \label{ct}}
\end{table}
It is customary to set $C_1 = \{ e \}$, thus in the first column the
dimensions $n_\alpha$ of the irreps are read
off because $\chi^{(\alpha)}(e) = n_\alpha$. 
Furthermore, the usual convention is that the first irrep is
the trivial irrep $a \mapsto 1$ $\forall a \in G$. 
Therefore, the first line has 1 in every entry.
Finally, irreps in a character table are ordered according to
increasing dimensions $n_\alpha$.

From equation~(\ref{o2}) we know that the line vectors 
\begin{equation} 
\left( \sqrt{\frac{c_1}{\mbox{ord}\, G}} \,
\chi^{(\alpha)}_1, \ldots,
\sqrt{\frac{c_n}{\mbox{ord}\, G}} \, \chi^{(\alpha)}_n
\right),
\end{equation}
form an orthonormal basis of $\mathbbm{C}^n$. Consequently, also 
the column vectors
\begin{equation}
\sqrt{\frac{c_k}{\oo G}} \left( \begin{array}{c}
\chi^{(1)}_k \\ \vdots \\ \chi^{(n)}_k 
\end{array} \right) 
\quad (k=1,\ldots,n)
\end{equation}
define an orthonormal basis whose orthonormality conditions can be written as
\begin{equation}\label{charo1}
\sum_\alpha \left( \chi^{(\alpha)}_k \right)^*
\chi^{(\alpha)}_\ell = \frac{\oo G}{c_k}\, \delta_{k\ell}.
\end{equation}
The orthogonality relations~(\ref{o2}) and
(\ref{charo1}) are very useful for the construction of the cha\-racter table.

The benefit of character tables shows up when one wants to decompose a
reducible representation into its irreducible constituents.
Let us assume that a representation $D$ of a group $G$ is given. 
For a finite group any representation can be written as a sum
over its irreducible constituents, i.e.
$D = \bigoplus_{\alpha} m_\alpha D^{(\alpha)}$ where the $m_\alpha$ 
denote the multiplicities with which the irreps $D^{(\alpha)}$ occur in $D$.
Therefore, the character $\chi_D$ of a reducible representation is the
corresponding sum
\begin{equation}
\chi_D = \sum_\alpha m_\alpha \chi^{(\alpha)}.
\end{equation}
Then, with the orthogonality relation~(\ref{characterorthogonality1}) we find 
\begin{equation}\label{ma}
m_\alpha = \langle \chi^{(\alpha)} | \chi_D \rangle
\end{equation}
and 
\begin{equation}
\langle \chi_D | \chi_D \rangle = \sum_\alpha m_\alpha^2.
\end{equation}
The latter relation yields the following theorem.
\begin{theorem}\label{irred}
A necessary and sufficient condition for a representation $D$ to be
irreducible is $\langle\chi_{D}\vert\chi_{D}\rangle = 1$. 
\end{theorem}
\noindent
Relation~(\ref{ma}) is particularly useful for tensor products because
the character 
of the tensor product $D^{(\alpha)} \otimes D^{(\beta)}$ is given by the
product of the characters of $D^{(\alpha)}$ and $D^{(\beta)}$:
\begin{equation}
\chi^{(\alpha \otimes \beta)}(a) = 
\chi^{(\alpha)}(a) \times \chi^{(\beta)}(a).
\end{equation}
Consequently, the multiplicity $m_\gamma$ of an irrep $D^{(\gamma)}$
in the tensor product is given by 
\begin{equation}\label{mg}
m_\gamma = 
\langle \chi^{(\gamma)} | \chi^{(\alpha)} \times \chi^{(\beta)} \rangle.
\end{equation}

From the character table one can also read off if an irrep is
faithful. According to equation~(\ref{chidim}), the kernel of an irrep
$D^{(\alpha)}$ with dimension $n_\alpha$ 
consists of all classes $C_i$ with $\chi^{(\alpha)}_i = n_\alpha$.
Thus an irrep is faithful if and only if in the corresponding line 
in the character table the number $n_\alpha$ occurs only once, namely
in the column corresponding to $C_1 = \{ e \}$.

\subsubsection{Some remarks on the construction of irreducible
  representations} 
\label{remarks}

To find all irreps of a given finite group can be a formidable task,
in particular, if the group is large. 
However, for small groups it is often possible to
find all irreps by some straightforward procedures starting from a
known faithful irrep. Such an irrep is usually provided by the group
itself if it is a matrix group.

Let us assume now that this is the case, i.e. the group $G$ consists
of an irreducible set of $d \times d$ matrices $a$.  
Then, departing from the irrep $a \mapsto a$ we obtain immediately the
following further irreps:
\begin{itemize}
\item the complex conjugate irrep $a \mapsto a^*$, 
\item the one-dimensional irreps $a \mapsto (\det a)^k$ with 
      $k=0,\pm 1, \pm 2,\ldots$,
\item the $d$-dimensional irreps 
      $a \mapsto (\det a)^k \, a$ and 
      $a \mapsto (\det a)^k \, a^*$.
\end{itemize}
For a specific group, each of these procedures above could lead to equivalent
irreps and may thus be useless. 
For instance, for subgroups of $SU(n)$ the representation 
$a \mapsto \det a$ will always be the trivial one.
In any case, with our requirements on the group $G$
we know at least two inequivalent irreps: 
the defining irrep $a \mapsto a$ and 
the trivial irrep $a \mapsto 1$.
In this context theorem~\ref{n2} is very important and useful, in
particular, if we know the number of classes and, therefore, the
number of inequivalent irreps: It gives a handle on the dimensions of
missing irreps in conjunction with theorem~\ref{go}; sometimes it even 
allows to determine the dimensions of the missing irreps; if the
number of classes is not known it tells us when the task of finding
all inequivalent irreps is completed.

The procedures mentioned above can be generalized in a
straightforward manner. 
Often one has the defining irrep of a matrix group at disposal
and some of the one-dimensional irreps $\one_i$. 
Denoting the defining irrep with dimension $d$ by $\mathbf{d}_\mathrm{def}$,
then an obvious first attempt, which is often successful, to find
further $d$-dimensional irreps is to consider
\begin{equation}\label{1xd}
\one_i \otimes \mathbf{d}_\mathrm{def} \quad \mbox{and} \quad 
\one_i \otimes \mathbf{d}_\mathrm{def}^*.
\end{equation}

Note that, trivially, every one-dimensional representation of a group $G$ is
irreducible. If the symbol $\chi^{(\one_i)}$ denotes the character
of a one-dimensional irrep $\one_i$, the values of the character,
$\chi^{(\one_i)}(a)$, are obviously identical
with the $1 \times 1$ representation matrices. 
If the presentation of a group is available, the usage of the
presentation is the quickest way to discover \emph{all} one-dimensional
irreps~\cite{Ramond}. If the presentation is not available but one
knows a couple of one-dimensional irreps 
$\one_i$, $\one_j$,\ldots, one can always try to
obtain additional ones via the procedures
\begin{equation}
a \mapsto \left( \chi^{(\one_i)}(a) \right)^k, \quad
a \mapsto \left( \chi^{(\one_i)}(a) \right)^k \left( \chi^{(\one_j)}(a)
\right)^l, \ldots, 
\end{equation}
where $k$, $l$,\ldots are integers.

Other irreps may be obtained by reduction of
tensor products of the known irreps via the \textit{Clebsch--Gordan series}
\begin{equation}
D^{(\alpha)} \otimes D^{(\beta)} = \bigoplus_\gamma m_\gamma D^{(\gamma)}.
\end{equation}
The elements of the matrices which perform the basis transformation
from the tensor product basis to the basis in which the tensor product
decays into its irreducible constituents are called
\textit{Clebsch--Gordan coefficients}. Of course, 
if the character table of the group
is already known one can use equation~(\ref{mg}) 
to find the multiplicities $m_\gamma$.

Frequently one has to deal with the tensor product of an irrep
$D$ of dimension $n$ with itself. Such a tensor product is always reducible, because
the \textit{antisymmetric subspace} spanned by\footnote{Here we have denoted
the Cartesian basis vectors in
$\mathbbm{C}^n$ by $e_k$ with $k=1,...,n$.}
\begin{equation}
\frac{1}{\sqrt{2}}(e_i \otimes e_j - e_j \otimes e_i), 
\quad i,j=1,\ldots,n \quad (i < j)
\end{equation}
and the \textit{symmetric subspace} spanned by
\begin{equation}
\frac{1}{\sqrt{2}}(e_i \otimes e_j + e_j \otimes e_i),
\quad i,j=1, \ldots, n \quad (i < j)
\quad \mbox{and} \quad e_k \otimes e_k, \quad k = 1, \ldots, n
\end{equation}
are invariant under the action of $D\otimes D$.
 The dimensions of these two subspaces
are $n(n-1)/2$ and $n(n+1)/2$, respectively.
Another important case is the tensor product $D\otimes (D^{-1})^T$ for which
\begin{equation}
\frac{1}{\sqrt{n}}\sum_{j=1}^n e_j\otimes e_j
\end{equation}
spans an invariant subspace. Evidently the corresponding irrep is the trivial representation.
Note that for unitary representations $(D^{-1})^T = D^{\ast}$. In the following we will always assume
$D$ to be unitary.

Since there are three generations of fermions, three-dimensional
irreps play a prominent role. Three-dimensional irreps are either real or
complex, but not pseudoreal. We first consider the case that the
matrices $a$ form a \emph{real} irrep $\three$.
For real matrices $a$ 
we have $\three \otimes \three = \three \otimes \three^*$. According to
the above discussion we know three invariant subspaces, namely the 
subspace spanned by
\begin{equation}\label{b1}
\frac{1}{\sqrt{3}} 
\left( e_1 \otimes e_1 + e_2 \otimes e_2 + e_3 \otimes e_3
\right),
\end{equation}
the antisymmetric subspace spanned by
\begin{equation}\label{b3}
\frac{1}{\sqrt{2}} \left( e_2 \otimes e_3 -  e_3 \otimes e_2 \right),
\quad
\frac{1}{\sqrt{2}} \left( e_3 \otimes e_1 -  e_1 \otimes e_3 \right),
\quad
\frac{1}{\sqrt{2}} \left( e_1 \otimes e_2 -  e_2 \otimes e_1 \right),
\end{equation}
and the symmetric subspace. Since the vector~(\ref{b1})
is contained in the symmetric subspace, its orthogonal complement
in the symmetric subspace, which is the five-dimensional space
of traceless symmetric tensors, must be invariant too. Thus we have
arrived at the decomposition
\begin{equation}\label{33=136}
\three \otimes \three = \one \oplus \three' \oplus \five.
\end{equation}
Clearly, the $\one$ and the $\three'$ on the right-hand side of
equation~(\ref{33=136}) are irreps and given by 
\begin{equation}
\one: \; a \mapsto 1, \quad 
\three': \; a \mapsto (\det a)\, a,
\end{equation}
respectively. 
Any new irrep in $\three \otimes \three$, 
not yet obtained from the $\three$ by the procedures discussed above,
must reside in the five-dimensional space. If
this space is irreducible under the action of $G$, we have found a
five-dimensional irrep in this way. Or else we 
possibly obtain new irreps by reduction of this space. 

For complex three-dimensional irreps we can proceed in a similar manner
and we obtain
\begin{equation}
\three \otimes \three = \three'' \oplus \six, \quad
\three \otimes \three^{\ast} = \one \oplus \eight.
\end{equation}
The bases for the one- and three-dimensional spaces on the right-hand side of
these relations are again given by equations~(\ref{b1}) and~(\ref{b3}),
respectively. The respective irreps are
\begin{equation}\label{3''}
\one: \; a \mapsto 1, \quad 
\three'': \; a \mapsto (\det a)\, a^*.
\end{equation}
New irreps might be obtained by reduction of the six- and
eight-dimensio\-nal spaces. The six-dimensional space corresponds to symmetric
tensors. A standard construction of the basis of the eight-dimensional space
is given by
\begin{equation}
\frac{1}{\sqrt{2}} \sum_{i,j=1}^3 \lambda^a_{ij}\, e_i \otimes e_j
\quad \mbox{with} \quad a = 1,\ldots,8
\end{equation}
where the $\lambda^a$ are the Gell-Mann matrices.

Now we move on to some more theoretical means for obtaining irreps. 
Suppose there are two groups $G$ and $H$ with a 
homomorphism $\varphi: G \rightarrow H$ such that $\varphi(G) = H$. 
Then the kernel of $\varphi$, 
$\mbox{ker}\, \varphi = \{ a \in G| \varphi(a) = e_H \}$, is a normal
subgroup of $G$ and $\oo \mbox{ker}\, \varphi \times \oo H = \oo G$.
If we have an
irrep $D^{(\alpha)}$ of $H$, then $D^{(\alpha)} \circ \varphi$ induces a
representation of $G$ with the character 
$\chi^{(\alpha)} \circ \varphi$. Due to theorem~\ref{irred} and 
\begin{eqnarray}
\lefteqn{\langle \chi^{(\alpha)} \circ \varphi | \chi^{(\alpha)} \circ
  \varphi \rangle_G =} \nonumber \\ && 
\frac{1}{\oo G}\,\sum_{a \in G} 
\left| \chi^{(\alpha)}(\varphi(a)) \right|^2  
= 
\frac{\oo \mbox{ker}\, \varphi}{\mathrm{ord}\, G}\, 
\sum_{b\in H} \left| \chi^{(\alpha)}(b) \right|^2 = 
\langle \chi^{(\alpha)} | \chi^{(\alpha)} \rangle_H = 1, 
\end{eqnarray}
we obtain that $D^{(\alpha)} \circ \varphi$ is irreducible.
Thus we have found the following theorem.
\begin{theorem}\label{factorgrouprep}
Let $\varphi$ be a homomorphism from the finite
group $G$ onto $H$. Then every irrep $D^{(\alpha)}$ of $H$ induces 
an irrep $D^{(\alpha)} \circ \varphi$ of $G$. 
Applied to factor groups we see that if $N$ is a normal subgroup of
$G$ and $\varphi$ the canonical homomorphism $a \mapsto aN$,
then every irrep $D^{(\alpha)}$ of the factor group $G/N$
induces an irrep $D^{(\alpha)} \circ \varphi$ of $G$. 
\end{theorem}
\noindent
Consequently, the irreps of factor groups can be considered as
irreps of the full group. For a proper 
normal subgroup $N$ we have $1 < \oo (G/N)< \oo G$,
thus the representations $D^{(\alpha)} \circ \varphi$ will be non-faithful
representations of $G$. Therefore, a group for which all
non-trivial irreps are faithful cannot possess
a proper normal subgroup. Also the converse is true: if a group
possesses a non-faithful (and non-trivial) irreducible representation
$D^{(\alpha)}$, the kernel $\mathrm{ker}\,D^{(\alpha)}$ is a non-trivial normal
subgroup. Thus we have found a criterion for simple groups. 
\begin{theorem}
A finite group is simple if and only if all its non-trivial
irreps are faithful. 
\end{theorem}

Theorem~\ref{factorgrouprep} provides a tool to construct
non-faithful irreps of finite groups. In particular, when 
a group possesses a long principal series---see definition~\ref{princ}, 
many irreps can be constructed via the irreps of $G/N_i$. 
However, a principal series has an even stronger property.
Consider the sequence of factor groups 
\be\label{factorgroupseries}
\{e\} \cong G/N_m,\enspace G/N_{m-1},\ldots, G/N_{2},\enspace G/N_1,
\enspace G,
\ee
which is arranged in ascending group order. Then, 
for every pair of indices $j,k$ with $1 \leq j < k \leq m$ there is a natural
homomorphism given by
\be
\varphi_{jk}: 
\begin{array}{c} G/N_j \rightarrow G/N_k, \\
                  aN_j \mapsto     aN_k. 
\end{array}
\ee
Therefore, in the sequence~(\ref{factorgroupseries}) the
irreps of any $G/N_i$ are irreps of \textit{all} groups to the right of it. 
An example for the usage of principal series is found at the end of
section~\ref{D6}. For more examples we refer the reader to~\cite{GL10}.

\section{The permutation groups $S_n$ and $A_n$}
\label{chapterpermutation}
\subsection{Cycles and classes}
The symmetric group $S_n$ is the group of permutations of $n$ different
objects. In the following we will always regard a permutation as a mapping of
the set of numbers $\{ 1,2,\ldots,n \}$ onto itself.
Thus every element $p \in S_n$ can be written as a scheme of numbers
\begin{equation}\label{p}
p = \left( \begin{array}{cccc} 
1   & 2   & \cdots & n \\
p_1 & p_2 & \cdots & p_n 
\end{array} \right),
\end{equation}
where each number $1, \ldots, n$ occurs exactly once in the second line.
This notation implies the operation
$1 \to p_1$, $2 \to p_2$, etc., i.e. the numbers of the first line are mapped
onto the corresponding numbers below. There are $n!$ such mappings, therefore, 
\begin{equation}
\oo S_n = n!\hspace{0.7pt}.
\end{equation}
Obviously, if we permute the columns in equation~(\ref{p}), the mapping
$p$ does not change. Therefore, for any
\begin{equation}
s = \left( \begin{array}{cccc} 
1   & 2   & \cdots & n \\
s_1 & s_2 & \cdots & s_n 
\end{array} \right) \in S_n
\end{equation}
we can write
\begin{equation}\label{p'}
p = \left( \begin{array}{cccc} 
s_1   & s_2   & \cdots & s_n \\
p_{s_1} & p_{s_2} & \cdots & p_{s_n} 
\end{array} \right). 
\end{equation}
This freedom of arranging the numbers allows to write the inverse of
the permutation $p$ of equation~(\ref{p}) as 
\begin{equation}\label{p-1}
p^{-1} = \left( \begin{array}{cccc} 
p_1 & p_2 & \cdots & p_n \\
1   & 2   & \cdots & n 
\end{array} \right).
\end{equation}

We can also present permutations as \emph{cycles}. 
A cycle of length $r$ ($1 \leq r \leq n$) is a mapping
\begin{equation}\label{cycle}
(a_1 \to a_2 \to a_3 \to \cdots \to a_r \to a_1) \equiv 
(a_1a_2a_3 \cdots a_r)
\end{equation}
such that the $a_1$, \ldots, $a_r$ are different numbers between $1$ and
$n$; any number which does not occur in the cycle is mapped onto itself. 
All cycles of length one represent the identical mapping, i.e.\ they 
correspond to the unit element of $S_n$. 
Moreover, it is immaterial with which number a cycle starts off, i.e.
\begin{equation}
(a_1a_2a_3 \cdots a_r) = (a_ra_1a_2 \cdots a_{r-1}) =  
(a_{r-1}a_ra_1 \cdots a_{r-2}) = \cdots .
\end{equation}
Evidently, the following statement holds true.
\begin{theorem}
Every permutation is a unique product of cycles which have no common
elements. 
\end{theorem}
\noindent
For instance, 
\begin{equation}
\left( \begin{array}{cccccc} 
1 & 2 & 3 & 4 & 5 & 6 \\
6 & 4 & 3 & 1 & 2 & 5 
\end{array} \right) = (16524)(3).
\end{equation}
The order in which these cycles are arranged is irrelevant because cycles
which have no common element commute. 

We can also write a cycle~(\ref{cycle}) in the notation of equation~(\ref{p'}).
If $a_{r+1}, \ldots, a_n$ are the numbers which do not occur in the cycle
then 
\begin{equation}\label{a}
(a_1a_2a_3 \cdots a_r) \equiv
\left( \begin{array}{ccccccc} 
a_1 & a_2 & \cdots & a_r & a_{r+1} & \cdots & a_n \\
a_2 & a_3 & \cdots & a_1 & a_{r+1} & \cdots & a_n
\end{array} \right).
\end{equation}
For finding the classes of $S_n$ it suffices to compute $psp^{-1}$
where $s$ is a cycle and $p$ a general permutation.
We can write the cycle $s$ as in equation~(\ref{a}). Now we only need to
present a general permutation $p$ in a suitable way:
\begin{equation}
p = 
\left( \begin{array}{ccccccc} 
a_1    & a_2    & \cdots & a_r     & a_{r+1}   & \cdots & a_n \\
p_{a_1} & p_{a_2} & \cdots & p_{a_r} & p_{a_{r+1}} & \cdots & p_{a_n } 
\end{array} \right).
\end{equation}
The inverse of $p$ is obtained by exchanging the two lines in $p$. Now
we readily compute
\begin{equation}
psp^{-1} = (p_{a_1} p_{a_2} \cdots p_{a_r}).
\end{equation}
We have obtained an important result: Under conjugation with a
permutation $p$ a cycle $s$ of 
length $r$ remains a cycle of length $r$ and its entries are obtained
by applying $p$ to entries of $s$. Consequently, 
the classes of $S_n$ are characterized by the cycle
structure~\cite{Hamermesh}. 
\begin{theorem}\label{sn-cycles}
The classes of $S_n$ consist of the permutations with the same cycle
structure.
\end{theorem}
\noindent
Let us apply this theorem to $S_4$. 
There are five different cycle structures denoted generically by
\begin{equation}\label{S4}
(a)(b)(c)(d), \quad (a)(b)(cd), \quad (ab)(cd), \quad (a)(bcd), \quad (abcd).
\end{equation}
The theorem tells us that these cycle structures correspond to the five
classes. Thus, $S_4$ has five inequivalent irreps.

There are several ways to introduce the notion of even and odd
permutations. One possibility is to consider the
function~\cite{Murnaghan} 
\begin{equation}
\Delta(x_1,\ldots,x_n) = \prod_{1 \leq i < j \leq n} (x_i - x_j)
\end{equation}
of $n$ variables $x_1, \ldots, x_n$. Any permutation of these
variables can, at most, change the sign of $\Delta(x_1,\ldots,x_n)$. 
\begin{define}\label{even-odd} 
A permutation $p$ is called \emph{even} if $\Delta$ is invariant under the
action of $p$. If $\Delta$ changes the sign, $p$ is called \emph{odd}.
The sign of $\Delta$ under the action of $p$ is denoted by 
$\mbox{sgn}(p)$.
\end{define}
A \emph{transposition} is defined as a cycle of length $r=2$. A
transposition $(a_1a_2)$ simply corresponds to the operation
of exchanging two numbers: $a_1 \leftrightarrow a_2$. 
It is a bit tedious to show that a transposition is odd.
Every permutation can be decomposed into transpositions, as can be
seen by the decomposition of a cycle of length $r$:
\begin{equation}
(a_1a_2a_3 \cdots a_r) = (a_1a_r) \cdots (a_1a_3)(a_1a_2).
\end{equation}
Note that that the right-hand side, as a sequence of mappings, has to
be read from right to left, whereas by convention a cycle is read from
left to right. 
A decomposition into transpositions is not unique but from 
definition~\ref{even-odd} it follows that even (odd) permutations are
decomposed into an even (odd) number of transpositions. Thus we arrive at
the following result.
\begin{theorem}
A cycle of even (odd) length is an odd (even) permutation.
In general, a permutation is even (odd) if it can be represented as a
product of an even (odd) number of transpositions.
\end{theorem}

All elements of a class are either even or odd, and the set of even
permutations is a subgroup of $S_n$.
\begin{define}
The \emph{alternating group} $A_n$ is defined as the group of even
permutations of $n$ different objects.
\end{define}
\begin{theorem}
$A_n$ is a normal subgroup of $S_n$ with $\oo A_n = n!/2$.
\end{theorem}
\noindent
Obviously, the relations 
\begin{equation}
S_2 \cong \zz_2, \quad A_3 \cong \zz_3
\end{equation}
hold. So $S_3$ is the first non-trivial symmetric group and $A_4$ the
first non-trivial alternating group. 
All elements of $A_n$ can be written as a product of three-cycles. 
This follows from the relations
$(a_1a_2)(a_3a_4) = (a_1a_3a_2)(a_1a_3a_4)$ and 
$(a_1 a_2) (a_1 a_3) = (a_1 a_3 a_2)$ where the numbers $a_1$, \ldots,
$a_4$ are all different.

The following group is an
important subgroup of $S_4$.
\begin{define}
\emph{Klein's four-group} is defined as the set of permutations
\begin{equation}
K = \{ e,\, (12)(34),\, (13)(24),\, (14)(23) \}.
\end{equation}
\end{define}
\noindent
Denoting the elements of Klein's four-group by $e,k_1,k_2,k_3$, the
following relations hold: $k_i^2 = e$, $k_i k_j = k_j k_i = k_l$ with 
$i \neq j \neq l \neq i$. Therefore, $K$ is Abelian
and $K \cong \zz_2 \times \zz_2$.
The following theorem lists all proper normal subgroups of the
symmetric and alternating groups.
\begin{theorem} \textbf{Normal subgroups of $A_n$ and $S_n$.} 
\begin{itemize}
\item
$A_n$ is simple for $n > 4$.
\item
The only proper normal subgroup of $A_4$ is $K$.
\item
The only proper normal subgroup of $S_n$ for $n=3$ and $n>4$ is
$A_n$.
\item
$S_4$ has two proper normal subgroups, $A_4$ and $K$, and possesses, therefore,
  the principal series $\{ e \} \lhd K \lhd A_4 \lhd S_4$.
\end{itemize}
\textit{Proof:} The proof of the first statement can for instance be
found in~\cite{Ramond} (p.190). The second and fourth statement can be
proved by direct computation. The third statement follows from the
simplicity of $A_n$. 
\end{theorem}
\noindent
The symmetric group has the structure
$S_n \cong A_n \rtimes \zz_2$ because any element $p \in S_n$ can be
written as a product $p = st$ with $s \in A_n$ and a fixed
transposition $t \in S_n$ . For instance one can choose $t = (12)$.

The following theorem lists useful properties of irreps of $S_n$---see
for instance the discussion in~\cite{Hamermesh}.
\begin{theorem}\label{Snirreps}
The irreps of $S_n$ have the following properties:
\begin{itemize}
\item 
All irreps are real.
\item 
There are exactly two one-dimensional irreps:
$p \mapsto 1$ and $p \mapsto \mbox{sign}(p)$.
\item
For $n=3$ and $n > 4$ all irreps, except those of dimension one, are
faithful.
\end{itemize}
\end{theorem}

Earlier we found the relation between the cycle structures and the
classes of $S_n$. For $A_n$ the following theorem, 
proven in~\cite{Murnaghan}, almost completely solves the analogous problem.
\begin{theorem}\label{A-cycles}
The classes of $A_n$ are obtained from those of $S_n$ in the following
way: All classes of $S_n$ with even permutations are also classes of
$A_n$, except those which consist exclusively of cycles of 
unequal odd length. Each of the latter classes of $S_n$ refines in $A_n$ into
two classes of equal size. 
\end{theorem}
\noindent
Let us illustrate this theorem with $S_5$ and $A_5$. In the case of
$S_5$ the possible cycle structures are 
\begin{equation}
\begin{array}{llll}
(a)(b)(c)(d)(e), & (a)(b)(c)(de), & (a)(b)(cde), & (a)(bcde) \\ 
(abcde), & (ab)(cde), & (a)(bc)(de). &
\end{array}
\end{equation}
Therefore, the number of inequivalent irreps of $S_5$ is seven. Now we
remove all cycle structures which correspond to odd permutations and
arrive at
\begin{equation}\label{ca5}
(a)(b)(c)(d)(e), \quad (a)(b)(cde), \quad (abcde), \quad (a)(bc)(de).
\end{equation}
According to theorem~\ref{A-cycles} we have to single out the classes
with cycles of unequal odd length. The first class of
equation~(\ref{ca5}), the class of the 
unit element, can of course not refine; 
in the language of the theorem the cycle structure 
has five cycles of equal length one. The second class has only
cycles of odd length, but two cycles of length one; thus it cannot
refine either. The third class has one cycle with odd length five,
thus the class of the type $(abcde)$ refines into two classes in $A_5$. 
The cycle structure of the last class of equation~(\ref{ca5}) has two
cycles of even length two, thus it does not refine. 
We conclude that $A_5$ has five inequivalent irreps.

\subsection{$S_3$, $A_4$, $S_4$ and $A_5$ as subgroups of $SO(3)$}
\label{SSS}
Of the non-trivial symmetric and alternating groups only $S_3$, $A_4$,
$S_4$ and $A_5$ can be considered as finite subgroups of the three-dimensional
rotation group~\cite{Miller}. 
The precise meaning of this statement is that only for the listed
groups there is a faithful representation in terms of $3 \times 3$
rotation matrices. 

Let us start with the trivial group $A_3 \cong \zz_3$ which is
generated by the the cyclic permutation $s = (123)$. Since $s^3 = e$
this permutation must be represented by a rotation of $120^\circ$. We
choose the $z$-axis as rotation axis and represent $s$ as
\begin{equation}\label{sS3}
s \mapsto 
\renewcommand{\arraystretch}{1.3}
\left( \begin{array}{rrr}
-\frac{1}{2} & -\frac{\sqrt{3}}{2} & 0 \\
\frac{\sqrt{3}}{2} & -\frac{1}{2} & 0 \\
0 & 0 & \hphantom{-}1 
\end{array}\right).
\end{equation}
In order to generate $S_3$, we pick a transposition, say $t=(12)$,
and take into account that $t s t^{-1} = s^2 = s^{-1}$. 
Then $t$ can be represented by 
\begin{equation}\label{tS3}
t \mapsto 
\renewcommand{\arraystretch}{1.3}
\left( \begin{array}{rrr}
0 & \hphantom{-}1 & 0 \\ 1 & 0 & 0 \\ 0 & 0 & -1 
\end{array}\right),
\end{equation}
which completes the representation of $S_3$ we have searched for.
We emphasize that the 33-entry in the matrix of equation~(\ref{tS3}) has to be
$-1$ in order to have a positive determinant.
Note that in our construction we have exploited that   
\begin{equation}\label{s3=z3z2}
S_3 \cong \zz_3 \rtimes \zz_2.
\end{equation}

In the case of $A_4$ it is advantageous to use Klein's four-group~$K$  
as the starting point~\cite{GLL}. The possible cycle structures are, 
apart from that of the unit element, only 
$(ab)(cd)$ and $(a)(bcd)$---see equation~(\ref{S4})
where the odd cycle structures have to be removed. This suggests to
use $K$ and one three-cycle,
say $s = (123)$, to generate $A_4$. 
Denoting the non-trivial elements of $K$ by $k_1,k_2,k_3$ and defining
$k_1 = (12)(34)$, we have the following relations:
\begin{equation}\label{s4k}
k_1^2 = k_2^2 = k_3^2 = e, \quad s^3 = e, \quad
s k_1 s^{-1} = (14)(23) =: k_2, \quad 
s k_2 s^{-1} = (13)(24) =: k_3.
\end{equation}
From these relations it follows immediately that we can represent
$k_1$ and $s$ by 
\begin{equation}\label{AE}
\renewcommand{\arraystretch}{1.3}
k_1 \to 
\left( \begin{array}{rrr}
1 & 0 & 0 \\ 0 & -1 & 0 \\ 0 & 0 & -1 
\end{array}\right) =: A,
\quad
s \to
\left( \begin{array}{rrr}
0 & 1 & 0 \\ 0 & 0 & 1 \\ 1 & 0 & 0
\end{array}\right) =: E.
\end{equation}
After this choice, $k_2$ and $k_3$ are fixed by equation~(\ref{s4k}):
\begin{equation}\label{Ak2}
\renewcommand{\arraystretch}{1.3}
k_2 \to 
\left( \begin{array}{rrr}
-1 & 0 & 0 \\ 0 & -1 & 0 \\ 0 & 0 & \hphantom{-}1 
\end{array}\right),
\quad
k_3 \to
\left( \begin{array}{rrr}
-1 & 0 & 0 \\ 0 & \hphantom{-}1 & 0 \\ 0 & 0 & -1
\end{array}\right).
\end{equation}
Note that the last equation correctly reproduces $k_2 k_3 = k_1$. In this
construction we have used that 
\begin{equation}
A_4 \cong K \rtimes \zz_3
\end{equation}
and that $K$, as an
Abelian group, can be represented by diagonal matrices.

The representation of $A_4$ in terms of rotation matrices can easily be
extended to $S_4$. We add a transposition, say $t = (12)$, which
fulfills the relations
\begin{equation}
t^2 = e, \quad tk_1t^{-1} = k_1, \quad tk_2t^{-1} = k_3, \quad
tst^{-1} = s^2.
\end{equation}
Obviously, these relations are satisfied with\footnote{The minus signs
  are required for obtaining a positive determinant.}
\begin{equation}\label{Rt}
t \to 
\left( \begin{array}{rrr}
-1 & 0 & 0 \\ 0 & 0 & -1 \\ 0 & -1 & 0
\end{array} \right) =: R_t.
\end{equation}
In summary, we have used the decomposition
\begin{equation}\label{s4kz3z2} 
S_4 \cong (K \rtimes \zz_3) \rtimes \zz_2 \cong K \rtimes S_3
\end{equation}
to represent $S_4$ as a subgroup of $SO(3)$.
Note that, since we started from
Klein's four group represented as diagonal matrices---see
equations~(\ref{AE}) and~(\ref{Ak2}), the representation of the generators $s$
and $t$ of the subgroup $S_3$ of $S_4$ 
differs from the one given in equations~(\ref{sS3}) and~(\ref{tS3}). 
However, these two representations of $S_3$ are equivalent which is shown by
explicit construction in appendix~\ref{XS3}.

To represent $A_5$ with its 60 elements as a rotation group is much more
tricky. For this purpose it is useful to take advantage of a suitable
presentation~\cite{Miller}. The following permutations constitute a set of
generators of $A_5$:
\begin{equation}\label{set}
s = (123), \quad k_1 = (12)(34), \quad k_4 = (12)(45).
\end{equation}
To sketch a proof of this statement we note that
\begin{equation}
k_1 s^2 k_1 = (124), \quad k_1 k_4 = (345).
\end{equation}
Thus we have the three-cycles $(123)$, $(124)$ and $(345)$ at our
disposal. By explicit computation one finds that all 20 three-cycles
of $A_5$ can be generated from these three by two processes:
quadrature and conjugation of one element by another. Since any even
permutation can be represented as a product of three-cycles, this
proves the statement. The three permutations of
equation~(\ref{set}) fulfill the relations
\begin{equation}\label{rela5}
s^3 = k_1^2 = k_4^2 = e, \quad (s k_1)^3 = (k_1 k_4)^3 = (s k_4)^2 = e.
\end{equation}
These relations define a presentation of $A_5$~\cite{Miller}.
Therefore, it suffices to find
rotation matrices which obey the relations of equation~(\ref{rela5}). 
We denote the representation matrices by 
\begin{equation}
s \to E, \quad k_1 \to A, \quad k_4 \to W,
\end{equation}
where $E$ and $A$ are given by equation~(\ref{AE}). With this choice,
the relations involving only $s$ and $k_1$ in equation~(\ref{rela5})
are already satisfied. It remains to find a rotation matrix $W$ such that
\begin{equation}\label{W}
W^2 = (AW)^3 = (EW)^2 = \bone.
\end{equation}
This set of equations is solved in appendix~\ref{A5W}.
There are two solutions for $W$~\cite{Miller}  
corresponding to the matrices 
\begin{equation}\label{WW'}
W = \frac{1}{2} \left( \begin{array}{ccc}
-1 & \mu_2 & \mu_1 \\ \mu_2 & \mu_1 & -1 \\ \mu_1 & -1 & \mu_2
\end{array} \right)
\quad \mbox{and} \quad
W' = \frac{1}{2} \left( \begin{array}{ccc}
-1 & \mu_1 & \mu_2 \\ \mu_1 & \mu_2 & -1 \\ \mu_2 & -1 & \mu_1
\end{array} \right),
\end{equation}
where
\begin{equation}\label{mu12}
\mu_1 = \frac{-1 + \sqrt{5}}{2}, \quad
\mu_2 = \frac{-1 - \sqrt{5}}{2}.
\end{equation}
In summary, up to basis transformations, there are two 
representations of $A_5$ as rotation matrices which differ in the
representation of $k_4$:
\begin{equation}\label{repa5}
s \to E, \quad k_1 \to A, \quad k_4 \to W \;\; \mbox{or} \;\; W'.
\end{equation}
The two representations are irreducible and inequivalent. The first property
follows from $E$ and $A$ alone, which are generators of an $A_4$ subgroup of
$A_5$. The second property is a consequence of the same fact: A similarity
transformation leading from the $W$ irrep to the $W'$ irrep must leave $E$ and
$A$ invariant; thus it is proportional to the unit matrix and cannot transform
$W$ into $W'$.

Finally, 
it is interesting to perform a check of our result by computing the rotation
angle corresponding to a five-cycle. Taking the $W$ representation of
equation~(\ref{repa5}) and computing
\begin{equation}\label{AWE}
k_1 k_4 s = (12453) \to AW\!E = 
\frac{1}{2} \left( \begin{array}{ccc}
\mu_1 & -1 & \mu_2 \\ 1 & -\mu_2 & -\mu_1 \\ -\mu_2 & -\mu_1 & 1
\end{array} \right),
\end{equation}
the rotation angle $\psi$ corresponding to $(12453)$ is given by the equation
$2 \cos \psi + 1 = \mbox{Tr}\,(AW\!E)$---see
equation~(\ref{alphan}). Indeed we find 
$\psi = 72^\circ = 360^\circ/5$. In the case of the $W'$ representation of
equation~(\ref{repa5}) we obtain $\psi' = 144^\circ = 2 \times 72^\circ$.

\section{The finite subgroups of $SU(2)$ and $SO(3)$}
\label{su2so3}

\subsection{The homomorphism from $SU(2)$ onto $SO(3)$}
\label{SU2SO3homomorphism}

Every rotation in $\mathbbm{R}^3$ is 
characterized by its
rotation axis and rotation angle. In the following we will
always use the convention that the rotation axis is attached to the
coordinate origin and its direction given by a unit vector 
$\vec n$, and that the rotation angle $\alpha$ lies in the interval 
$0 \leq \alpha < 2\pi$. Then, taking into account the right-hand rule,
the rotation matrix $R(\alpha,\vec n)$ is given by
	\be\label{SO3calc}
	R(\alpha,\vec{n})_{kl} = \cos\alpha\,\delta_{kl} +
        (1-\cos\alpha)\, n_k n_l - (\sin\alpha) \, n_j \varepsilon_{jkl}, 
	\ee
where $\varepsilon_{jkl}$ is totally antisymmetric with $\varepsilon_{123}=1$. 
Note that because of 
\begin{equation}\label{R2}
R(2\pi - \alpha, -\vec n) = R(\alpha,\vec n),
\end{equation}
every rotation $R \neq \bone$ can be expressed in two ways
by rotation axis and angle. 
Furthermore, for every rotation matrix the relations 
\begin{equation}
R^T(\alpha,\vec n) = R(\alpha,-\vec n) = R(2\pi-\alpha,\vec n) = 
R^{-1}(\alpha,\vec n)
\quad \mbox{and} \quad 
\det R(\alpha,\vec n) = 1
\end{equation}
hold. Vice versa, linear algebra tells us that to every orthogonal 
$3 \times 3$ matrix with determinant one we can find a unit vector
$\vec n$ and an angle $\alpha$ such that it is represented in the
form~(\ref{SO3calc}). 
Thus the group $SO(3)$ of orthogonal $3 \times 3$ matrices $R$ with
$\det R = 1$ is identical with the group of rotations in
three-dimensional space. Using equation~(\ref{SO3calc}),
for $R \in SO(3)$ one finds the relations
	\be\label{alphan}
	\cos\alpha=\frac{1}{2} \left( \mathrm{Tr}\,R - 1
        \right)
\quad \mbox{and} \quad
        (\sin\alpha)\,n_j= 
	-\frac{1}{2}\, \varepsilon_{jkl} R_{kl}.
	\ee
According to equation~(\ref{R2}), if
$\alpha \neq 0$ or $\pi$, equation~(\ref{alphan}) has always
two solutions, corresponding to the same rotation. 
For $\alpha = \pi$ one has to use the original
equation~(\ref{SO3calc}) for obtaining $\vec n$.

Now we come to the homomorphism from $SU(2)$ onto $SO(3)$. As discovered
by the ma\-the\-matician Felix Klein, every $U \in SU(2)$ induces a
rotation on the vectors $\vec x \in \mathbbm{R}^3$ via the construction
\begin{equation}\label{UR}
U (\vec\sigma \cdot \vec x)U^\dagger = \vec\sigma \cdot (R\vec x),
\end{equation}
where the $\sigma_j$ are the Pauli matrices and 
$\vec \sigma \cdot \vec x \equiv \sigma_j x_j$. 
Given a rotation matrix $R(\alpha,\vec n)$, one can
solve equation~(\ref{UR}) for $U$. The solution is 
\begin{equation}\label{SU2calc}
U(\alpha,\vec{n}) = 
\cos(\alpha/2)\, \bone_2 - i\,\sin(\alpha/2)\,\vec{n}\cdot\vec{\sigma}.
\end{equation}
The second solution is given by 
	\be
	U(\alpha+2\pi,\vec{n})=-U(\alpha,\vec{n}).
	\ee
The construction~(\ref{UR}) makes plain that the mapping 
\begin{equation}\label{phi}
\phi: \begin{array}{ccc}
SU(2) & \to & SO(3) \\
\pm U( \alpha, \vec n) & \mapsto & R( \alpha, \vec n)
\end{array}
\end{equation}
is a homomorphism, i.e. $\phi(U_1U_2) = \phi(U_1) \phi(U_2)$. Moreover, 
one can deduce from equation~(\ref{UR}) that the kernel of this
homomorphism is 
\begin{equation}
\mbox{ker}\, \phi = \{ \bone_2, -\bone_2 \}.
\end{equation}
This means that to every $SO(3)$-matrix $R$ there are exactly two
$SU(2)$-matrices $U$ which differ by the sign only. 

The homomorphism~(\ref{UR}) can also be understood by considering $SU(2)$ and
$SO(3)$ as Lie groups. The Lie algebras of these two groups are isomorphic,
due to the following correspondence of their generators:
\begin{equation}
\frac{\sigma_j}{2} \leftrightarrow T_j \quad (j=1,2,3),
\end{equation}
where the generators of $SO(3)$ are given by
\begin{equation}
(T_j)_{kl}=\frac{1}{i}\varepsilon_{jkl}.
\end{equation}
In terms of these generators, $U(\alpha,\vec n)$ and $R(\alpha,\vec n)$  
can be written in exponential form 
	\be 
	U(\alpha,\vec{n}) = \exp \left(-i\alpha
        \vec{n}\cdot\frac{\vec{\sigma}}{2}\right)
\quad \mbox{and} \quad
	R(\alpha,\vec{n}) = \exp \left( -i\alpha \vec{n}\cdot\vec{T} \right),	
	\ee
respectively. 

Now we are in a position to discuss the relationship between subgroups of
$SO(3)$ and subgroups of $SU(2)$. As discussed before, for
every $a \in SO(3)$ the equation $\phi(x) = a$, where $\phi$ is the
homomorphism defined in equation~(\ref{phi}), has two solutions which only
differ by a sign. In the 
following we will denote these solutions by $\pm\tilde{a}$. Let us now assume
that we have 
a finite subgroup $G=\{a_1,\ldots,a_m\}$ of $SO(3)$. Then the inverse image 
of $G$ under $\phi$,  
\be\label{dc}
\widetilde{G} := \phi^{-1}(G) = \{\pm\tilde{a}_1,\ldots,\pm\tilde{a}_m\},
\ee
is a group of order $2m$ which is called 
the \emph{double cover}\footnote{In the literature the double cover of
  a group $G$ is frequently denoted by $G'$ instead of $\widetilde{G}$.}
of $G$---see for instance~\cite{Ramond,frampton}. 
Note that this  
construction yields a one-to-one correspondence between finite subgroups of
$SO(3)$ and finite subgroups of $SU(2)$ with 
$\{\bone_2,-\bone_2\}$ contained in the center. 

On the other hand, one may ask the question whether 
there are subgroups $\mathcal{G}$ of $SU(2)$ which are isomorphic to 
$\phi(\mathcal{G})$,
i.e.\ $SU(2)$ subgroups whose center does not contain $-\bone_2$. 
The answer is yes, but such groups are
trivial: Anticipating the results of section~\ref{finitesubgroupsofSO3} one
can show that such groups are isomorphic to $\zz_n$ with $n$ odd. 
Their generator can be written as 
$\diag (e^{-i\beta},\, e^{i\beta}\,)$ with $\beta = 2\pi/n$. 

To summarize, if we know all finite subgroups of 
$SO(3)$ and their double covers, then, up to isomorphisms, we know \emph{all}
finite subgroups of $SU(2)$. To compute the double cover of a subgroup $G$ of
$SO(3)$, we only need to know the rotation angles and rotation axes of its
elements. Then the explicit form of the homomorphism~$\phi$, 
i.e.\ equations~(\ref{SU2calc}) and~(\ref{phi}), gives a prescription for
computing $\widetilde G$.

\subsection{Conjugacy classes of three-dimensional rotation groups}
\label{SO3classrules}

Before we analyze the finite subgroups of $SO(3)$ in detail we want to give a
simple set of rules which allow to determine the conjugacy classes of finite
three-dimensional rotation groups, i.e. of finite subgroups of $SO(3)$.

Let $R_1=R(\alpha,\,\vec{n})$ and $R_2$ be elements of $SO(3)$. Then, 
using equation~(\ref{SO3calc}), it is an easy task to demonstrate that
	\be\label{SO3conjugacy}
	R_2\,R(\alpha,\vec{n})\,R_2^{-1} = R(\alpha,R_2\vec{n}).
	\ee
Now we consider a finite subgroup $G$ of $SO(3)$.
Equation~(\ref{SO3conjugacy}) suggests the following
definition~\cite{Hamermesh}: 
\begin{itemize}
 \item Two rotation axes $\vec{a}$ and $\vec{b}$ of elements of $G$ are called 
   \emph{equivalent} if there exists a rotation $R\in G$ such that
	\be
	R\vec{a}=\vec{b}.
	\ee
 \item A rotation axis $\vec{a}$ is called 
   \emph{two-sided} if there exists a rotation $R\in G$ such that
	\be
	R\vec{a}=-\vec{a}.
	\ee
	If there is no such $R$ then $\vec{a}$ is called a \emph{one-sided}
	rotation axis.
\end{itemize}
Then, using equation~(\ref{SO3conjugacy}), we find the following rules
which determine the conjugacy classes of $G$:
\begin{enumerate}
 \item Rotations through the same angle about equivalent axes are equivalent.
 \item Rotations through $\alpha$ and $2\pi-\alpha$ about a two-sided
	axis are equivalent.
\end{enumerate}
Applications of these rules will follow in the next section.

\subsection{The finite subgroups of $SO(3)$ and their double covers}
\label{finitesubgroupsofSO3}

According to~\cite{Hamermesh} \emph{every} finite subgroup of $SO(3)$ belongs
to one of the following sets:
\begin{itemize}
 \item \textit{The uniaxial groups}. These are the groups of rotations about one
         axis. 
 \item The \textit{dihedral groups} $D_n$. These are the rotation groups 
       which leave the planar regular polygons with $n$ edges invariant. 
 \item The \textit{rotation symmetry groups of the regular polyhedra}
   (Platonic solids). 
		Though there are five regular polyhedra, they lead to only
                three different rotation groups~\cite{Hamermesh}, namely the
                \textit{tetrahedral}, \textit{octahedral} and 
		\textit{icosahedral} group. The rotation symmetry group of the
                cube is identical 
		with the octahedral group, and the rotation symmetry group of
                the dodecahedron is 
		identical with the icosahedral group.
	\end{itemize}
In the following we will call a symmetry axis $n$-fold if it corresponds to 
rotations through an angle $2\pi/n$ and multiples thereof.

\subsubsection{The uniaxial groups}

Without loss of generality we choose the $z$-axis as the rotation axis. The
group of rotations through the angles $d\eta$ ($d=0,\ldots,n-1$) is generated
by 
	\be\label{uniaxialgen}
	R(\eta,\,\vec{e}_z)=\bmat
	\cos\eta & -\sin\eta & 0\\
	\sin\eta & \cos\eta & 0\\
	0 & 0 & 1
	\emat,
	\ee
where $\eta=2\pi/n$. Evidently, the uniaxial groups are Abelian. The
representation defined by equation~(\ref{uniaxialgen}) is the 
sum of a two-dimensional representation and the trivial representation. The
two-dimensional representation 
	\be
	R(\eta,\,\vec{e}_z)\mapsto 
	\bmat
	\cos\eta & -\sin\eta\\
	\sin\eta & \cos\eta
	\emat
	\ee
can be reduced by the similarity transformation
	\be\label{similarityX}
	X^{\dagger}
	\bmat
	\cos\eta & -\sin\eta\\
	\sin\eta & \cos\eta
	\emat
	X=
	\bmat
	e^{-i\eta} & 0\\
	0 & e^{i\eta}
	\emat,	
	\ee
where
	\be\label{X}
	X=\frac{1}{\sqrt{2}}\bmat
	1 & i\\
	i & 1
	\emat.
	\ee
Thus the uniaxial group generated by $R(\eta, \vec e_z)$ of 
equation~(\ref{uniaxialgen}) is
isomorphic to the cyclic group $\zz_n$ generated by $e^{i\eta}$. The double
cover of 
the uniaxial group is generated by
	\be
	U(\eta,\vec{e}_z)=\bmat
	e^{-i\eta/2} & 0\\
	0 & e^{i\eta/2}
	\emat
	\ee
and is thus isomorphic to $\zz_{2n}$. 
In summary, uniaxial groups are isomorphic to cyclic groups.

\subsubsection{The dihedral groups and their double covers}

\begin{figure}
\begin{center}
  \epsfig{scale=0.5,file=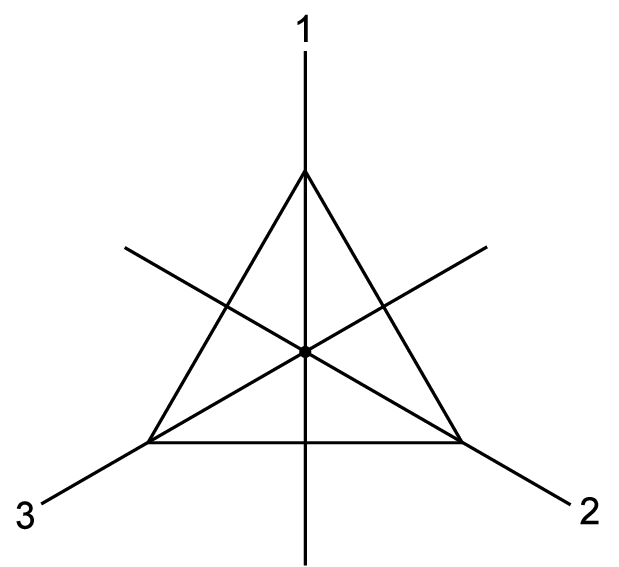}
  \caption{The symmetry axes of an equilateral triangle. The $z$-axis
    (perpendicular 
to the plane of projection) is a two-sided three-fold axis. The three two-fold
axes labeled~1, 2 and~3 are equivalent one-sided axes.}\label{D3sketch}
\end{center}
\end{figure}

\begin{figure}
\begin{center}
  \epsfig{scale=0.5,file=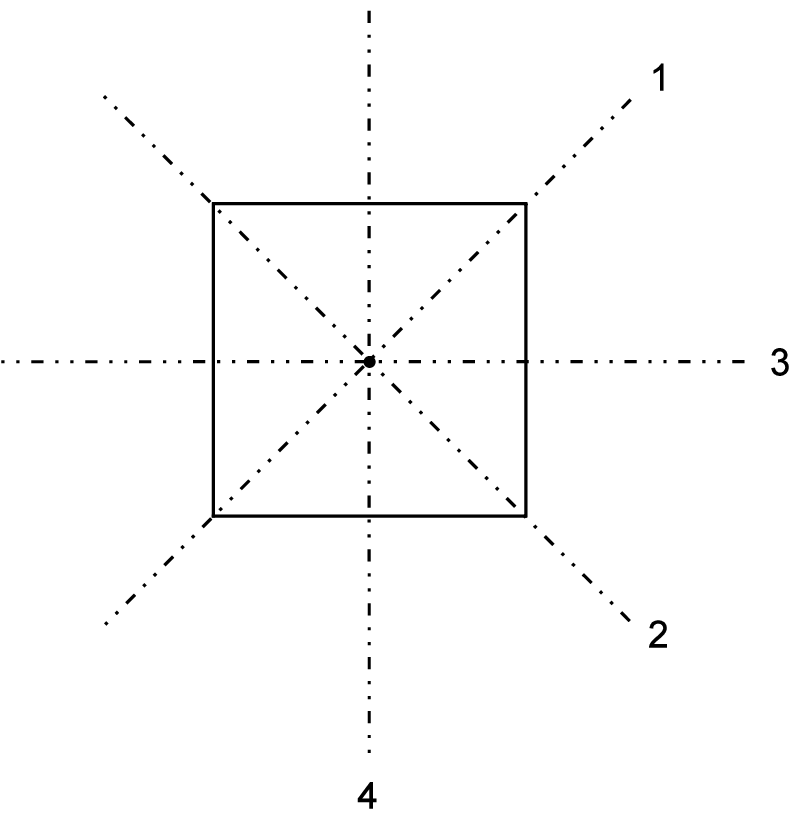}
  \caption{The symmetry axes of a square. The $z$-axis (perpendicular
to the plane of projection) is a two-sided four-fold axis. The axes 1
and 2 are equivalent, and so are 3 and 4. All axes 1 to 4 are 
two-sided and two-fold.}\label{D4sketch}
\end{center}
\end{figure}

\paragraph{The dihedral group $D_n$:}
For the discussion of $D_n$---see also~\cite{blum}---we choose the following
geometrical setting: The regular polygon with $n$ edges lies in the
$xy$-plane, with its center being the origin and 
the $y$-axis being a symmetry axis. 
Since the structure of $D_n$ depends on whether $n$ is
odd or even, 
we present two figures, one for $D_3$ (see figure~\ref{D3sketch}) and
one for $D_4$ (see figure~\ref{D4sketch}). 
Concerning the rotation axes, we use the notions introduced in
section~\ref{SO3classrules}. One-sided axes are drawn as
solid lines, two sided axes as dashed-dot-dotted lines.
The rotations corresponding to the $n$-fold rotation axis, which is
the $z$-axis in our convention, are generated by
	\be\label{Dngeneratora}
	a:=R(\eta,\,\vec{e}_z)=\bmat
	\cos\eta & -\sin\eta & 0\\
	\sin\eta & \cos\eta & 0\\
	0 & 0 & 1
	\emat,
	\ee
where $\eta=2\pi/n$. The rotation about the $y$-axis 
is a rotation through $\pi$ and given by
	\be\label{Dngeneratorb}
	b:=R(\pi,\vec{e}_y)=\bmat
	-1 & 0 & 0\\
	0 & 1 & 0\\
	0 & 0 & -1
	\emat.
	\ee
The group $D_n$ is generated by the two elements
$a$ and $b$. Since 
	\be\label{baba}
	bab^{-1}=a^{-1}
	\ee
the group generated by $a$ is a normal subgroup of $D_n$ and
isomorphic to $\zz_n$. The second generator $b$ generates a
$\zz_2$. Therefore, with equation~(\ref{baba}) we find\footnote{Note
  that sometimes in the literature the dihedral 
  group of order $2n$ is called $D_{2n}$ instead of $D_n$.} 
	\be
	D_n \cong \zz_n \rtimes \zz_2
\quad \mbox{and} \quad \oo D_n = 2n.
	\ee
A presentation of $D_n$ is thus given by
	\be\label{Dnpresentation}
	a^n=b^2=(ab)^2=\bone.
	\ee
From this presentation we read off that for $n=2$ the generators $a$ and
$b$ commute, therefore, we find $D_2 \cong \zz_2 \times \zz_2$.
For $n=3$ the presentation~(\ref{Dnpresentation}) agrees with the
presentation of $S_3$---see the discussion leading to
equation~(\ref{s3=z3z2}). Thus we find 
$D_3 \cong S_3$, which has been used as a flavour group already very
early---see for instance~\cite{S3}. Also $D_4$, $D_5$ and higher dihedral
groups have been used for this purpose~\cite{D4,D5,D7-14}.

We can now determine the
classes of $D_n$ using the rules presented in
section~\ref{SO3classrules}. It is useful to distinguish between $n$
odd and $n$ even.
\paragraph{$D_n$ with $n$ odd:}
For any group the identity element
forms a class $C_1=\{e\}$ of its own. 
Since the $z$-axis is a two-sided rotation axis,
all classes corresponding to this axis must contain also the inverses
of each element in the class. Since for $n$ odd there 
are no elements of the form $a^k$ ($k=1,\ldots,n-1$) which are their
own inverse, we find $\frac{n-1}{2}$ 
classes consisting of two elements each:
	\be
	C_2=\{a,\,a^{n-1}\},\; C_3=\{a^2,\,a^{n-2}\},\ldots,
        C_{(n+1)/2}=\{a^{(n-1)/2},\,a^{(n+1)/2}\}. 
	\ee
All remaining axes are equivalent and two-fold. Thus we obtain
just one further class containing one element for each remaining axis:
	\be
	C_{(n+3)/2}=\{b,\, ab,\, a^2 b,\ldots, a^{n-1}b\}.
	\ee
The $\frac{n+3}{2}$ classes we have found contain $2n$ elements, therefore,
the set of classes is complete.

Using the presentation~(\ref{Dnpresentation}), it is very easy to find
the irreps of $D_n$. For the one-dimensional irreps we deduce
that $a^n, a^2, b^2 \mapsto 1$. Since $n$ is odd, $a \mapsto 1$ is compelling.
Therefore, there are two one-dimensional irreps given by
\begin{equation}\label{1-nodd}
\one^{(q)}: \quad a \mapsto 1, \quad b \mapsto (-1)^q \quad (q=0,1).
\end{equation}
The higher dimensional irreps can be constructed in the following way.
Due to
	\be\label{bakb}
	ba^kb^{-1}=a^{-k}=(a^k)^{-1}
	\ee
the mapping $a \mapsto a^k$, $b \mapsto b$ defines a representation. 
Moreover, equation~(\ref{bakb}) tells us that 
representations with $k$ and $n-k$ are equivalent. Therefore, removing
the third columns and third rows in $a$ and $b$ of
equations~(\ref{Dngeneratora}) and~(\ref{Dngeneratorb}), respectively, 
we obtain the following two-dimensional inequivalent irreps:
	\be\label{nodd-2}
	\two^{(k)}: \quad a\mapsto 
	\bmat
	\cos(k\eta) & -\sin(k\eta)\\
	\sin(k\eta) & \cos(k\eta)
	\emat
	,\quad b\mapsto
	\bmat
	-1 & 0\\
	\hphantom{-}0 & 1	
	\emat
\quad \mbox{with} \quad k = 1, \ldots, \frac{n-1}{2}.
	\ee
There are no further irreps because the number of irreps we have found
equals the number of classes.
The character table of $D_n$ for~$n$ odd is depicted in table~\ref{charDnodd}. 
In order to enhance the legibility of the table we add in parentheses
to each class symbol one of its elements. 
\begin{table}
\begin{center}
\begin{tabular}{|c|ccc|}\hline
$D_n$ ($n$ odd)  & $C_1(e)$ & $C_{m+1}(a^m)$ & $C_{(n+3)/2}(b)$ \\ 
(\# $C$) & $(1)$ & $(2)$ & $(n)$ \\ 
$\oo(C)$ & $1$ & $r_m$ & $2$ \\ \hline
$\one^{(0)}$  & $1$ & $1$ & $1$ \\
$\one^{(1)}$  & $1$ & $1$ & $-1$ \\
$\two^{(k)}$  & $2$ & $2\cos(km\eta)$ & $0$ \\ 
\hline
\end{tabular}
\end{center}
\caption{The character table of $D_n$ ($n$
  odd) with $k,m=1,\ldots,\frac{n-1}{2}$. The order of $a^m$ is $r_m =
  n/\mathrm{gcd}(m,n)$, where $\mathrm{gcd}(m,n)$ is the greatest
  common divisor of $m$ and $n$. \label{charDnodd}} 
\end{table}
\paragraph{$D_n$ with $n$ even:}
The treatment of $D_n$ with $n$ even is very similar to the case of $n$
odd. We start again with the class structure. 
From the $n$-fold axis we obtain again classes
consisting of two elements which are inverse to each other. In
addition, since $n$ is even, we will have one 
element which is its own inverse, namely $a^{n/2}$ which is the rotation
through $\pi$ about the $z$-axis---see figure~\ref{D4sketch}. 
Thus, apart from $C_1=\{e\}$, we find the classes
	\be
	C_2=\{a,\,a^{n-1}\},\, C_3=\{a^2,\,a^{n-2}\},\,\ldots,\,
	C_{n/2}=\{a^{(n-2)/2},\,a^{(n+2)/2}\},\, C_{(n+2)/2}=\{a^{n/2}\}.
	\ee
For $n$ odd, we saw that the two-fold symmetry axes were all
equivalent.  This is not true for $n$ even: 
the $n/2$ axes which connect the vertices are equivalent,
and the $n/2$ axes which intersect the edges are equivalent. This can be
gathered from a look at figure~\ref{D4sketch}. The fact that there are two
sets of two-fold axes such that one set is inequivalent with the other one is
reflected in the classes where now two classes are associated with  the
two-fold axes: 
	\be
	C_{(n+4)/2}=\{b,\, a^2b,\ldots, a^{n-2}b\} 
\quad \mbox{and} \quad
        C_{(n+6)/2}=\{ab,\, a^3 b,\ldots, a^{n-1}b\}. 
	\ee

Again we use the presentation~(\ref{Dnpresentation}) for finding the
irreps. For one-dimensional irreps both $a^n$ and $a^2$ are mapped onto one,
just as for $n$ odd. However, since $n$ is even, also $a \mapsto -1$ is
possible. Therefore, there are four one-dimensional irreps:
\begin{equation}\label{1-neven}
\one^{(p,q)}: \quad a \mapsto (-1)^p, \quad b \mapsto (-1)^q
\quad (p,q = 0,1).
\end{equation}
The two-dimensional irreps have the same structure as before in
equation~(\ref{nodd-2}). Now 
we deduce that the values $k=1,\ldots,n/2$ give rise to inequivalent
representations. However, $\two^{(n/2)}$ is reducible and decays into 
$\one^{(1,0)}$ and $\one^{(1,1)}$. Therefore, we end up with 
	\be\label{Dnevenirreps}
        \two^{(k)}: \quad a \mapsto
	\bmat
	\cos(k\eta) & -\sin(k\eta)\\
	\sin(k\eta) & \cos(k\eta)
	\emat
	,\quad b\mapsto
	\bmat
	-1 & 0\\
	0 & 1	
	\emat, \quad k=1,\ldots,\frac{n}{2}-1.
	\ee
The character table of $D_n$ ($n$ even) can be found in table~\ref{charDneven}.
\begin{table}
\begin{center}
\begin{tabular}{|c|ccccc|}\hline
$D_n$ ($n$ even) & $C_1(e)$ & $C_{m+1}(a^m)$ & $C_{(n+2)/2}(a^{n/2})$ &
    $C_{(n+4)/2}(b)$ & $C_{(n+6)/2}(ab)$ \\ 
(\# $C$) & $(1)$ & $(2)$ & $(1)$ & $(n/2)$ & $(n/2)$ \\ 
$\oo(C)$ & $1$ & $r_m$ & $2$ & $2$ & $2$\\ \hline
$\one^{(0,0)}$  & $1$ & $1$ & $1$ & $1$ & $1$\\
$\one^{(0,1)}$  & $1$ & $1$ & $1$ & $-1$ & $-1$ \\
$\one^{(1,0)}$  & $1$ & $(-1)^m$ & $(-1)^{n/2}$ & $1$ & $-1$\\
$\one^{(1,1)}$  & $1$ & $(-1)^m$ & $(-1)^{n/2}$ & $-1$ & $1$ \\
$\two^{(k)}$ & $2$ & $2\cos(km\eta)$ & $2(-1)^k$ & $0$ & $0$ \\ 
\hline
\end{tabular}
\end{center}
\caption{The character table of $D_n$ ($n$
  even) with $k,m=1,\ldots,\frac{n}{2}-1$. The order of $a^m$ is $r_m =
  n/\mathrm{gcd}(m,n)$, where $\mathrm{gcd}(m,n)$ is the greatest
  common divisor of $m$ and $n$. \label{charDneven}} 
\end{table}
\paragraph{Tensor products of irreps of $D_n$:}
Here we only deal with tensor products of the two-dimensional
irreps. For other tensor products and more details we refer the reader
to~\cite{blum}. Tensor products of the two-dimensional
irreps are reduced according to 
\begin{equation}\label{2222}
\two^{(k)} \otimes \two^{(l)} = \two^{(k-l)} \oplus \two^{(k+l)}.
\end{equation}
The corresponding basis is given by
\begin{equation}
\begin{array}{cc}
F_1 = \frac{1}{\sqrt{2}} \left( e_1 \otimes e_2 - e_2 \otimes e_1
\right), &
F_2 = \frac{1}{\sqrt{2}} \left( e_1 \otimes e_1 + e_2 \otimes e_2
\right), \\
F'_1 = \frac{1}{\sqrt{2}} \left( e_1 \otimes e_2 + e_2 \otimes e_1
\right), &
F'_2 = -\frac{1}{\sqrt{2}} \left( e_1 \otimes e_1 - e_2 \otimes e_2
\right),
\end{array}
\end{equation}
where $\{ F_1, F_2 \}$ belongs to $\two^{(k-l)}$ and 
$\{ F'_1, F'_2 \}$ to $\two^{(k+l)}$. 
In order to identify $\two^{(k-l)}$ and $\two^{(k+l)}$
with the irreps of the character table\footnote{Obviously the definition of
$\two^{(k)}$ is meaningful for all $k\in\mathbbm{Z}$.} one may use
\begin{equation}
\two^{(-j)}\cong \two^{(n-j)}\cong \two^{(j)}.
\end{equation}
For special cases the
two-dimensional representations on the right-hand side of
equation~(\ref{2222}) decay into one-dimensional irreps.
For $n$ odd one has
\begin{equation}
\two^{(0)} \cong \one^{(0)} \oplus \one^{(1)},
\end{equation}
and for $n$ even one finds
\begin{equation}
\two^{(0)} \cong \one^{(0,0)} \oplus \one^{(0,1)},\quad
\two^{(n/2)} \cong \one^{(1,0)} \oplus \one^{(1,1)}.
\end{equation}
For a complete discussion see~\cite{Ramond,blum}.
\paragraph{The double cover of $D_n$:}
According to the general discussion of double covers,\footnote{Note
  that in~\cite{Ramond} $\widetilde{D}_n$ is called $Q_{2n}$. Another
  name common in the literature is $Q_{4n}$.} the order of
$\widetilde D_n$ is twice the order of $D_n$:
\begin{equation}
\oo \widetilde D_n = 4n.
\end{equation}
Using equation~(\ref{SU2calc}), we can compute the 
generators of the double cover $\widetilde{D}_n$ from
equations~(\ref{Dngeneratora}) and~(\ref{Dngeneratorb}). The result is
	\be\label{Dndoublegenerators}
	\tilde{a}:=U(\eta,\vec{e}_z)=\bmat
	e^{-i\eta/2} & 0\\
	0 & e^{i\eta/2}
	\emat,\quad \tilde{b}:=U(\pi,\vec{e}_y)=
	\bmat
	0 & -1\\
	1 & 0
	\emat.
	\ee
Note that $\widetilde D_1 \cong \zz_4$. For $n=2$ one finds 
$\tilde a = -i \sigma_3$ and $\tilde b = -i \sigma_2$. Therefore,
\begin{equation}
\widetilde D_2 = \{ \pm \bone_2, \pm i\sigma_1,  \pm i\sigma_2,  \pm
i\sigma_3 \}
\end{equation}
with the Pauli matrices $\sigma_j$. This group is also known as 
quaternion group.

Using the above generators, the presentation~(\ref{Dnpresentation}) of $D_n$
can be generalized to a presentation of $\widetilde{D}_n$:
	\be\label{Dndoublepresentation}
	\tilde{a}^{2n}=\tilde{b}^4=\tilde{b}\tilde{a}\tilde{b}^{-1}\tilde{a}=
	\tilde{a}^n\tilde{b}^{-2}=\bone_2.
	\ee
The last relation tells us that $\tilde{a}^n = \tilde{b}^2$ is in the
center of $\widetilde{D}_n$.
From equation~(\ref{Dndoublepresentation}) we find
	\be
	\tilde{b}\tilde{a}\tilde{b}^{-1}=\tilde{a}^{-1}\quad\mbox{and}\quad
	\tilde{a}\tilde{b}\tilde{a}^{-1}=\tilde{a}^2\tilde{b},
	\ee
from which we obtain the classes
	\be
	\begin{split}
	& C_{1}=\{e\},\\
	& C_2=\{\tilde{a},\,\tilde{a}^{2n-1}\},\,\ldots,\,
	  C_{n}=\{\tilde{a}^{n-1},\,\tilde{a}^{n+1}\},\\
	& C_{n+1}=\{\tilde{a}^{n}\}=\{\tilde{b}^2\},\\
	&
          C_{n+2}=\{\tilde{b},\,\tilde{a}^2\tilde{b}, \ldots, 
          \tilde{a}^{2n-2}\tilde{b}\},\\  
	& C_{n+3}=\{\tilde{a}\tilde{b},\, \tilde{a}^3\tilde{b},
          \ldots,
           \tilde{a}^{2n-1}\tilde{b}\}.
	\end{split}
	\ee

Now we want to determine all irreps of $\widetilde D_n$. To do so we
keep in mind that because of the homomorphism 
$\widetilde D_n \to D_n$ all irreps of $D_n$ are also irreps of 
$\widetilde D_n$. 
As usual, the one-dimensional irreps are determined
by using the presentation and exploiting the fact that in a one-dimensional
representation all group elements commute. 
With $\tilde a \mapsto \rho_a$ and $\tilde b \mapsto \rho_b$, 
($\rho_a, \rho_b \in \mathbbm{C}$), 
equation~(\ref{Dndoublepresentation}) reduces to
	\be
	\rho_a^{2n} = \rho_b^4 = \rho_a^{2} =
	\rho_a^n \rho_b^{-2} = 1.
	\ee
The solutions depend on whether $n$ is even or odd. For $n$ even we
find exactly the four one-dimensional irreps of $D_n$ given in 
equation~(\ref{1-neven}). For $n$ odd  
we find, in addition to the two one-dimensional irreps of $D_n$---see
equation~(\ref{1-nodd}), two additional irreps given by
$\rho_a = -1$, $\rho_b = \pm i$. 
For the two-dimensional irreps we can proceed as in the case of
$D_n$. The irrep $\two^{(1)}$ is defined by the
generators~(\ref{Dndoublegenerators}). Further irreps $\two^{(k)}$ are
obtained by taking powers of the generator $\tilde a$. However, for
${\tilde a}^n = {\tilde b}^2 = -\bone_2$, only $k$ odd is allowed in
order to satisfy ${\tilde a}^{kn} = -\bone_2$. The
remaining two-dimensional irreps are those of $D_n$, which we bring to
a form analogous to $\two^{(1)}$ by using the
similarity transformation~(\ref{similarityX}).

In summary, for $n$ odd we have the following irreps:
	\be\label{Dndoubleoddirreps}
	\begin{split}
	& \one^{(q)}: \quad \tilde{a}\mapsto 1,\quad \tilde{b}\mapsto
          (-1)^q, \quad q = 0,1, \\
	& {\one'}^{(q)}: \quad \tilde{a}\mapsto -1,\quad
          \tilde{b}\mapsto (-1)^q i, \quad q = 0,1, \\
	& \two^{(k)}: \quad \tilde{a}\mapsto
	\bmat
	e^{-ik\eta/2} & 0\\
	0 & e^{ik\eta/2}
	\emat, \quad
	\tilde{b}\mapsto
	\bmat
	0 & -1\\
	1 & 0
	\emat, \quad k=1,\,3, \ldots, n-2, \\
	& \two^{(k)}: \quad \tilde{a}\mapsto
	\bmat
	e^{-ik\eta/2} & 0\\
	0 & e^{ik\eta/2}
	\emat, \quad
	\tilde{b}\mapsto
	\bmat
	0 & -i\\
	i & 0
	\emat,
	\quad k=2,\,4, \ldots, n-1.
	\end{split}
	\ee
For $n$ even the irreps are quite similar:
	\be\label{Dndoubleevenirreps}
	\begin{split}
	& \one^{(p,q)}: \quad \tilde{a}\mapsto (-1)^p,
          \quad \tilde{b}\mapsto (-1)^q, \\
	& \two^{(k)}: \quad \tilde{a}\mapsto
	\bmat
	e^{-ik\eta/2} & 0\\
	0 & e^{ik\eta/2}
	\emat, \quad
	\tilde{b}\mapsto
	\bmat
	0 & -1\\
	1 & 0
	\emat, \quad k=1,\,3,\ldots, n-1, \\ 
	& \two^{(k)}: \quad \tilde{a}\mapsto
	\bmat
	e^{-ik\eta/2} & 0\\
	0 & e^{ik\eta/2}
	\emat, \quad
	\tilde{b}\mapsto
	\bmat
	0 & -i\\
	i & 0
	\emat,
	\quad k=2,\,4, \ldots, n-2.
	\end{split}
	\ee
Now we can write down the character tables of $\widetilde{D}_n$ (see 
tables~\ref{charDndoubleodd} and~\ref{charDndoubleeven}). 
\begin{table}
\begin{center}
\begin{tabular}{|c|ccccc|}\hline
\rule{0pt}{15pt}$\widetilde{D}_n$ ($n$ odd) & $C_1(e)$ & 
$C_{m+1}(\tilde{a}^m)$ & $C_{n+1}(\tilde{a}^{n})$ & 
$C_{n+2}(\tilde{b})$ & $C_{n+3}(\tilde{a}\tilde{b})$ \\
(\# $C$) & $(1)$ & $(2)$ & $(1)$ & $(n)$ & $(n)$ \\ 
$\oo(C)$ & $1$ & $r'_m$ & $2$ & $4$ & $4$\\ \hline
$\one^{(0)}$  & $1$ & $1$ & $1$ & $1$ & $1$\\
$\one^{(1)}$  & $1$ & $1$ & $1$ & $-1$ & $-1$ \\
${\one'}^{(0)}$  & $1$ & $(-1)^m$ & $-1$ & $i$ & $-i$\\
${\one'}^{(1)}$  & $1$ & $(-1)^m$ & $-1$ & $-i$ & $i$ \\
$\two^{(k)}$  & $2$ & $2\cos(km\eta/2)$ & $2(-1)^k$ & $0$ & $0$ \\ 
\hline
\end{tabular}
\end{center}
\caption{The character table of $\widetilde{D}_n$ ($n$
  odd) with $k,m=1,\ldots,n-1$.  The order of $a^m$ is $r'_m =
  2n/\mathrm{gcd}(m,2n)$, where $\mathrm{gcd}(m,2n)$ is the greatest
  common divisor of $m$ and $2n$. \label{charDndoubleodd}} 
\end{table}
\begin{table}
\begin{center}
\begin{tabular}{|c|ccccc|}\hline
\rule{0pt}{15pt}$\widetilde{D}_n$ ($n$ even) & 
$C_1(e)$   & $C_{m+1}(\tilde{a}^m)$ & 
$C_{n+1}(\tilde{a}^{n})$ & $C_{n+2}(\tilde{b})$ &
$C_{n+3}(\tilde{a}\tilde{b})$ \\ 
(\# $C$) & $(1)$ & $(2)$ & $(1)$ & $(n)$ & $(n)$ \\ 
$\oo(C)$ & $1$ & $r'_m$ & $2$ & $4$ & $4$\\ \hline
$\one^{(0,0)}$  & $1$ & $1$ & $1$ & $1$ & $1$\\
$\one^{(0,1)}$  & $1$ & $1$ & $1$ & $-1$ & $-1$ \\
$\one^{(1,0)}$  & $1$ & $(-1)^m$ & $1$ & $1$ & $-1$\\
$\one^{(1,1)}$  & $1$ & $(-1)^m$ & $1$ & $-1$ & $1$ \\
$\two^{(k)}$  & $2$ & $2\cos(km\eta/2)$ & $2(-1)^k$ & $0$ & $0$ \\ 
\hline
\end{tabular}
\end{center}
\caption{The character table of $\widetilde{D}_n$ ($n$
  even) with $k,m=1,\ldots,n-1$. The order of $a^m$ is $r'_m =
  2n/\mathrm{gcd}(m,2n)$, where 
  $\mathrm{gcd}(m,2n)$ is the greatest
  common divisor of $m$ and 
  $2n$. \label{charDndoubleeven}} 
\end{table}
One can make an interesting comparison between $D_{2n}$ and $\widetilde D_n$ ($n$ even),
which have the same number of elements; these groups have almost the same
character tables, differing only in the order of the group elements, i.e. in
the third line. 

Using the character tables one can deduce all tensor products.
In the tensor product 
\begin{equation}\label{2222t}
\two^{(k)} \otimes \two^{(l)} = \two^{(k+l)} \oplus \two^{(k-l)},
\end{equation}
where one has to take into account the equivalence relations
\begin{equation}
\two^{(-j)} \cong \two^{(2n-j)} \cong \two^{(j)},
\end{equation}
the invariant subspaces are spanned by 
\begin{equation}
\{ e_1 \otimes e_1,\; e_2 \otimes e_2 \} 
\quad \mbox{and} \quad
\{ e_1 \otimes e_2,\; e_2 \otimes e_1 \}.
\end{equation}
The first space corresponds to $\two^{(k+l)}$ and the second one to 
$\two^{(k-l)}$. For special cases the two-dimensional irreps on the
right-hand side of equation~(\ref{2222t}) decay into one-dimensional
irreps, namely 
\begin{equation}
\two^{(0)} = 
\one^{(0)} \oplus \one^{(1)},\quad \two^{(n)} = \one'^{(0)} \oplus \one'^{(1)}
\end{equation}
for $n$ odd and 
\begin{equation}
\two^{(0)} \cong \one^{(0,0)} \oplus \one^{(0,1)},\quad
\two^{(n)} \cong \one^{(1,0)} \oplus \one^{(1,1)}.
\end{equation}
for $n$ even. The latter relations one could also have obtained by
taking into account that, for $n$ even, $\widetilde D_n$ and $D_{2n}$ have the
same characters, and thus also the same tensor products.
For further details see~\cite{Ramond,blum}.

\subsubsection{The tetrahedral group and its double cover}
\label{chaptertetrahedral}

The tetrahedral group $T$ is the rotation symmetry group of the regular
tetrahedron. In order to construct the conjugacy classes we need all symmetry
axes of the tetrahedron. There are two groups of symmetry axes:
\begin{itemize}
 \item Four equivalent one-sided three-fold axes. These
	axes connect a vertex with the center of the opposite face.
 \item Three equivalent two-sided two-fold axes. These axes
	pass through the centers of two opposite edges of the
	tetrahedron.
\end{itemize}
\begin{figure}
\begin{center}
  \epsfig{scale=0.5,file=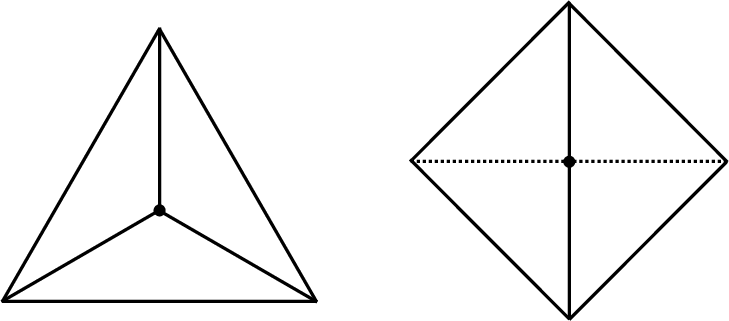}
  \caption{The two types of symmetry axes of the regular tetrahedron. 
	In the left picture the axis perpendicular to the plane of projection
	is a one-sided three-fold axis. In the picture on the right the axis
	perpendicular to the plane of projection is two-sided and two-fold.}
	\label{tetrahedronsketch}
\end{center}
\end{figure}
Using the rules presented in section~\ref{SO3classrules} we find the following
classes of the tetrahedral group:
	\begin{itemize}
	 \item The class of the identity element \\ $C_1=\{e\}$,
	 \item the class of rotations through $120^{\circ}$ about the four
           three-fold axes \\
           $C_{2}=\{a_1,\, a_2,\, a_3,\, a_4\}$,
	 \item the class of rotations through $240^{\circ}$ about the four
           three-fold axes \\ 
           $C_{3}=\{a_1^2,\, a_2^2,\, a_3^2,\, a_4^2\}$, 
	 \item the class of rotations about the three two-fold axes \\  
           $C_4=\{b_1,\,b_2,\,b_3\}$.
	\end{itemize}
Using corollary~\ref{generatorcorollary} we deduce that $T$ is generated by
the two rotations $a_1$ and $b_1$, which we rename as $a$ and $b$,
respectively. 

\begin{figure}
\begin{center}
  \epsfig{scale=0.4,file=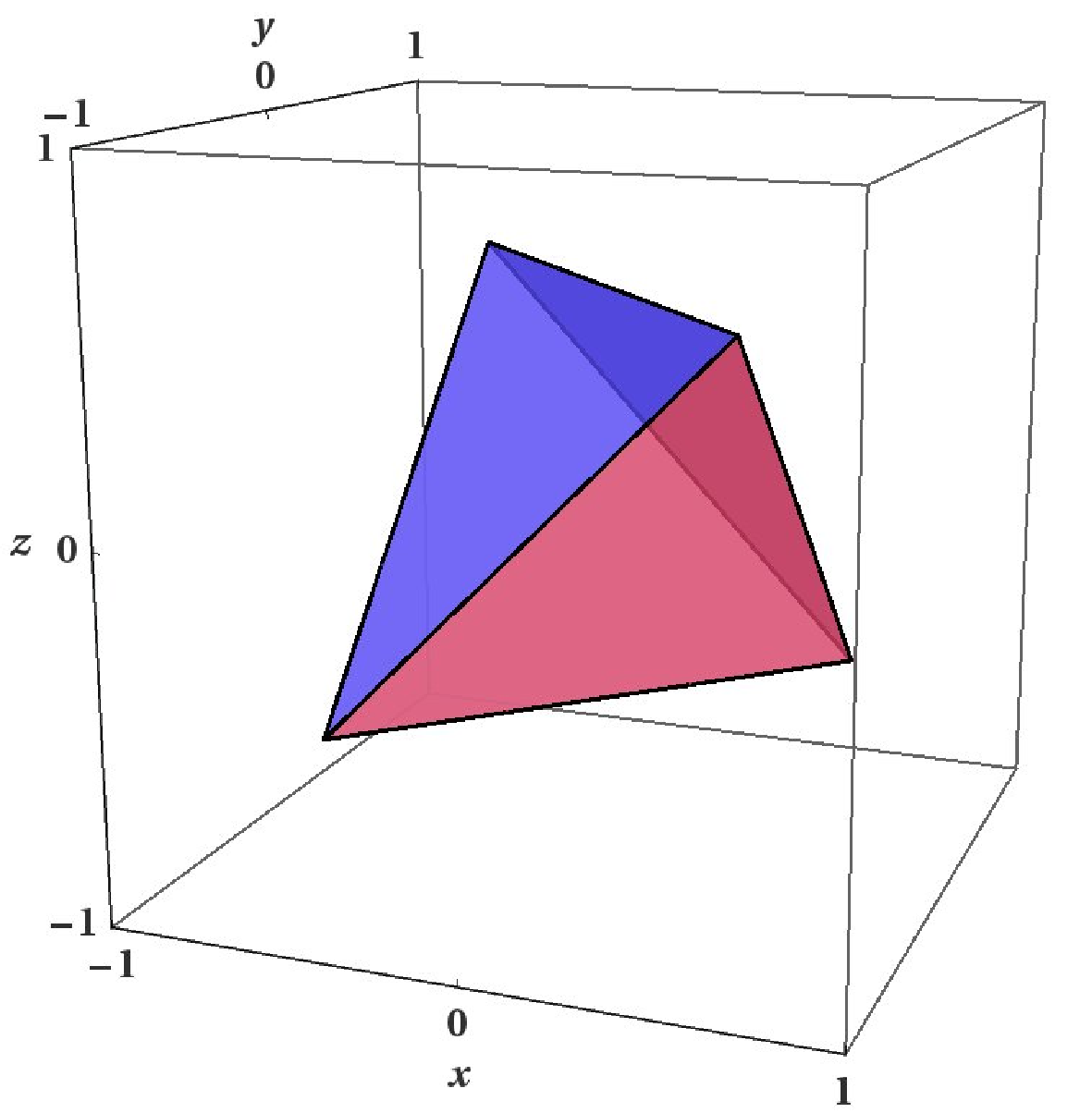}
  \caption{Three-dimensional plot of the tetrahedron with the vertices
  of equation~(\ref{t-vertices}).}
\label{tetrahedron1} 
\end{center}
\end{figure}

Now we want to demonstrate that $T$ is isomorphic to $A_4$. It suffices
to show that $T$ contains the matrices $A$ and $E$ of equation~(\ref{AE}). Of
course, this statement depends on the position and orientation of the
tetrahedron in space. 
Let us assume that the two-fold rotation axis of $b$ is the
$x$-axis and that the rotation axis of $a$ is given by
	\be\label{n}
	\vec{n} = -\frac{1}{\sqrt{3}}\bmat
	1\\1\\1
	\emat.
	\ee
Then we obtain
	\be\label{ab}
	a := R(2\pi/3,\vec{n})=\bmat
	0 & 1 & 0\\
	0 & 0 & 1\\
	1 & 0 & 0	
	\emat
\quad \mbox{and} \quad
	b := R(\pi,\vec{e}_x)=\bmat
	1 & 0 & 0\\
	0 & -1 & 0\\
	0 & 0 & -1	
	\emat,
	\ee
which agree with $E$ and $A$, respectively. Considering $\vec n$ of
equation~(\ref{n}) as one vertex of the tetrahedron, then by successive
application of $a$ and $b$ on $\vec n$ one obtains the remaining three-fold
axes and, therefore, the remaining vertices. Consequently, in our geometric
setting the tetrahedron has the vertices
\begin{equation}\label{t-vertices}
\frac{1}{\sqrt{3}} \left( \begin{array}{r} -1 \\ -1 \\ -1
\end{array} \right), \quad
\frac{1}{\sqrt{3}} \left( \begin{array}{r} 1 \\ 1 \\ -1
\end{array} \right), \quad
\frac{1}{\sqrt{3}} \left( \begin{array}{r} -1 \\ 1 \\ 1
\end{array} \right), \quad
\frac{1}{\sqrt{3}} \left( \begin{array}{r} 1 \\ -1 \\ 1
\end{array} \right),
\end{equation}
and $b$, as given in equation~(\ref{ab}), is indeed a symmetry rotation
through $180^\circ$ of the tetrahedron defined by these vertices---see
figure~\ref{tetrahedron1}.  
Thus we have demonstrated that the rotation symmetry group $T$ of the
tetrahedron is generated by the rotation matrices~(\ref{ab}) and that $T$ is,
therefore, by virtue of equation~(\ref{AE}) isomorphic to $A_4$.
A three-dimensional model of a tetrahedron greatly
facilitates the above considerations. 
In summary, comparing with equations~(\ref{AE}) and~(\ref{Ak2})
we find that 
	\be
	s=(123)\mapsto a,\quad
	k_1=(12)(34)\mapsto b
	\ee
provides the isomorphism between $A_4$ and $T$.

Now we express the elements of the classes in terms of $a$ and $b$: 
	\be
\begin{array}{l}
	C_{1}=\{e\}, \quad
	C_2=\{a,\, bab^{-1},\, ab,\, ba\}, \\
C_3=\{a^2,\, ba^2b^{-1},\, (ab)^2,\, (ba)^2\}, \quad 
C_{4}=\{b,\,aba^{-1},\,a^2 ba^{-2}\}. 
\end{array}
\ee
The classes $C_1$ and $C_4$ are straightforward, the other two 
can be found by using conjugation with $b$ and 
the relation
$(ab)a(ab)^{-1}=ba$.

A presentation of $A_4$ is given by equation~(\ref{s4k}) together with 
$k_2 k_3=k_1$. 
Expressing these equations in terms of the generators $a$ and $b$, we
obtain the well-known presentation of $T\cong A_4$:
	\be\label{Tpresentation}
	a^3=b^2=(ab)^3=\bone.
	\ee
Using the presentation, we easily find the one-dimensional irreps
	\be\label{T1dimirreps}
	\one^{(p)}: \quad a\mapsto \omega^p,\quad b\mapsto 1
\quad \mbox{with} \quad p = 0,1,2
	\ee
and $\omega=\exp(2\pi i/3)$.
These together with the defining irrep
	\be
	\three:\quad a\mapsto \bmat
	0 & 1 & 0\\
	0 & 0 & 1\\
	1 & 0 & 0	
	\emat,\quad b\mapsto
	\bmat
	1 &  0 & 0 \\
	0 & -1 & 0 \\
	0 & 0 & -1	
	\emat
	\ee
comprise all irreps of $T$. Its character table is shown in
table~\ref{Tchar}. At present, the tetrahedral group is the most
extensively used flavour 
group in the lepton sector---see for instance~\cite{ma,A4} and the references
in~\cite{review}. 
\begin{table}
\begin{center}
\begin{tabular}{|c|cccc|}\hline
$T\cong A_4$  & $C_1$   & $C_2$ &  $C_3$ & $C_4$ \\
(\# $C$) & $(1)$ & $(4)$ & $(4)$ & $(3)$ \\ 
$\oo(C)$ & $1$ & $3$ & $3$ & $2$ \\ \hline
$\one^{(0)}$  & $1$ & $1$ & $1$ & $1$ \\
$\one^{(1)}$  & $1$ & $\omega$ & $\omega^2$ & $1$ \\
$\one^{(2)}$  & $1$ & $\omega^2$ & $\omega$ & $1$\\
$\three$  & $3$ & $0$ & $0$ & $-1$ \\ 
\hline
\end{tabular}
\end{center}
\caption{The character table of $T\cong A_4$ with $\omega=\exp(2\pi i/3)$.
\label{Tchar}}
\end{table}

Finally, we discuss the famous tensor product
\begin{equation}\label{TxT}
\three \otimes \three = 
\one^{(0)} \oplus \one^{(1)} \oplus \one^{(2)} \oplus \three \oplus \three,
\end{equation}
which is the basis of all model building with $A_4$~\cite{ma,review,A4}.
The proof of this relation consists of providing a basis where 
equation~(\ref{TxT}) is explicit. For the one-dimensional irreps such a basis
is given by
\begin{equation}\label{BT}
\begin{array}{cccl}
\one^{(0)}: &
\mathcal{B}_0 &=& \frac{1}{\sqrt{3}} \left( 
e_1 \otimes e_1 + e_2 \otimes e_2 + e_3 \otimes e_3 \right), \\
\one^{(1)}: &
\mathcal{B}_1 &=& \frac{1}{\sqrt{3}} \left( 
e_1 \otimes e_1 + \omega e_2 \otimes e_2 + \omega ^2 e_3 \otimes e_3
\right), \\
\one^{(2)}: & 
\mathcal{B}_2 &=& \frac{1}{\sqrt{3}} \left( 
e_1 \otimes e_1 + \omega ^2 e_2 \otimes e_2 + \omega e_3 \otimes e_3 \right).
\end{array}
\end{equation}
For the two identical three-dimensional irreps on the right-hand side of
equation~(\ref{TxT}) one has a lot of freedom in the choice of basis. 
For instance, one can choose 
\begin{equation}
e_2 \otimes e_3,\; e_3 \otimes e_1,\; e_1 \otimes e_2 
\quad \mbox{and} \quad
e_3 \otimes e_2,\; e_1 \otimes e_3,\; e_2 \otimes e_1,
\end{equation}
or the symmetrized and antisymmetrized versions of these bases.

Also the double cover $\widetilde{T}$ of the tetrahedral group
has been used for model building---see e.g.~\cite{aranda,hagedorn}. It is
generated by 
	\be\label{tildeTgenerators}
	\tilde{a}: = U(2\pi/3,\vec{n})=\frac{1}{2}
\left( \begin{array}{rr}
	1+i & 1+i\\
	-1+i & 1-i 
\end{array} \right)
	\quad\mbox{and}
	\quad
	\tilde{b}:= -U(\pi,\vec{e}_x)=\bmat
	0 & i \\
	i & 0
	\emat,
	\ee
where $\vec n$ is given by equation~(\ref{n}). Note the minus sign in the
definition of $\tilde b$. It allows to formulate the presentation as~\cite{GL10}
\begin{equation}
{\tilde b}^4 = {\tilde a}^3 {\tilde b}^{-2} = 
( \tilde a \tilde b )^3 = e.
\end{equation}
Switching to a basis in which $\tilde{a}$ is diagonal we
obtain the irrep
	\be
	\two^{(0)}:\quad \tilde{a}\mapsto\bmat
	-\omega & 0\\
	0 & -\omega^2
	\emat,\quad
	\tilde{b}\mapsto
	\bmat
	-\frac{i}{\sqrt{3}} & \sqrt{\frac{2}{3}}\\
	-\sqrt{\frac{2}{3}} & \frac{i}{\sqrt{3}}
	\emat.
	\ee
Two further inequivalent two-dimensional irreps are obtained by
\begin{equation}
\label{2p}
\two^{(p)} := \one^{(p)} \otimes \two^{(0)}
\quad \mbox{with} \quad p = 1,2,
\end{equation}
with the $\one^{(p)}$ being the one-dimensional irreps of $T$.
Thus we have constructed three inequivalent two-dimensional irreps of 
$\widetilde T$. The remaining irreps are given by the irreps of $T$ via the
homomorphism $\tilde a \mapsto a$, $\tilde b \mapsto b$. 
According to 
	\be
	3\times 1^2 + 3\times 2^2 + 3^2 = 24 = \oo\widetilde{T},
	\ee
theorem~\ref{n2} tells us that these are all irreps of
$\widetilde{T}$. For the tensor products we refer the reader
to~\cite{hagedorn}. 

In order to construct the character table we need the conjugacy
classes. Since any homomorphism maps conjugacy classes onto conjugacy
classes, by looking at the classes of $T$ we immediately
find the following classes of 
$\widetilde{T}$:
	\be
	C_1(\bone_2),\enspace C_2(-\bone_2),\enspace C_3(\tilde{a}),
	\enspace C_6(\tilde{a}^2),\enspace C_7(\tilde{b}).
	\ee
Here we did not only number the classes but also characterized them by
indicating one element in each class. 
Since in the defining irrep~(\ref{tildeTgenerators}), which is equivalent
to $\two^{(0)}$, we have $\mathrm{Tr}\,\tilde{a}\neq 0$, it is evident
that $\tilde{a}$ and $-\tilde{a}$ must be in different classes.
The same holds
for $\tilde{a}^2$ and $-\tilde{a}^2$. Therefore, the two missing
classes have to be\footnote{Note that $\tilde{a}^2$ and $-\tilde{a}$, though
  they have equal trace, cannot be equivalent because this would lead to
  $C_2(a)=C_3(a^2)$ in $T$ which is not the 
  case.}
	\be
	C_4(-\tilde{a})\quad\mbox{and}\quad C_6(-\tilde{a}^2).	
	\ee 
The character table of $\widetilde{T}$ is presented in table~\ref{Ttildechar}.
\begin{table}
\begin{center}
\begin{tabular}{|c|ccccccc|}\hline
\rule{0pt}{15pt}$\widetilde{T}$  & $C_1(\bone_2)$   &
$C_2(-\bone_2)$ & 
$C_3(\tilde{a})$ & $C_4(-\tilde{a})$ & $C_5(\tilde{a}^2)$ & 
$C_6(-\tilde{a}^2)$ & $C_7(\tilde b)$ \\
(\# $C$) & $(1)$ & $(1)$ & $(4)$ & $(4)$ & $(4)$ & $(4)$ & $(6)$ \\ 
$\oo(C)$ & $1$ & $2$ & $6$ & $3$ & $3$ & $6$ & $4$\\ \hline
$\one^{(0)}$  & $1$ & $1$ & $1$ & $1$ & $1$ & $1$ & $1$\\
$\one^{(1)}$  & $1$ & $1$ & $\omega$ & $\omega$ & $\omega^2$ & $\omega^2$ &
$1$ \\ 
$\one^{(2)}$  & $1$ & $1$ & $\omega^2$ & $\omega^2$ & $\omega$ & $\omega$ &
$1$ \\ 
$\two^{(0)}$  & $2$ & $-2$ & $1$ & $-1$ & $-1$ & $1$ & $0$\\
$\two^{(1)}$  & $2$ & $-2$ & $\omega$ & $-\omega$ & $-\omega^2$ & $\omega^2$ &
$0$\\ 
$\two^{(2)}$  & $2$ & $-2$ & $\omega^2$ & $-\omega^2$ & $-\omega$ & $\omega$ &
$0$\\ 
$\three$  & $3$ & $3$ & $0$ & $0$ & $0$ & $0$ & $-1$\\
\hline
\end{tabular}
\end{center}
\caption{The character table of $\widetilde{T}$ with
  $\omega=\exp(2\pi i/3)$. 
\label{Ttildechar}}
\end{table}
With this table and equation~(\ref{psrc}) it is straightforward to
check that $\two^{(0)}$ is a pseudoreal representation while
$\two^{(1)}$ and $\two^{(2)}$ are complex. These facts can also be
understood from the point of view of $SU(2)$: The matrices of the
irrep $\two^{(0)}$ have determinant $+1$ and can, therefore, be
conceived as a finite subgroup of $SU(2)$. 
Since $SU(2)$ is pseudoreal, all subgroups must be pseudoreal as well.
On the other hand, from equation~(\ref{2p}) it is clear that 
the irreps $\two^{(1)}$ and $\two^{(2)}$ cannot be identified with the
fundamental two-dimensional representation of $SU(2)$, but instead have
to arise from the decomposition of larger $SU(2)$ irreps~\cite{Ramond}.

\subsubsection{The octahedral group and its double cover}

The octahedral group $O$ is the rotation symmetry group of the regular
octahedron and the cube~\cite{Speiser}. 
In order to obtain the classes with the help of the
rules presented in section~\ref{SO3classrules}, we 
list the different types of rotation axes of the octahedron.

An octahedron has six vertices, twelve edges and eight faces; the latter
number is responsible for the name of this solid. Using a
three-dimensional model of an octahedron, one can easily find the different
types of axes and their properties:
\begin{itemize}
 \item Type 1: Axes connecting two opposite vertices. The octahedron 
	 has three equivalent four-fold axes of this type. These axes are
         two-sided and lead to two conjugacy classes. The conjugacy class of
         rotations through $\pm 90^\circ$ has six elements, while the conjugacy
         class of rotations through $180^\circ$ has three elements. 
 \item Type 2: Axes passing through the centers of two opposite edges.
	This type leads to six equivalent two-fold two-sided axes giving rise to
	a single conjugacy class containing six elements. 
 \item Type 3: Axes passing through the centers of two opposite faces. This
        type of axes comprises four equivalent three-fold axes. Since
        these axes are two-sided, they lead to only one class containing
        eight elements, namely rotations through $\pm 120^{\circ}$. 
\end{itemize}
Summarizing, we list the classes of the octahedron:
	\be\label{octahedralclassstructure}
	\begin{matrix}
	\mbox{axis type} & 0 & 1 & 1 & 2 & 3 \\
	     (\# C)& (1) & (6) & (3) & (6) & (8) \\
	\mathrm{ord}\,C & 1 & 4 & 2 & 2 & 3
	\end{matrix}	
	\ee
The class of the identity element is denoted by axis type~0.

\begin{figure}
\begin{center}
  \epsfig{scale=0.4,file=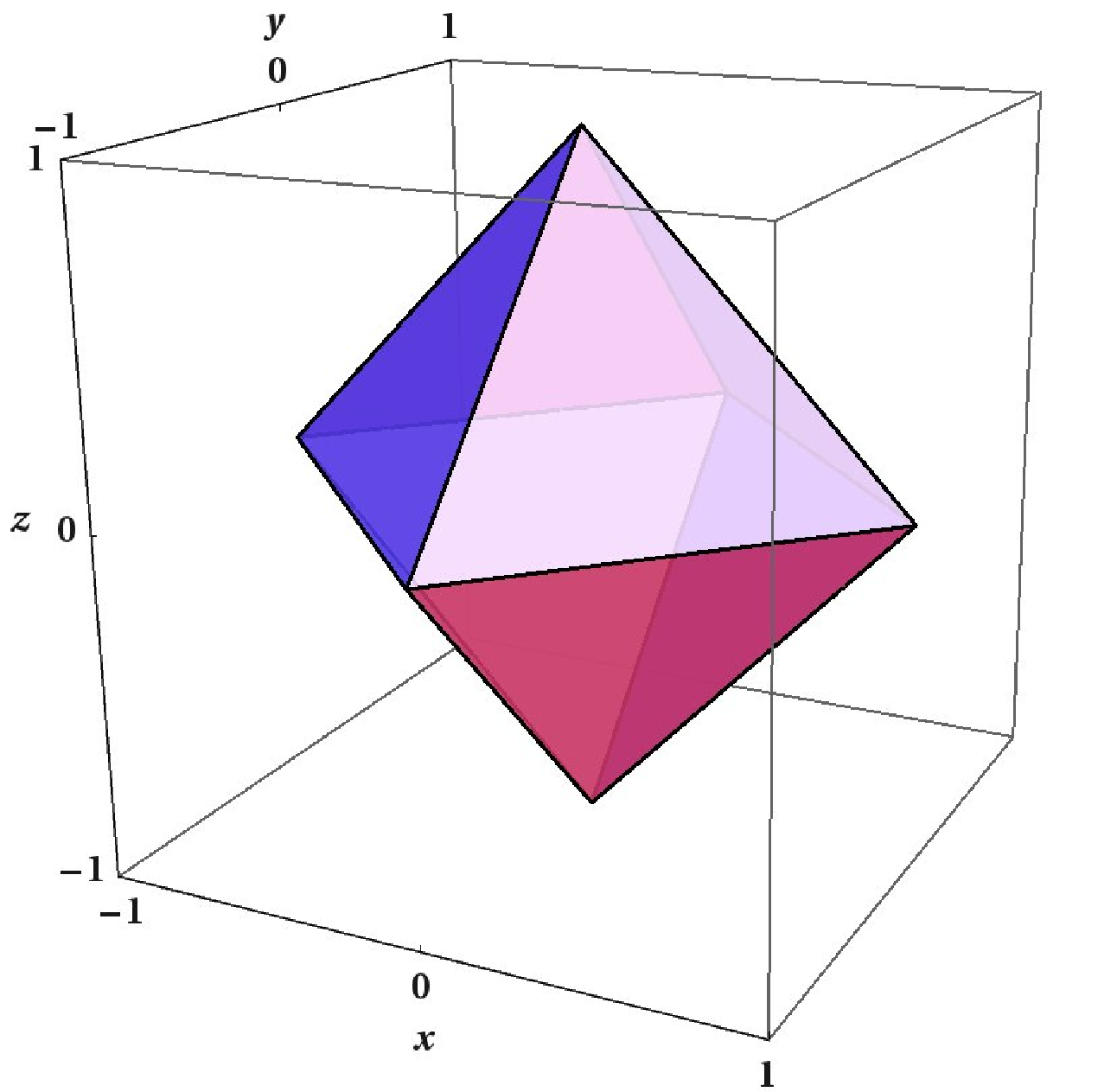}
  \caption{Three-dimensional plot of the octahedron with the vertices
$\pm \vec e_x$, $\pm \vec e_y$, $\pm \vec e_z$.}
\label{octahedron} 
\end{center}
\end{figure}

In section~\ref{chaptertetrahedral} we have learned that the tetrahedral
group is isomorphic to the permutation group $A_4$.  
Now we want to show that $O$ is isomorphic to $S_4$. In section~\ref{SSS} we
have represented $S_4$ as a rotation group, generated by the matrices $A$ and
$E$ of equation~(\ref{AE}) and the matrix $R_t$ of equation~(\ref{Rt}).
Therefore, also 
\begin{equation}\label{Rx}
E^ {-1} A E R_t =: R_x = 
\left( \begin{array}{ccc}
1 & 0 & 0 \\ 0 & 0 & -1 \\ 0 & 1 & 0 
\end{array} \right)
\end{equation}
belongs to the rotational representation of $S_4$. Obviously, $R_x$ is a
rotation about the $x$-axis through $90^\circ$. This inspires us to position
the vertices of the octahedron at the points $\pm \vec e_x$,
$\pm \vec e_y$, $\pm \vec e_z$. This octahedron is depicted in
figure~\ref{octahedron}. 
In this way, $R_x$ is a symmetry
rotation through $90^\circ$ 
about the four-fold axis $\vec e_x$ of the octahedron. But then,
trivially, also $A = R_x^2$ is a symmetry rotation through $180^\circ$ about
the same axis, and $E$ rotates the octahedron about $120^\circ$ through the
axis given by equation~(\ref{n}). Having seen that $A$, $E$ and $R_x$, and
thus also $R_t$, are
symmetry rotations of the octahedron defined above via its vertices and
knowing that $A$, $E$ and $R_t$ generate $S_4$ as a rotation group, we have
indeed demonstrated the isomorphism $O \cong S_4$.
Explicitly, it is given by
\begin{equation}\label{isoS4-O}
s = (123) \mapsto E, \quad k_1 = (12)(34) \mapsto A, \quad
t = (12) \mapsto R_t.
\end{equation}

Having established the isomorphism~(\ref{isoS4-O}), 
in the following, we will---for the sake of simplicity---use the
notation of $S_4$ as introduced in section~\ref{chapterpermutation}. 
In section~\ref{SSS} it was shown that $S_4$ is generated by $A_4$, which is
itself generated by $k_1=(12)(34)$ and $s=(123)$, and the 
transposition $t=(12)$.
Theorem~\ref{sn-cycles} tells us that the classes of $S_n$ correspond
to the different cycle structures. Equation~(\ref{S4}) shows the five
different cycle structures of the elements of $S_4$. 
We can model a four-cycle after $R_x$: 
\begin{equation}\label{rxr}
r := s^{-1} k_1 s t = (1423).
\end{equation}
We denote the classes of $S_4$,
in the order of equation~(\ref{S4}), by specifying one element of each class: 
	\be
	C_1(e),\enspace C_2(t),\enspace C_3(k_1),\enspace C_4(s),\enspace
        C_5(r). 
	\ee
Using the generators of $S_4$ it is easy to calculate $(\# C)$ and
$\mathrm{ord}\,C$ to see the correspondence with
the list of symmetry rotations~(\ref{octahedralclassstructure})---see
also table~\ref{charS4}. Also $S_4$ is a favoured group for model
building--see for instance~\cite{S4}

Now we compute the irreps of $O \cong S_4$. 
We already have the representation of $S_4$ as a three-dimensional rotation
group, i.e.\ as the octahedral group. Denoting this irrep by $\three$, 
in summary it is given by
	\be
	\three:\quad 
	t\mapsto R_t = \bmat
	-1 & 0 & 0\\
	0 & 0 & -1\\
	0 & -1 & 0
	\emat,\quad
	k_1 \mapsto A = 
	\bmat
	1 & 0 & 0\\
	0 & -1 & 0\\
	0 & 0 & -1	
	\emat,\quad
	s\mapsto E = \bmat
	0 & 1 & 0\\
	0 & 0 & 1\\
	1 & 0 & 0	
	\emat.
	\ee
Theorem~\ref{Snirreps} tells us that there are only two
one-dimensional irreps, namely
	\be
	\begin{array}{clcc}
	\one:  & t\mapsto 1, & k_1 \mapsto 1, & s \mapsto 1,\\
	\one': & t\mapsto -1, & k_1\mapsto 1, & s \mapsto 1.
	\end{array}
	\ee
Then we obtain another three-dimensional irrep via
	\be
	\three':=\one' \otimes \three.
	\ee
Since $S_4$ possesses five classes, one irrep is missing. 
Using theorem~\ref{n2} we find its dimension to be two:
	\be
	24=2\times 1^2 + 2\times 3^2 + d^2 \;\Rightarrow\; d=2.
	\ee
The origin of this two-dimensional irrep can be understood from
equation~(\ref{s4kz3z2}). Since Klein's four-group $K$ is
a normal subgroup of $S_4$, all irreps of
	\be
	S_4/K\cong \zz_3\rtimes\zz_2\cong S_3
	\ee
are irreps of $S_4$ too~\cite{GLL}. The identity element of $S_4/K$ is $K$,
thus $k_1\in K$ must be mapped onto $\bone_2$. The representation matrices
of $s$ and $t$ follow from the representation of $S_3$ as a
three-dimensional rotation group---see equations~(\ref{sS3}) and~(\ref{tS3}):
	\be\label{s4-2}
	\two:\quad t\mapsto\bmat
	0 & 1\\
	1 & 0
	\emat,\quad
	k_1\mapsto\bmat
	1 & 0\\
	0 & 1
	\emat,\quad
	s\mapsto\frac{1}{2}\bmat
	-1 & -\sqrt{3}\\
	\sqrt{3} & -1
	\emat.
	\ee
With a basis transformation one can make the representation matrix of $s$
diagonal without altering the other two representation matrices: 
	\be\label{s4-2'}
	\two:\quad t\mapsto\bmat
	0 & 1\\
	1 & 0
	\emat,\quad
	k_1\mapsto\bmat
	1 & 0\\
	0 & 1
	\emat,\quad
	s\mapsto \bmat
	\omega & 0 \\
	0 & \omega^2
	\emat
\quad \mbox{with} \quad \omega = e^{2\pi i/3}.
	\ee
Sometimes this form of the $\two$ is useful. 
We now have all information needed in order to compute the
character table of $S_4$, which is depicted in table~\ref{charS4}.
We want to mention that $S_4$ is quite popular as a flavour group---see for
instance~\cite{GLL,S4} and references in~\cite{ludl10}. 
\begin{table}
\begin{center}
\begin{tabular}{|c|ccccc|}\hline
$O\cong S_4$  & $C_1(e)$   & $C_2(t)$ &  $C_3(k_1)$ & $C_4(s)$ & $C_5(r)$ \\
(\# $C$) & $(1)$ & $(6)$ & $(3)$ & $(8)$ & $(6)$ \\ 
$\oo(C)$ & $1$ & $2$ & $2$ & $3$ & $4$ \\ \hline
$\one$  & $1$ & $1$ & $1$ & $1$ & $1$ \\
$\one'$  & $1$ & $-1$ & $1$ & $1$ & $-1$ \\
$\two$ & $2$ & $0$ & $2$ & $-1$ & $0$ \\
$\three$  & $3$ & $-1$ & $-1$ & $0$ & $1$ \\ 
$\three'$  & $3$ & $1$ & $-1$ & $0$ & $-1$ \\ 
\hline
\end{tabular}
\end{center}
\caption{The character table of $O\cong S_4$.
\label{charS4}}
\end{table}

As an example for tensor products of irreps of $S_4$ we compute 
$\three \otimes \three$. Taking advantage of the character table~\ref{charS4}
we obtain
\begin{equation}\label{OxO}
\three \otimes \three = \one \oplus \two \oplus \three \oplus \three'.
\end{equation}
Since $T$ is a subgroup of $O$, equation~(\ref{OxO}) must be related
to equation~(\ref{TxT}). Indeed, 
with the vectors defined in equation~(\ref{BT}), the first two summands on the
right-hand side of equation~(\ref{OxO}) are obtained in the following way:
\begin{equation}
\one: \;\; \mathcal{B}_0, \quad
\two: \;\; \mathcal{B}_1,\; \mathcal{B}_2.
\end{equation}
With this basis the $\two$ emerges in the form~(\ref{s4-2'}), which
clearly shows that the $\two$ is composed of $\one^{(1)}$ and 
$\one^{(2)}$ of $T$. 
For the three-dimensional irreps one now has to use the symmetrized
and antisymmetrized bases:
\begin{equation}
\frac{1}{\sqrt{2}} \left( e_2 \otimes e_3 \pm e_3 \otimes e_2 \right), 
\quad
\frac{1}{\sqrt{2}} \left( e_3 \otimes e_1 \pm e_1 \otimes e_3 \right),
\quad
\frac{1}{\sqrt{2}} \left( e_1 \otimes e_2 \pm e_2 \otimes e_1 \right).
\end{equation}
The plus sign refers to the $\three'$ and the minus sign to the $\three$.

Now we move on to $\widetilde O$, the double cover of $O$.
The generators of $\widetilde O$ are 
	\be\label{DoubleS4generators}
	\tilde{t}:=U(\pi,\vec{n}_2),\quad
        \tilde{k}_1:=-U(\pi,\vec{e}_x), \quad 
	\tilde{s}:=U(2\pi/3,\vec{n}_3)
        \ee
with
\begin{equation}
\vec{n}_2 = \frac{1}{\sqrt{2}} \left(
\begin{array}{c} 0 \\ 1 \\ -1 \end{array} \right), \quad
\vec{n}_3 = -\frac{1}{\sqrt{3}} \left(
\begin{array}{c} 1 \\ 1 \\ 1 \end{array} \right).
\end{equation}
Equation~(\ref{DoubleS4generators}) can be conceived as the defining
irrep of $\widetilde O$ which explicitly is given by
	\be\label{S4definingirrep}
	\widetilde\two:\quad
	\tilde t = \frac{1}{\sqrt{2}}\bmat
	i & -1\\
	1 & -i
	\emat,\quad
	\tilde{k}_1 =
	\bmat
	0 & i\\
	i & 0
	\emat,\quad 
	\tilde{s} = \frac{1}{2}
\left( \begin{array}{rc}
	1+i  & 1+i\\
	-1+i & 1-i
\end{array} \right).
	\ee
In the further discussion we will also utilize the elements 
\begin{equation}
\tilde r = {\tilde s}^{-1} \tilde k_1 \tilde s \tilde t = 
\frac{1}{\sqrt{2}} \left( 
\begin{array}{rr} 1 & -i \\ -i & 1 
\end{array} \right) 
\quad \mbox{and} \quad
\tilde k_2 = \tilde s \tilde k_1 {\tilde s}^{-1} = 
\left( \begin{array}{cc} i & 0 \\ 0 & -i 
\end{array} \right).
\end{equation}
In order to find all irreps of $\widetilde O$ and the character table,
we need to know the classes. We proceed in the 
same way as we did for the classes of $\widetilde{T}$.
We take over the classes 
$C_1(\bone_2)$, $C_3(\tilde{t})$, $C_4(\tilde{k}_1)$,
$C_5(\tilde{s})$ and $C_7(\tilde{r})$ from the octahedral group.
Because of 
	\be
	\mathrm{Tr}\,\bone_2\neq 0,\enspace \mathrm{Tr}\,\tilde{s}\neq 0,
	\enspace \mathrm{Tr}\,\tilde{r} \neq 0
	\ee
we find in the case of $\widetilde O$ three more classes:
	\be\label{S4doublenewclasses}
	C_2(-\bone_2),\enspace C_6(-\tilde{s})\quad\mbox{and}\quad
        C_8(-\tilde{r}). 
	\ee
The relations
\begin{equation}
\tilde k_1 \tilde t {\tilde k_1}^{-1} = -\tilde t
\quad \mbox{and} \quad
\tilde k_2 \tilde k_1 {\tilde k_2}^{-1} = -\tilde k_1
\end{equation}
demonstrate that there are no further classes. Since $\widetilde{O}$
possesses eight conjugacy classes, it also possesses eight inequivalent
irreps. We already know six of them, namely the five irreps of $O$
and the defining irrep~(\ref{S4definingirrep}). The dimensions of the
unknown irreps can be computed by means of theorem~\ref{n2}:
	\be
	48=2\times 1^2 + 2\times 2^2 + 2\times 3^2 + d_7^2 + d_8^2
\; \Rightarrow \;
	d_7=2,\quad d_8=4.
	\ee
The missing two-dimensional irrep is given by 
\begin{equation}
{\widetilde\two}': =  \one' \otimes \widetilde\two. 
\end{equation}
Note that we keep the notation $\one'$ for the one-dimensional
non-trivial irrep of $O$. Keeping also the notation for the $\two$, 
one can check that the tensor product $\two \otimes \widetilde\two$
fulfills 
	\be
	\langle \chi^{(\two \otimes \widetilde\two)}\vert
	\chi^{(\two \otimes \widetilde\two)}
	\rangle = 1.
        \ee
Then we know from theorem~\ref{irred} that
\begin{equation}
\widetilde\four: = \two \otimes \widetilde\two
\quad \mbox{is irreducible}.
\end{equation}
Having obtained all classes and irreps we can compute the
character table which is presented in table~\ref{S4doublechar}.
\begin{table}
\begin{center}
\begin{tabular}{|c|cccccccc|}\hline
\rule{0pt}{15pt}$\widetilde{O}$  & $C_1(\bone_2)$ &
$C_2(-\bone_2)$ & 
$C_3(\tilde{t})$ &  $C_4(\tilde{k}_1)$ & $C_5(\tilde{s})$ & $C_6(-\tilde{s})$ &
$C_7(\tilde{r})$  & $C_8(-\tilde{r})$\\
(\# $C$) & $(1)$ & $(1)$ & $(12)$ & $(6)$ & $(8)$ & $(8)$ & $(6)$ & $(6)$ \\ 
$\oo(C)$ & $1$ & $2$ & $4$ & $4$ & $6$ & $3$ & $8$ & $8$\\ \hline
$\one$  & $1$ & $1$ & $1$ & $1$ & $1$ & $1$ & $1$ & $1$\\
$\one'$  & $1$ & $1$ & $-1$ & $1$ & $1$ & $1$ & $-1$ & $-1$\\
$\two$  & $2$ & $2$ & $0$ & $2$  & $-1$ & $-1$ & $0$ & $0$\\
$\widetilde\two$  & $2$ & $-2$ & $0$ & $0$  & $1$ & $-1$ & $\sqrt{2}$
& $-\sqrt{2}$\\ 
${\widetilde\two}'$  & $2$ & $-2$ & $0$ & $0$  & $1$ & $-1$ &
$-\sqrt{2}$ & $\sqrt{2}$ \\ 
$\three$ & $3$ & $3$ & $-1$ & $-1$  & $0$ & $0$ & $1$ & $1$\\
$\three'$ & $3$ & $3$ & $1$ & $-1$  & $0$ & $0$ & $-1$ & $-1$\\
$\widetilde{\four}$ & $4$ & $-4$ & $0$ & $0$ & $-1$ & $1$ & $0$ & $0$\\
\hline
\end{tabular}
\end{center}
\caption{The character table of $\widetilde{O}$.
The irreps without tilde correspond to the irreps of $O$.
\label{S4doublechar}}
\end{table}

\subsubsection{The icosahedral group and its double cover}
\label{icodouble}
The icosahedral group $I$ is the rotation symmetry group
of both the regular icosahedron and the regular
dodecahedron~\cite{Speiser}. 
In order to construct the classes of the icosahedral group, we
once more make use of the rules presented in section~\ref{SO3classrules}.
The regular icosahedron possesses twelve vertices, thirty edges and twenty
faces, which are unilateral triangles. The denomination of this solid follows
from the Greek word for ``twenty.''
Using a three-dimensional model of an icosahedron, one can easily
find the different types of rotation axes and their properties:
\begin{itemize}
 \item Type 1: Axes pointing through two opposite vertices. The icosahedron
        has six equivalent five-fold axes of this type. Since these axes are
        two-sided, with any element also its inverse belongs to the same
        class. Therefore, we 
	find two classes, the class of rotations through $\pm72^{\circ}$ and the
	class of rotations through $\pm144^{\circ}$, containing twelve
        elements each. 
 \item Type 2: Axes connecting the centers of two opposite edges.
	Fifteen equivalent two-fold two-sided axes are of this type, 
        giving rise to a single conjugacy class containing fifteen elements. 
 \item Type 3: Axes pointing through the
	centers of two opposite faces. This type of axes corresponds to ten
        equivalent three-fold axes. One finds that these axes are also
        two-sided, so we obtain a single conjugacy class consisting of twenty
        elements which are rotations through $\pm 120^{\circ}$.
\end{itemize}
Thus we arrive at the following list of classes of $I$:
	\be\label{icosahedralclassstructure}
	\begin{matrix}
	\mbox{axis type} & 0 & 1 & 1 & 2 & 3 \\
	(\# C)& (1) & (12) & (12) & (15) & (20) \\
	\mathrm{ord}\,C & 1 & 5 & 5 & 2 & 3
	\end{matrix}	
	\ee
The class of the identity element is denoted by axis type~0.
From the above list we find $\oo I=60$. 

In the previous section we have already seen that 
the tetrahedral and the octahedral groups are isomorphic to the permutation
groups $A_4$ and $S_4$, respectively. Furthermore, in section~\ref{SSS}, we
have derived a representation of $A_5$ as a three-dimensional rotation group of
order 60. Since also the classes~(\ref{icosahedralclassstructure}) match the
classes of $A_5$---see equation~(\ref{ca5}) and the subsequent discussion
there, it is suggestive to assume that $A_5$ is isomorphic to $I$. 
That this is the case can be proven by
constructing the vertices of the icosahedron from the 
rotational representation of $A_5$. This is done in appendix~\ref{A5I}. 

Having established that the rotation symmetry group $I$ of a
regular icosahedron is isomorphic to $A_5$, it is then clear from the
material in section~\ref{SSS} that
the concrete realization of this isomorphism is given by 
	\be\label{isoIA5}
	s=(123) \mapsto E, \quad k_1=(12)(34) \mapsto A, \quad 
        k_4=(12)(45) \mapsto W,
	\ee
where $A$ and $E$ are defined in equation~(\ref{AE}), and $W$ is defined in
equation~(\ref{WW'}).
\begin{figure}
\begin{center}
  \epsfig{scale=0.4,file=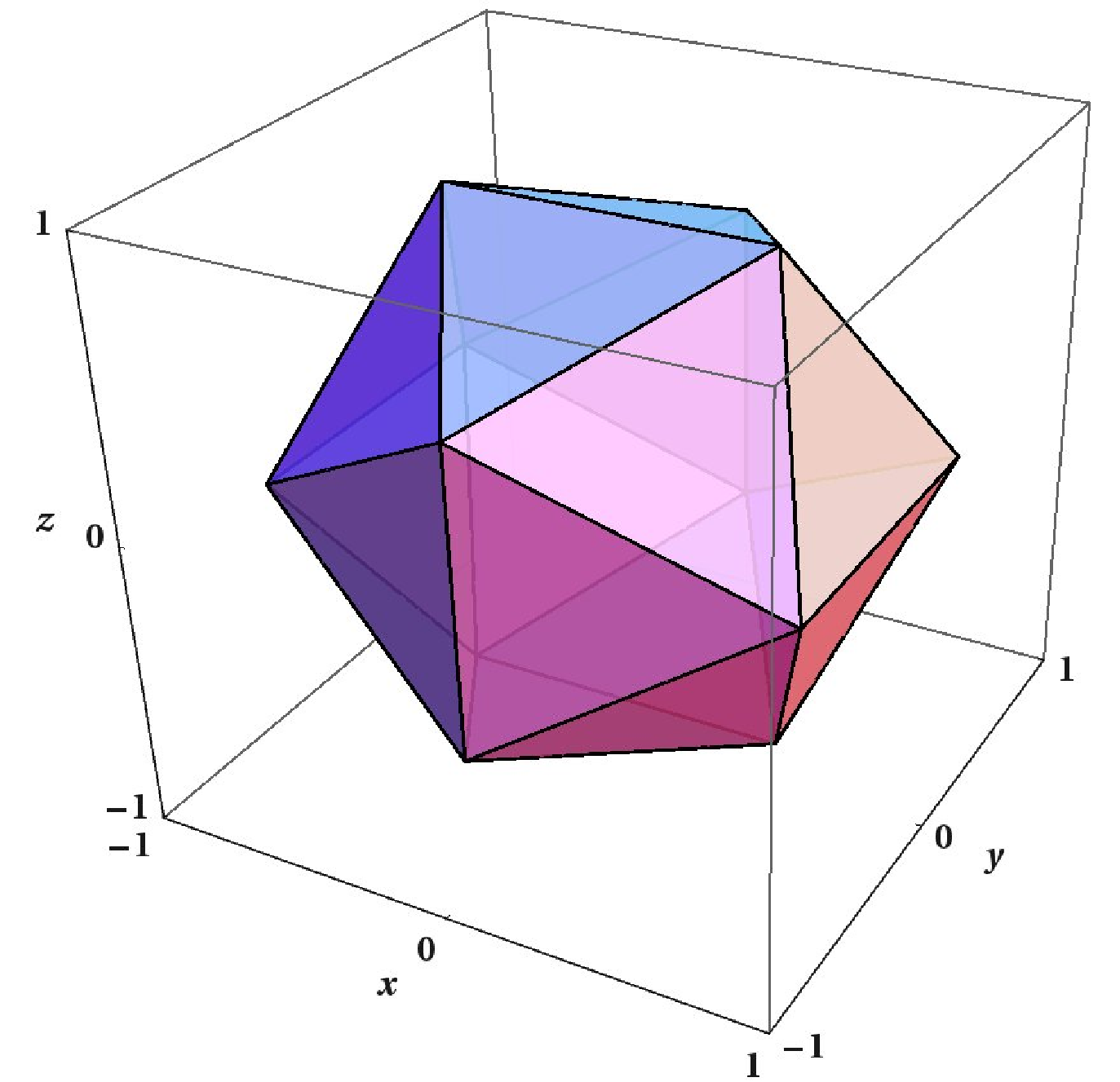}
  \caption{Three-dimensional plot of the icosahedron obtained by interpreting
    the vectors of equation~(\ref{verI}) as its vertices.}
\label{icosahedron} 
\end{center}
\end{figure}

Now we want to construct 
the irreps and the character table of $A_5$. 
First we determine the classes. 
According to the discussion following equation~(\ref{ca5})
we immediately find the two non-trivial classes $C_2(s)$ and $C_3(k_1)$.
The other two classes are the classes consisting of five-cycles.
Since $k_1 k_4 s=(12453)$ and $k_4 k_1 s=(12543)$ are related via
	\be
	(45)(12453)(45)^{-1}=(12543) 
	\ee
with $(45)\not\in A_5$, the two remaining classes are 
	\be
	C_4(k_1 k_4 s)\quad\mbox{and}\quad C_5(k_4 k_1 s).
	\ee
Because $s$, $k_1$ and $k_4$ generate an element of each
conjugacy class, they generate the whole group---see 
corollary~\ref{generatorcorollary}. In this way we have found
an alternative proof that $A_5$ is generated by the
permutations~(\ref{isoIA5}). The two inequivalent three-dimensional
representations of $A_5$ constructed in section~\ref{SSS} are given by
	\be
	\begin{array}{cccl}
	\three: & s\mapsto E, & k_1\mapsto A, & k_4\mapsto W, \\
	\three': & s\mapsto E, & k_1\mapsto A, & k_4\mapsto W',
	\end{array}
	\ee
where $A$, $E$ and $W$, $W'$ are defined in equations~(\ref{AE})
and~(\ref{WW'}), respectively. The dimensions of the missing non-trivial
irreps follow from
	\be
	60=1^2 + 2\times 3^2 + d_4^2 + d_5^2 \;\Rightarrow\; d_4=4,
        \enspace d_5=5.
	\ee
The five-dimensional irrep corresponds to the five-dimensional
$SO(3)$-irrep, which can be computed from equation~(\ref{33=136}).
Therefore, in terms of irreps of $A_5$ we find 
	\be\label{33135}
	\three \otimes \three = \one \oplus \three \oplus \five.
	\ee
The bases of the one and three-dimensional invariant subspaces are provided
by equations~(\ref{b1}) and~(\ref{b3}), respectively. Using these bases one
can easily reduce $\three\otimes\three$ and construct the $\five$ explicitly.
However, here we confine ourselves to the character $\chi^{(\textbf{5})}$,
which---according to equation~(\ref{33135})---is given by
	\be
	\chi^{(\five)}=\chi^{(\three)}\times\chi^{(\three)}-
	\chi^{(\three)}-\chi^{(\one)}.
	\ee
Using the relation~(\ref{ma}), one finds that of the irreps we have already
constructed only the $\five$ is contained in $\three\otimes\three'$.
Therefore, the other irrep in the tensor product can only be the
$\four$. Its character is thus given by 
	\be
	\chi^{(\four)}=\chi^{(\three)}\times\chi^{(\three')}-
	\chi^{(\five)}.
	\ee
Having completed the computation of characters, we are in a position to write
down the character table of $I \cong A_5$---see table~\ref{charA5}. 
This flavour group has for instance been used in~\cite{A5} for model building.
\begin{table}
\begin{center}
\begin{tabular}{|c|ccccc|}\hline
$I\cong A_5$  & $C_1(e)$   & $C_2(s)$ &  $C_3(k_1)$ & $C_4(k_1 k_4 s)$ &
  $C_5(k_4 k_1 s)$ \\ 
(\# $C$) & $(1)$ & $(20)$ & $(15)$ & $(12)$ & $(12)$ \\ 
$\oo(C)$ & $1$ & $3$ & $2$ & $5$ & $5$ \\ \hline
$\one$  & $1$ & $1$ & $1$ & $1$ & $1$ \\
$\three$  & $3$ & $0$ & $-1$ & $-\mu_2$ & $-\mu_1$ \\
$\three'$  & $3$ & $0$ & $-1$ & $-\mu_1$ & $-\mu_2$ \\
$\four$ & $4$ & $1$ & $0$ & $-1$ & $-1$\\
$\five$ & $5$ & $-1$ & $1$ & $0$ & $0$\\
\hline
\end{tabular}
\end{center}
\caption{The character table of $I\cong A_5$ with
$\mu_1=\frac{-1+\sqrt{5}}{2}$, $\mu_2=\frac{-1-\sqrt{5}}{2}$.
\label{charA5}}
\end{table}

Now we discuss the double cover $\widetilde I$.
Using equation~(\ref{alphan}), we compute the rotation
axes and angles for $W$ and $W'$. We obtain $\alpha = 180^\circ$ for both $W$
and $W'$, and the axes
\begin{equation}
\vec w = \frac{1}{2} \left( \begin{array}{c} 1 \\ \mu_2 \\ \mu_1 
\end{array} \right)
\quad \mbox{and} \quad
{\vec w}' = \frac{1}{2} \left( \begin{array}{c} 1 \\ \mu_1 \\ \mu_2
\end{array} \right),
\end{equation}
respectively. Inserting this result into equation~(\ref{SU2calc}),
we define
	\be
	\tilde{k}_4:= U(\pi,\vec w)=
	\frac{1}{2}\bmat
	-i\mu_1 & -\mu_2-i\\
	\mu_2-i & i\mu_1
	\emat
        \ee
and
        \be
	\tilde{k}_4':=U(\pi,{\vec w}')=
	\frac{1}{2}\bmat
	-i\mu_2 & -\mu_1-i\\
	\mu_1-i & i\mu_2
	\emat.
	\ee
Together with $\tilde{s}$ and $\tilde k_1$ defined in 
equation~(\ref{DoubleS4generators}), 
we obtain two faithful inequivalent two-dimensional irreps of
$\widetilde{I}$: 
	\be\label{DoubleA52irreps}
	\begin{array}{cccl}
	\widetilde{\two}: & \tilde{s}\mapsto\tilde{s}, &
        \tilde{k}_1\mapsto\tilde{k}_1, & 
	\tilde{k}_4\mapsto\tilde{k}_4, \\
	\widetilde{\two}': & \tilde{s}\mapsto\tilde{s}, &
        \tilde{k}_1\mapsto\tilde{k}_1, &
	\tilde{k}_4\mapsto\tilde{k}_4'.
	\end{array}
	\ee
The classes are easily deduced from the classes of $A_5$---see
table~\ref{charA5}. In this way, we find 
$C_1(\bone_2)$, $C_2(-\bone_2)$, $C_3(\tilde s)$, $C_4(-\tilde s)$.
Defining
	\be
	\tilde y: = \tilde{s}\tilde{k}_1\tilde{s}^{-1}=\bmat
	i &  0 \\
	0 & -i
	\emat,
	\ee
we deduce that
	\be
	\tilde y \tilde{k}_1 {\tilde y}^{-1}=-\tilde{k}_1 \;\Rightarrow\;
        C_5(\tilde{k}_1) = C_5(-\tilde{k}_1).
	\ee
Since the traces of $\tilde{k}_1 \tilde{k}_4 \tilde{s}$ and 
$\tilde{k}_4 \tilde{k}_1 \tilde{s}$ are non-zero, 
the classes corresponding to the two remaining classes of $A_5$
do not coincide with their negatives. 
Thus there are nine conjugacy classes and consequently also nine
inequivalent irreps. We already know seven of them, namely the five irreps of
$A_5$ and the two irreps~(\ref{DoubleA52irreps}). We denote the dimensions
of the two remaining irreps by $d_7$ and $d_9$, already anticipating
their positions in the character table. They are determined by
	\be
	120 = 1^2 + 2\times 3^2 + 4^2 + 5^2 + 2\times 2^2 + d_7^2 + d_9^2
	\;\Rightarrow\, d_7=4,\enspace d_9=6.
	\ee
Using the characters of the already constructed irreps, one can compute that
\begin{equation}
\widetilde{\six}:= \widetilde{\two}'\otimes\three \cong
\widetilde{\two}\otimes\three' 
\end{equation}
is irreducible. Finally, the
character of $\widetilde{\four}$ can be found from
	\be
	\widetilde{\two}\otimes\three=\widetilde{\two}\oplus\widetilde{\four}.
	\ee
The character table of $\widetilde{I}$ is presented in 
table~\ref{A5doublechar}. For the tensor products of irreps
see~\cite{hashimoto}. An alternative discussion of $\widetilde I$
is presented in~\cite{luhn168}.  
\begin{table}
\begin{center}
\rotatebox{90}{\parbox{1.2\textwidth}{
\begin{tabular}{|c|ccccccccc|}\hline
\rule{0pt}{16pt}$\widetilde{I}$  & $C_1(\bone_2)$ &
$C_2(-\bone_2)$ & $C_3(\tilde{s})$ & $C_4(-\tilde{s})$ &
 $C_5(\tilde{k}_1)$ & $C_6(\tilde{k}_1 \tilde{k}_4 \tilde{s})$\! &
 $C_7(-\tilde{k}_1 \tilde{k}_4 \tilde{s})$\! & 
$C_8(\tilde{k}_4 \tilde{k}_1 \tilde{s})$\! &
$C_9(-\tilde{k}_4 \tilde{k}_1 \tilde{s})$ \\
(\# $C$) & $(1)$ & $(1)$ & $(20)$ & $(20)$ & $(30)$ & $(12)$ & $(12)$
 & $(12)$ & $(12)$\\ 
$\oo(C)$ & $1$ & $2$ & $6$ & $3$ & $4$ &
$10$ & $5$ & $10$ & $5$\\ \hline
$\one$  & $1$ & $1$ & $1$ & $1$ & $1$ & $1$ & $1$ & $1$ & $1$ \\
$\widetilde\two$ & $2$ & $-2$ & $1$ & $-1$ & $0$ & $-\mu_2$ & $\mu_2$ &
$-\mu_1$ & $\mu_1$ \\
${\widetilde\two}'$ & $2$ & $-2$ & $1$ & $-1$ & $0$ & $-\mu_1$ & $\mu_1$ &
$-\mu_2$ & $\mu_2$ \\
$\three$ & $3$ & $3$ & $0$ & $0$ & $-1$ & $-\mu_2$ & $-\mu_2$ & $-\mu_1$ &
  $-\mu_1$ \\
$\three'$ & $3$ & $3$ & $0$ & $0$ & $-1$ & $-\mu_1$ & $-\mu_1$ & $-\mu_2$ &
  $-\mu_2$ \\
$\four$ & $4$ & $4$ & $1$ & $1$ & $0$ & $-1$ & $-1$ & $-1$ & $-1$\\
$\widetilde{\four}$ & $4$ & $-4$ & $-1$ & $1$ & $0$ & $1$ & $-1$ & $1$ & $-1$\\
$\five$ & $5$ & $5$ & $-1$ & $-1$ & $1$ & $0$ & $0$ & $0$ & $0$\\
$\widetilde{\six}$ & $6$ & $-6$ & $0$ & $0$ & $0$ & $-1$ & $1$ & $-1$ & $1$\\
\hline
\end{tabular}
}}
\end{center}
\caption{The character table of $\widetilde{I}$ with
$\mu_1=\frac{-1+\sqrt{5}}{2}$, $\mu_2=\frac{-1-\sqrt{5}}{2}$. The irreps
  without tilde correspond to the irreps of $I$.
\label{A5doublechar}}
\end{table}

\section{The finite subgroups of $SU(3)$}
\label{finitesu3}
\subsection{Classification}
\label{classification}
According to~\cite{Miller,fairbairn} the finite subgroups of $SU(3)$ can be
classified into the following types:
\begin{enumerate}
\renewcommand{\theenumi}{\Alph{enumi}}
\item
Groups of diagonal matrices  corresponding to Abelian groups. 
\item
Groups corresponding to linear transformations of two variables. Up to basis
transformations such $SU(3)$-matrices have the form
\begin{equation}
\left( \begin{array}{cc}
\det A^* & 0_{1 \times 2} \\ 0_{2 \times 1} & A 
\end{array} \right)
\mbox{with} \quad A \in U(2).
\end{equation}
\item
The groups $C(n,a,b)$ generated by the matrices
\begin{equation}\label{F}
E = \begin{pmatrix}
0 & 1 & 0 \\ 0 & 0 & 1 \\ 1 & 0 & 0 \\ \end{pmatrix}
\quad \mbox{and} \quad
F(n,a,b) = \begin{pmatrix}
\eta^a & 0 & 0 \\ 0 & \eta^b & 0 \\ 0 & 0 & \eta^{-a-b} \\ \end{pmatrix},
\end{equation}
where $n$ is a positive integer, $\eta=\exp(2\pi i/n)$ and integers
$a$, $b$ with $0 \leq a,b \leq n-1$. 
\item
The groups $D(n,a,b; d,r,s)$ generated by $E$ and $F(n,a,b)$ of
equation~(\ref{F}) and 
\begin{equation}
\tilde G(d,r,s) = 
\left( \begin{array}{ccc}
\delta^r & 0 & 0 \\ 0 & 0 & \delta^s \\ 0 & -\delta^{-r-s} & 0
\end{array} \right),
\end{equation}
where $d$ is a positive integer, $\delta = \exp(2\pi i/d)$ and integers
$r$, $s$ with $0 \leq r,s \leq d-1$. 
\item
The ``exceptional'' finite subgroups of $SU(3)$, denoted by
$\Sigma(60)$, $\Sigma(168)$, ${\Sigma(36 \times 3)}$, 
$\Sigma(72 \times 3)$, $\Sigma(216 \times 3)$ and $\Sigma(360 \times 3)$.
\end{enumerate}
For completeness we should add to the groups of type~E 
the direct products 
$\Sigma(60) \times \zz_3$ and $\Sigma(168) \times \zz_3$. 
A brief history of the classification of the finite subgroups of $SU(3)$
can be found in the introduction of~\cite{CommentsSU3}. 

\paragraph{Groups of type~A:}
Such groups are Abelian and, therefore, they must have a
structure conforming to theorems~\ref{AbelianStructure1} 
and~\ref{AbelianStructure2}. However, knowing that these groups are
subgroups of $SU(3)$ a more specific statement can be made, as  
proven in~\cite{CommentsSU3}. 
\begin{theorem}\label{asubsu3}
Every Abelian finite subgroup $G$
of $SU(3)$ is isomorphic to $\zz_m \times \zz_n$ where $m$ is the
maximum of the orders of the elements of $G$ and $n$ is a divisor of
$m$.
\end{theorem}

\paragraph{Groups of type~B:}
These groups contain all $SU(2)$ subgroups and the dihedral
groups, which have been discussed already in section~\ref{su2so3}. 
In addition, they comprise the 
genuine $U(2)$ subgroups, not treated in this review.

\paragraph{Groups of type~C:}
The structure of groups of type~C can be understood in the following way. 
We denote the subgroup of diagonal matrices of $C(n,a,b)$ by $N(n,a,b)$.
Because of 
\begin{equation}\label{abc}
E\, \diag (\alpha, \beta, \gamma) E^{-1} = \diag (\beta, \gamma, \alpha)
\end{equation}
it is obvious that this subgroup is a normal subgroup, whence it follows that  
every element $g \in C(n,a,b)$ can be written in the form 
$g = FE^k$, with $F \in N(n,a,b)$ and $k = 0,1,2$. 
It is easy to see that $N(n,a,b)$ is generated by 
\begin{equation}
F(n,a,b) 
\quad \mbox{and} \quad
E F(n,a,b) E^{-1} = \mbox{diag}\, ( \eta^b, \eta^{-a-b}, \eta^a ).
\end{equation}
Armed with this knowledge we can write the multiplication rule of two group
elements as 
\begin{equation}\label{Cmult}
\left( F_1 E^{k_1} \right) \left( F_2 E^{k_2} \right) = 
\left( F_1 E^{k_1} F_2 E^{-k_1} \right) E^{k_1+k_2}.
\end{equation}
Therefore, since $E^3 = \bone$, the groups of type C can be characterized as
\begin{equation}
\label{Csemi}
C(n,a,b) \cong N(n,a,b) \rtimes \zz_3,
\end{equation}
where the homomorphism of the semidirect product, 
$\varphi: \zz_3 \to \mbox{Aut}(N(n,a,b))$, is read off from
equation~(\ref{Cmult}): 
\begin{equation}
\varphi(E^k)F = E^kFE^{-k}.
\end{equation}
According to theorem~\ref{asubsu3}, there are positive integers $m$ and $p$
such that 
\begin{equation}
N(n,a,b) \cong \zz_m \times \zz_p.
\end{equation}
In~\cite{CommentsSU3} a set of rules has been derived which allows to
compute the integers $m$ and $p$ from $n,a,b$. The value of $m$ is found
to be the smallest positive natural number such that 
\begin{equation}
(\eta^a)^m = (\eta^b)^m = 1.
\end{equation}
The parameter $p$ can be determined following a prescription, which
involves another parameter $t$. 
Starting with $p'=1$ we go through all divisors of $m$ and check
whether there exists a $t \in \{1,...,m/p'-1\}$ such that
\begin{equation}
p'(b-at)\,\mathrm{mod}\, n=0 \quad \mbox{and} \quad
p'(a+b(1+t))\,\mathrm{mod}\, n=0
\end{equation}
The smallest $p'$ for which such a $t$ exists is our sought-for value
of $p$---for details see~\cite{CommentsSU3}. 

As proven in appendix~\ref{irrepsC},
the groups of type~C can only have irreps of dimension one and three.

The groups $C(n,a,b)$ contain the important series 
\begin{equation}\label{nn3}
C(n,1,0) \equiv \Delta (3n^2) \cong (\zz_n \times \zz_n) \rtimes \zz_3,
\end{equation}
which will be discussed in section~\ref{D3}. 
Another series is $T_n$, given by $C(n,1,a)$ with 
$1+a+a^2 = 0\, \mbox{mod}\,n$. This is possible only for specific
$n$, as we will discuss in section~\ref{Tn}.
The smallest group of type~C that does not belong to $\Delta(3n^2)$
and $T_n$ and that is not a direct product is
\begin{equation}
C(9,1,1) \cong ( \zz_9 \times \zz_3 ) \rtimes \zz_3
\end{equation}
with 81 elements~\cite{CommentsSU3}. Apart from $A_4 \cong \Delta(12)$ which
has already been discussed, among the groups of type~C which have been used as
flavour symmetries there are $\Delta(27)$~\cite{Delta27}, 
$\Delta(75)$~\cite{Delta75},
$T_7$~\cite{luhn7,smirnov,maT7} and $T_{13}$~\cite{T13}. 

\paragraph{Groups of type~D:}
To understand the structure of $D(n,a,b;d,r,s)$, it is useful to switch to a
new set of generators~\cite{ludl-dt,zwicky}. 
The derivation of this new set is performed in appendix~\ref{generatorsD}.
With 
\begin{equation}
F_r = 
\diag \left( -\delta^{-r},\,-\delta^{-r},\, \delta^{2r} \right), 
\quad
F'_t = \diag( \delta^{-t},\,\delta^{-t},\, \delta^{2t} )
\quad \mbox{with} \quad t = r-s
\end{equation}
and 
\begin{equation}\label{BM}
B =
\left( \begin{array}{rrr}
-1 & 0 & 0 \\ 0 & 0 & -1 \\ 0 & -1 & 0
\end{array} \right),
\end{equation}
it is given by 
\begin{equation}
F(n,a,b),\; F_r,\; F'_t,\; E,\; B.
\end{equation}
This set of generators clearly shows the structure of
the groups $D(n,a,b;d,r,s)$. 
The matrices $E$ and $B$ obviously generate a subgroup
isomorphic to $S_3$. Moreover, the group $D(n,a,b;d,r,s)$ has a normal subgroup
$N(n,a,b;d,r,s)$ of diagonal
matrices which consists of $F(n,a,b)$, $F_r$, $F'_t$, 
the matrices obtained from these three by similarity transformations with $E$
and $B$---see equations~(\ref{abc}) and
\begin{equation}
B\, \diag (\alpha, \beta, \gamma) B^{-1} = \diag (\alpha, \gamma, \beta),
\end{equation}
and all products thereof. 
Furthermore, it follows that every $g \in D(n,a,b;d,r,s)$ can be written as
\begin{equation}
g = FE^k B^l \quad \mbox{with} \quad F \in N(n,a,b;d,r,s),\;
k = 0,1,2,\;\; l = 0,1.
\end{equation}
So we find that groups of type~D have the structure of a semidirect product:
\begin{equation}
\label{Dsemi}
D(n,a,b;d,r,s) \cong N(n,a,b;d,r,s) \rtimes S_3.
\end{equation}
Finally we can exploit theorem~\ref{asubsu3} which tells us that
$N(n,a,b;d,r,s)$ is a direct product of two cyclic groups with
suitably chosen orders $m$ and $p$, where $p$ is a divisor of $m$.
Therefore, we end up with
\begin{equation}
D(n,a,b;d,r,s) \cong \left( \zz_m \times \zz_p \right) \rtimes S_3.
\end{equation}
The dimensions of the irreps of the groups of type~D are quite
restricted, only one, two, three and six 
are possible---see the proof
in appendix~\ref{irrepsD}. In essence this follows from the semidirect
product~(\ref{Dsemi}).

The groups of type~D comprise the well-known series
of dihedral-like groups
\begin{equation}\label{DD}
D(n,0,1;2,1,1) \equiv \Delta(6n^2) \cong 
\left( \zz_n \times \zz_n \right) \rtimes S_3,
\end{equation}
which will be discussed in section~\ref{D6}.
Apart from $\Delta(6) \cong S_3$ and $\Delta(24) \cong S_4$, a group of type~D
used as flavor symmetry is $\Delta(54)$~\cite{Delta54}.
The smallest group of type~D that is not a direct product or a group
of the $\Delta(6n^2)$-series is~\cite{CommentsSU3}
\begin{equation}
D(9,1,1;2,1,1) \cong \left( \zz_9 \times \zz_3 \right) \rtimes S_3.
\end{equation}

The first group, $\Sigma(60)$, of the ``exceptional'' finite subgroups
of $SU(3)$ is identical with the icosahedral group and thus isomorphic
to $A_5$ which has been discussed in sections~\ref{SSS}
and~\ref{icodouble}. It is the smallest non-Abelian simple group. The other five
``exceptional''  groups will not be discussed in this review. For
$\Sigma(168)$, which is the second smallest non-Abelian simple group, we
refer the reader to~\cite{luhn168,king}. The groups denoted by
$\Sigma(n \times 3)$ contain the center of $SU(3)$; an extensive
discussion of the groups with $n=36,\,72,\,216$ can be found in~\cite{GL10}.

\subsection{The series $\Delta (3n^2)$}
\label{D3}
We follow the discussion of~\cite{luhn3n2} where the notation 
\begin{equation}\label{acd}
a \equiv E, \quad 
c = \diag(\eta,\,1,\,\eta^{-1}), \quad
d = \diag(\eta^{-1},\,\eta,\,1)
\end{equation}
is used, with $\eta = \exp(2\pi i/n)$. 
This is a set of generators of 
$C(n,1,0) \equiv \Delta(3n^2)$ because
$c = F(n,1,0)$ and $d = E^{-1} F(n,1,0) E$.
Actually, we could skip
either $c$ or $d$ but keeping both makes the structure of $\Delta(3n^2)$ more
transparent, as pointed out in~\cite{luhn3n2}. One can readily verify that
the generators fulfill
\begin{equation}\label{presD3n2}
a^3 = c^n = d^n = e, \quad cd = dc, \quad
\quad aca^{-1} = c^{-1}d^{-1}, \quad ada^{-1} = c.
\end{equation}
Thus $c$ and $d$ commute and each of them generates a $\zz_n$, 
while the element $a$
generates a $\zz_3$ and acts on $\zz_n \times \zz_n$. Therefore,
$\Delta(3n^2)$ is the semidirect product of $\zz_n \times \zz_n$ with $\zz_3$
and its order is
\begin{equation}
\oo \Delta(3n^2) = 3n^2.
\end{equation}
We already know the first two members of this series: 
$\Delta(3) \cong \zz_3$ and $\Delta(12) \cong A_4$. The latter
statement is evident from equation~(\ref{AE}).

By means of equation~(\ref{presD3n2}) it is nearly trivial to find the
one-dimensional irreps. In these irreps $c$ and $d$ must be presented by the
same number and, moreover, $c^3 \mapsto 1$. Since also 
$c^n \mapsto 1$, we must distinguish if three is a
divisor of $n$ or not:
\begin{equation}\label{1-dimirreps}
\begin{array}{ccrcccl}
\frac{n}{3} \not\in \mathbbm{N} & \Rightarrow & \one^{(p)}: & 
a \mapsto \omega^p, & c \mapsto 1, & 
d \mapsto 1 & (p=0,1,2), \\
\frac{n}{3} \in \mathbbm{N} & \Rightarrow & \one^{(p,q)}: & 
a \mapsto \omega^p, & c \mapsto \omega^q, & d \mapsto \omega^q & (p,q=0,1,2).
\end{array}
\end{equation}
Thus, if three is a divisor of $n$, there are nine inequivalent one-dimensional
irreps, otherwise this number is three. 
Therefore, $\Delta(27)$ has nine, but $A_4 \cong \Delta(12)$ has only
three one-dimensional irreps.

Since $\Delta(3n^2)$ is a group of type~C, the remaining irreps must
all have dimension three. The number of three-dimensional irreps is
$(n^2-1)/3$ if three is not 
a divisor of $n$; otherwise their number
is $(n^2-3)/3$. 
The three-dimensional irreps are given by~\cite{luhn3n2} 
\begin{equation}
\three^{(k,l)}: \quad a \mapsto E, \quad 
c \mapsto \diag(\eta^l,\, \eta^k,\, \eta^{-k-l}), \quad
d \mapsto \diag(\eta^{-k-l},\, \eta^l,\, \eta^k)
\end{equation}
with $k,l, = 0,1,\ldots,n-1$. The defining irrep corresponds to 
$(k,l) = (0,1)$. The pair $(0,0)$ does not give an irrep; if three is a
divisor of $n$, also $(\frac{n}{3},\frac{n}{3})$ and 
$(\frac{2n}{3},\frac{2n}{3})$ must be excluded. Moreover, 
for some pairs $(k,l)$ one obtains equivalent
irreps. For a detailed discussion we refer the reader
to~\cite{luhn3n2}. 

We discuss the irreps of $\Delta(27)$ corresponding to $n=3$ in detail. 
Its nine one-dimensional irreps are known from
equation~(\ref{1-dimirreps}). A minimal set of generators for the defining
three-dimensional irrep is, for instance, given by
\begin{equation}\label{minimalD27}
\Delta(27): \quad
E \quad \mbox{and} \quad C:= \diag(1,\,\omega,\,\omega^2),
\end{equation}
where $E$ is defined in equation~(\ref{F}). Note that
equation~(\ref{minimalD27}) can be considered as the representation
$a \mapsto E$, $cd \mapsto C$. With $a cd a^{-1} = d^{-1}$ the
representation matrices of $c$ and $d$ can be reconstructed.
The classes are computed by using the relations
\begin{equation}\label{264}
E C E^{-1} = \omega C, \quad C E C^{-1} = \omega^2 E.
\end{equation}
It convenient to define the set of matrices
\begin{equation}\label{cz}
\mathcal{Z} \equiv \{ \bone,\,  \omega \bone,\, \omega^2 \bone \}
\end{equation}
which corresponds to the center of $SU(3)$. With 
the abbreviation 
$\mathcal{Z} \cdot g \equiv \{ g,\, \omega g,\, \omega^2 g \}$,
we can readily formulate the conjugacy classes~\cite{GL10}:
\begin{equation}\label{cDelta27}
\begin{array}{llll}
C_1 = \{ \bone \},& C_2 = \{ \omega \bone \},&
C_3 = \{ \omega^2 \bone \}, & \\ 
C_4 = \mathcal{Z} \cdot C,  &  C_5 = \mathcal{Z} \cdot C^2,  &
C_6 = \mathcal{Z} \cdot E,  &  C_7 = \mathcal{Z} \cdot E^2,  \\ 
C_8 = \mathcal{Z} \cdot C E, &  C_9 = \mathcal{Z} \cdot C^2 E^2, & 
C_{10} = \mathcal{Z} \cdot C^2 E,  & C_{11} = \mathcal{Z} \cdot CE^2.
\end{array}
\end{equation}
Therefore, there are eleven inequivalent irreps. We already know the nine
one-dimensional ones. The two remaining irreps must be
three-dimensional due to
\begin{equation}
9 \times 1^2 + 2 \times 3^2 = 27 
\end{equation}
and are given by the defining irrep and its complex conjugate:
\begin{equation}
\begin{array}{ccc}
\three: & E \mapsto E, & C \mapsto C, \\
\three^*: & E \mapsto E, & C \mapsto C^*.
\end{array}
\end{equation}
The character table of $\Delta(27)$ is found in table~\ref{d27ct}.
Note that the one-dimensional irreps $\one^{(p,q)}$ are obtained by the mappings
$E \mapsto \omega^p$, $C \mapsto \omega^{2q}$.
\begin{table}
\begin{center}
\begin{tabular}{|c|ccccccccccc|}
\hline 
$\Delta(27)$ & $C_1$ & $C_2$ & $C_3$ & $C_4$ & $C_5$ &
$C_6$ & $C_7$ & $C_8$ & $C_9$ & $C_{10}$ & $C_{11}$ \\
(\# $C$) & (1) & (1) & (1) & (3) & (3) & (3) & (3) & (3) & (3) & (3) & (3) \\
$\oo(C)$ & 1 & 3 & 3 & 3 & 3 & 3 & 3 & 3 & 3 &  3 & 3 \\
\hline 
$\mathbf{1}^{(0,0)}$
& $1$ & $1$ & $1$ & $1$ & $1$ & $1$ & $1$ & $1$ & $1$ & $1$ & $1$ \\
$\mathbf{1}^{(0,1)}$
& $1$ & $1$ & $1$ & $\omega^2$ & $\omega$ & $1$ & $1$ 
& $\omega^2$ & $\omega$  & $\omega$ & $\omega^2$ \\ 
$\mathbf{1}^{(0,2)}$
& $1$ & $1$ & $1$ & $\omega$ & $\omega^2$ & $1$ 
& $1$ & $\omega$ & $\omega^2$  & $\omega^2$ & $\omega$ \\ 
$\mathbf{1}^{(1,0)}$
& $1$ & $1$ & $1$ & $1$ & $1$ & $\omega$ 
& $\omega^2$ & $\omega$ & $\omega^2$  & $\omega$ & $\omega^2$ \\ 
$\mathbf{1}^{(1,1)}$
& $1$ & $1$ & $1$ & $\omega^2$ & $\omega$ & $\omega$ & $\omega^2$ 
& $1$ & $1$  & $\omega^2$ & $\omega$ \\ 
$\mathbf{1}^{(1,2)}$
& $1$ & $1$ & $1$ & $\omega$ & $\omega^2$ & $\omega$ & $\omega^2$ 
& $\omega^2$ & $\omega$  & $1$ & $1$ \\ 
$\mathbf{1}^{(2,0)}$
& $1$ & $1$ & $1$ & $1$ & $1$ & $\omega^2$ 
& $\omega$ & $\omega^2$ & $\omega$  & $\omega^2$ & $\omega$ \\ 
$\mathbf{1}^{(2,1)}$
& $1$ & $1$ & $1$ & $\omega^2$ & $\omega$ & $\omega^2$ & $\omega$ 
& $\omega$ & $\omega^2$& $1$ & $1$ \\ 
$\mathbf{1}^{(2,2)}$
& $1$ & $1$ & $1$ & $\omega$ & $\omega^2$ & $\omega^2$ & $\omega$ 
& $1$ & $1$  & $\omega$ & $\omega^2$ \\ 
$\three$ 
& $3$ & $3\omega$ & $3\omega^2$ 
& $0$ & $0$ & $0$ & $0$ & $0$ & $0$ & $0$ & $0$ \\
$\three^*$ 
& $3$ & $3\omega^2$ & $3\omega$ 
& $0$ & $0$ & $0$ & $0$ & $0$ & $0$ & $0$ & $0$ \\
\hline 
\end{tabular}
\caption{Character table of $\Delta(27)$ with 
$\omega = \frac{-1+i\sqrt{3}}{2}$.} \label{d27ct}
\end{center}
\end{table}

Finally we discuss tensor products of the irreps of $\Delta(27)$.
Equation~(\ref{264}) tells us that 
$\omega \bone \mapsto 1$ in every one-dimensional irrep. 
Using this observation and denoting the characters of the
one-dimensional irreps by $\chi^{(p,q)}$, we read off from
table~\ref{d27ct} that
\begin{equation}
\chi^{(p,q)} \times \chi^{(\three)} = \chi^{(\three)} 
\quad \mbox{and} \quad
\chi^{(p,q)} \times \chi^{(\three^*)} = \chi^{(\three^*)},
\end{equation}
whence we conclude
\begin{equation}
\one^{(p,q)} \otimes \three \cong \three, \quad
\one^{(p,q)} \otimes \three^* \cong \three^*.
\end{equation}
We discuss one example of this equivalence, namely $p=1$, $q=0$. In
this case the tensor product corresponds to the representation
\begin{equation}
E \mapsto \omega E, \quad C \mapsto C.
\end{equation}
Then the basis transformation which demonstrates the equivalence 
\begin{equation}\label{equiv1x3}
\one^{(1,0)} \otimes \three \cong \three
\end{equation}
is provided by the matrix $C$ due to $C (\omega E) C^{-1} = E$.

Concerning the tensor products of the three-dimensional irreps, we put
forward the relations
\begin{eqnarray}
\label{d27-33}
\three \otimes \three &=& \three^* \oplus \three^* \oplus \three^*, \\
\label{d27-3*3}
\three^* \otimes \three &=& \bigoplus_{p,q=1}^3 \one^{(p,q)}.
\end{eqnarray}
That in equation~(\ref{d27-33}) the~$\three^*$ occurs three times on
the right-hand side is easy to understand. One~$\three^*$ comes about
through the antisymmetrized basis~(\ref{b3}) and
equation~(\ref{3''}). However, in the case of the $\Delta(27)$ also the
symmetrized basis gives~$\three^*$. The third~$\three^*$ 
comes from
\begin{equation}
\three^*: \quad e_1 \otimes e_1,\;\; e_2 \otimes e_2,\;\; e_3 \otimes e_3.
\end{equation}
In order to prove equation~(\ref{d27-3*3}) we have to find nine
simultaneous eigenvectors to $E \otimes E$ and $C^* \otimes C$.
For the purpose of a convenient formulation of such eigenvectors we define 
an operation $k \mapsto \widehat k$ on $k \in \zz$ which denotes a shift
by multiples of three such that $\widehat k \in \{ 1,2,3 \}$. This means
that $\widehat k = k$ for $k=1,2,3$, $\widehat 0 = 3$, $\widehat 4 =
1$, and so on.  
With this notation the action of $E$ and $C$ on the Cartesian basis
vectors is formulated as 
\begin{equation}
E e_i = e_{\,\widehat{i-1}} \quad \mbox{and} \quad C e_i = \omega^{i-1} e_i,
\end{equation}
respectively.
We define nine vectors
\begin{equation}\label{B}
\mathcal{B}_{pq} = \frac{1}{\sqrt{3}}\, \sum_{i=1}^3
\omega^{p(i-1)}\,e_i \otimes e_{\,\widehat{i+q}} \quad (p,q = 0,1,2)
\end{equation}
in $\mathbbm{C}^3 \otimes \mathbbm{C}^3$. These vectors constitute a
generalization of equation~(\ref{BT}) with 
$\mathcal{B}_p \equiv \mathcal{B}_{p0}$.
It is not difficult to check that
they form an orthonormal basis and that
\begin{equation}\label{ECB}
\left(E \otimes E\right) \mathcal{B}_{pq} = \omega^p\, \mathcal{B}_{pq},
\quad
\left(C^* \otimes C\right) \mathcal{B}_{pq} = \omega^q\, \mathcal{B}_{pq},
\end{equation}
which proves statement~(\ref{d27-3*3}).
We note that the basis~(\ref{B}) will prove useful also in the context of
$\Delta(54)$---see section~\ref{D6}.

\subsection{The series $T_n$}
\label{Tn}
Now we deal with the series
\begin{equation}\label{Tnsemi}
C(n,1,a) \equiv T_n \cong \zz_n \rtimes \zz_3,
\end{equation}
which has been treated in~\cite{bovier,bovier1,fairbairn1}.
These groups exist only for very specific integers $n$. 
It was demonstrated in~\cite{bovier1} that for integers
$n$ which are a product of prime numbers
of the form $6k+1$ with $k \in \mathbbm{N}$, the equation
\begin{equation}\label{moda}
1+a+a^2 = 0\,\mbox{mod}\,n
\end{equation}
has at least one solution. The generators of $T_n$ are then
given by~\cite{bovier1,fairbairn1}
\begin{equation}
E \quad \mbox{and} \quad 
T = \diag ( \rho,\, \rho^a, \rho^{a^2} )
\quad \mbox{with} \quad \rho = e^{2\pi i/n}.
\end{equation}
The matrix $E$ is defined in equation~(\ref{F}).
The group $T_n$ does not contain the center of $SU(3)$ which is given by 
$\{ \bone,\, \omega\bone,\, \omega^2\bone \}$ with $\omega = \exp(2\pi i/3)$.
The generators $E$ and $T$ fulfill
\begin{equation}\label{presTn}
E^3 = T^n = \bone 
\quad \mbox{and} \quad
E T E^{-1} = \diag ( \rho^a,\, \rho^{a^2}, \rho ) = T^a.
\end{equation}
The second relation holds because from equation~(\ref{moda}) it
follows that $a^3 = 1\,\mbox{mod}\,n$. 
Moreover, this relation explains why no other cyclic group than the $\zz_n$
generated by $T$ is 
involved and why equation~(\ref{Tnsemi}) holds.
The groups $T_n$ are of type~C, therefore, their irreps have only
dimensions one and three. Moreover, from equation~(\ref{presTn}) it
follows that there are only three one-dimensional irreps, namely those
where $E$ is represented by a cubic root of unity and $T$ is
represented by one---see also appendix~\ref{irrepsC}. From this
consideration we also find that $T_n$ has $(n-1)/3$ inequivalent
three-dimensional irreps.

Equation~(\ref{presTn}) actually is a presentation of $T_n$ which
implies equation~(\ref{Tnsemi}) and, therefore,
\begin{equation}
\oo T_n = 3n.
\end{equation}
For the computation of the classes, the relations 
\begin{equation}\label{Tn1}
T E T^{-1} = T^{1-a} E, \quad  
T E^2 T^{-1} = T^{1-a^2} E^2
\end{equation}
are particularly useful.

The smallest prime number of the form $6k+1$ is 
$7 = 6 \times 1 + 1$; the group $T_7$ will be discussed in detail
later in this section. To get an idea for which $n$ a group $T_n$ exists,
we list the pairs of numbers $(n,a)$ for $n < 100$, taken 
from~\cite{ludl10}:
$(7,2)$, $(13,3)$, $(19,7)$, $(31,5)$, $(37,10)$, $(43,6)$, $(49,30)$,
$(61,13)$, $(67,29)$, $(73,8)$, $(79,23)$, $(91,9)$, $(91,16)$, $(97,35)$.
Of the 14 numbers $n$, all are prime numbers of the form $6k+1$, 
except $49 = 7 \times 7$ and $91 = 7 \times 13$, which are products of such
primes. Furthermore, $n=91$ is the first instance where two solutions for $a$
exist. Therefore, there are two non-isomorphic groups $T_{91}$ and
$T'_{91}$ with $3 \times 91$ 
elements.

Now we turn to a discussion of $T_7$ with 21
elements~\cite{luhn7,smirnov}, generated by 
\begin{equation}
E \quad \mbox{and} \quad T = \diag (\rho,\, \rho^2, \rho^4 ).
\end{equation}
With equations~(\ref{presTn}) and~(\ref{Tn1}) it is straightforward to compute
the classes:
\begin{equation}\label{clT7}
\begin{array}{l}
C_1 = \{ \bone \}, \;\; C_2 = \{E, TE, \ldots,T^6E \}, \;\; 
C_3 = \{E^2, TE^2, \ldots,T^6E^2 \}, \\ 
C_4 = \{T, T^2, T^4 \}, \;\; C_5 = \{T^3, T^5, T^6 \}.
\end{array}
\end{equation}
Since there are five classes, we know that there are five inequivalent
irreps. 
As discussed 
before, the one-dimensional irreps are given by
\begin{equation}
\one^{(p)}: \quad T \mapsto 1, \quad E \mapsto \omega^p 
\quad \mbox{with}\; 
\omega = e^{2\pi i/3}\quad (p=0,1,2).
\end{equation}
\begin{table}
\begin{center}
\begin{tabular}{|c|ccccc|}\hline
$T_7$ & $C_1$ & $C_2$ & $C_3$ & $C_4$ & $C_5$ \\
(\# $C$) & (1) & (7) & (7) & (3) & (3) \\
$\oo(C)$ & 1 & 3 & 3 & 7 & 7 \\\hline
$\one^{(0)}$ & 1 & 1          & 1          & 1         & 1 \\
$\one^{(1)}$ & 1 & $\omega$   & $\omega^2$ & 1         & 1 \\
$\one^{(2)}$ & 1 & $\omega^2$ & $\omega$   & 1         & 1 \\
$\three$     & 3 & 0          & 0          & $\zeta$   & $\zeta^*$ \\
$\three^*$   & 3 & 0          & 0          & $\zeta^*$ & $\zeta$ \\\hline
\end{tabular}
\end{center}
\caption{Character table of $T_7$ with
$\omega = \frac{-1+ i \sqrt{3}}{2}$, 
$\zeta = \frac{-1+ i \sqrt{7}}{2}$.
\label{ctT7}}
\end{table}
We also know that the defining three-dimensional representation is
irreducible. Therefore, we are in a position to compute the first four lines of
the character table~\ref{ctT7}. In the case of the~$\three$ we use
\begin{equation}
\mbox{Tr}\,T \equiv 
\zeta = \rho + \rho^2 + \rho^4 = \frac{-1+ i \sqrt{7}}{2}.
\end{equation}
The line pertaining to the $\three$ is genuinely complex. Therefore, its
complex conjugate $\three^*$ is not equivalent to $\three$. Therefore, the
last line in the character table is obtained by complex conjugation of the
fourth line. Now we can perform all kinds of consistency checks, for instance
a check of the orthogonality relations~(\ref{o2}).
It also instructive to check theorem~\ref{n2}, relating the dimensions of the
irreps with the order of the group: 
$3 \times 1^2 + 2 \times 3^2 = 21$.

Finally we discuss the tensor products of the three-dimensional irreps of
$T_7$. It is sufficient to consider $\three \otimes \three$ and 
$\three^* \otimes \three$. As usual, we denote Cartesian basis vectors
of $\mathbbm{C}^3$ by $e_j$ ($j=1,2,3$). 
This is a basis of eigenvectors of $T$, while 
$Ee_1 = e_3$, $Ee_2 = e_1$, $Ee_3 = e_2$. 
It is easy to reduce the tensor
product $\three \otimes \three$ into irreps:
\begin{equation}\label{33}
\begin{array}{cc}
\three:   & e_3 \otimes e_3,\; e_1 \otimes e_1,\; e_2 \otimes e_2, \\
\three^*: & e_2 \otimes e_3,\; e_3 \otimes e_1,\; e_1 \otimes e_2, \\
\three^*: & e_3 \otimes e_2,\; e_1 \otimes e_3,\; e_2 \otimes e_1.
\end{array}
\end{equation}
In these bases, the reduction of the tensor product proceeds via
$E \otimes E \mapsto E$, $T \otimes T \mapsto T$ for $\three$ 
and 
$T \otimes T \mapsto T^*$ 
for $\three^*$. 
Note that the $\three^*$ occurs twice in the tensor product,
therefore, instead
of the bases given in equation~(\ref{33}), one could also use 
symmetrized and antisymmetrized basis vectors. 
Turning to $\three^* \otimes \three$, it is easy to
check the following decomposition:
\begin{equation}
\begin{array}{cl}
\one^{(p)}: & \mathcal{B}_p \;\; (p=0,1,2), \\
\three:   & e_1 \otimes e_2,\; e_2 \otimes e_3,\; e_3 \otimes e_1, \\
\three^*: & e_2 \otimes e_1,\; e_3 \otimes e_2,\; e_1 \otimes e_3. 
\end{array}
\end{equation}
The vectors $\mathcal{B}_p$ are defined in equation~(\ref{BT}).
Comparing with equation~(\ref{ECB}), it is obvious that the basis for
the one-dimensional irreps is given by $\mathcal{B}_{p0}$.
In summary, we have found the following results:
\begin{eqnarray}
\three \otimes \three &=& \three \oplus \three^* \oplus \three^*, \\
\three^* \otimes \three &=& \one^{(0)} \oplus \one^{(1)} \oplus \one^{(2)}
\oplus \three \oplus \three^*.
\end{eqnarray}

If one considers $T_n$ with $n > 7$, 
one may wonder how the further three-dimensional irreps look
like. For this purpose it suffices to investigate $T_{13}$. This
group has four inequivalent three-dimensional irreps. Clearly, in all
these irreps $E$ is simply represented by itself. So the four irreps
differ only in the representation of $T$. It is easy to see that the mappings
$T \mapsto T,\,T^2, \, T^{-1},\, T^{-2}$ lead to the four sought-after
irreps. Note that $T \mapsto T^3$ gives an irrep equivalent to the
defining one.

\subsection{The series $\Delta (6n^2)$}
\label{D6}

The presentation of $D(n,0,1;2,1,1) \equiv \Delta(6n^2)$ is obtained from
equation~(\ref{presD3n2}) by adding an element $b$ corresponding to the
matrix~$B$ of equation~(\ref{BM}) and supplementing the corresponding
relations~\cite{luhn6n2}: 
\begin{equation}\label{presD6n2}
\begin{array}{llll}
a^3 = c^n = d^n = e, & cd = dc, &
aca^{-1} = c^{-1}d^{-1}, & ada^{-1} = c, \\
b^2 = (ab)^2 = e, & bcb^{-1} = d^{-1}, & bdb^{-1} = c^{-1}. &
\end{array}
\end{equation}
The elements $a$ and $b$ generate $S_3$. Due to equation~(\ref{DD}) 
it is evident that
\begin{equation}
\oo \Delta(6n^2) = 6n^2.
\end{equation}
The first two members of this series are $\Delta(6) \cong S_3$ and $\Delta(24)
\cong S_4$, as we can infer from the discussion in section~\ref{SSS}.

In appendix~\ref{irrepsD} we have shown that for groups of type~D, and
thus for $\Delta(6n^2)$, the possible
dimensions of the irreps are one, two, three and six.
Using the presentation it is straightforward to derive all
one-dimensional irreps. For the 
elements $a$ and $b$ the presentation requires
the mappings $b^2 \mapsto 1$, $a^3 \mapsto 1$ and $(ab)^2 \mapsto 1$, 
therefore, 
$a \mapsto 1$ and $b \mapsto \pm 1$. Moreover, $c$ and $d$ are mapped into the
same number and $c^2d \mapsto 1$ for the same reason as in $\Delta(3n^2)$, but
in addition $cd \mapsto 1$ is required; these requirements are fulfilled
simultaneously only if both $c$ and $d$ are mapped into 1. In summary, there
are only two one-dimensional irreps, the trivial one and
\begin{equation}\label{1'}
\one': \quad a,c,d \mapsto 1, \quad b \mapsto -1.
\end{equation}

Applying the results of appendix~\ref{irrepsD}, also the
two-dimensional irreps are quickly found for general $n$. 
First we discuss the irrep where the normal subgroup 
$\zz_n \times \zz_n$ is represented trivially---see
equation~(\ref{2-1}). It is given by 
\begin{equation}\label{21}
\two: \quad 
\quad
a \mapsto \left(
\begin{array}{cc} \omega & 0 \\ 0 & \omega^2
\end{array} \right), 
\quad
b \mapsto \left(
\begin{array}{cc} 0 & 1 \\ 1 & 0
\end{array} \right), 
\quad
c = d \mapsto \left( \begin{array}{cc} 1 & 0 \\ 0 & 1 
\end{array} \right),
\end{equation}
which simply corresponds to the two-dimensional irrep of $S_3$.
Indeed, consider equations~(\ref{sS3}) and~(\ref{tS3})
where the generators $s$ and $t$ of $S_3$ are
represented by rotation matrices and take the $2 \times 2$
submatrices thereof. Then the basis transformation
\begin{equation}
X
\left( \begin{array}{rr}\renewcommand{\arraystretch}{1.3}
-\frac{1}{2} & -\frac{\sqrt{3}}{2}  \\
\frac{\sqrt{3}}{2} & -\frac{1}{2}  \\
\end{array}\right) X^\dagger =
\left(
\begin{array}{cc} \omega & 0 \\ 0 & \omega^2
\end{array} \right),
\quad
X \left(
\begin{array}{cc} 0 & 1 \\ 1 & 0
\end{array} \right) X^\dagger =
\left(
\begin{array}{cc} 0 & 1 \\ 1 & 0
\end{array} \right),
\end{equation}
where $X$ is defined in equation~(\ref{X}),
reproduces exactly the representations of $a$ and $b$, respectively,
of the irrep~(\ref{21}). This irrep is present for any $n \geq 2$.

To find the further two-dimensional irreps we note that, according to
the derivation in appendix~\ref{irrepsD}, in such irreps 
$D(a)$ commutes with $D(c)$ and $D(d)$, where $D$ indicates
the representation. Then, using the third and fourth relation in
equation~(\ref{presD6n2}), it follows that $D(c) = D(d)$ and 
$D(c)^3 = \bone_2$. From the latter equation we deduce
that any eigenvalue $\lambda$ of $D(c)$ fulfills 
$\lambda^3 = \lambda^n = 1$. This is only possible if three is a
divisor of $n$. Therefore, in the case of $n$ being a multiple of
three it follows from the derivation in appendix~\ref{irrepsD} that,
in addition to the irrep~(\ref{21}), there are three further
two-dimensional irreps:
\begin{alignat}{7}
\label{23}
\two'&: \quad
&
a &\mapsto 
\left( \begin{array}{cc} \omega & 0 \\ 0 & \omega^2
\end{array} \right), 
& \quad
b &\mapsto \left(
\begin{array}{cc} 0 & 1 \\ 1 & 0
\end{array} \right),
& \quad
c = d 
& \mapsto 
\left(
\begin{array}{cc} \omega & 0 \\ 0 & \omega^2
\end{array} \right), 
\\
\label{24}
\two''&: 
&
a &\mapsto 
\left( \begin{array}{cc} \omega & 0 \\ 0 & \omega^2
\end{array} \right), 
&
b &\mapsto \left(
\begin{array}{cc} 0 & 1 \\ 1 & 0
\end{array} \right), 
& 
c = d &\mapsto 
\left(
\begin{array}{cc} \omega^2 & 0 \\ 0 & \omega
\end{array} \right), 
\\
\label{22}
\two'''&: 
&
a &\mapsto 
\left( \begin{array}{cc} 1 & 0 \\ 0 & 1 
\end{array} \right), 
&
b &\mapsto \left(
\begin{array}{cc} 0 & 1 \\ 1 & 0
\end{array} \right),
& 
c = d &\mapsto 
\left(
\begin{array}{cc} \omega & 0 \\ 0 & \omega^2
\end{array} \right). 
\end{alignat}

The three-dimensional irreps are easy to guess, taking into account
the presentation of $\Delta(6n^2)$, equation~(\ref{acd}) and the 
results of appendix~\ref{irrepsD}. There are $2(n-1)$ inequivalent
three-dimensional irreps given by~\cite{ludl-dt,luhn6n2} 
\begin{equation}
\three^{(k,\pm)}: \quad
a \mapsto E, \quad b \mapsto \pm B, \quad
c \mapsto \diag( \eta^k,\, 1,\, \eta^{-k}), \quad
d \mapsto \diag( \eta^{-k},\, \eta^k,\, 1)
\end{equation}
with $\eta = \exp (2\pi i/n)$ and $k=1, \ldots, n-1$.

From the discussion of the irreps of $\Delta(6n^2)$ up to now we
know already the number of one, two and three-dimensional irreps,
therefore, we can compute the number of six-dimensional inequivalent
irreps with the help of theorem~\ref{n2}. This allows to present the
following list~\cite{luhn6n2}:
\begin{equation}\label{dims}
\renewcommand{\arraystretch}{1.3}
\begin{array}{c|cccc}
\mbox{dim} & \; 1\; & \;2\; & 3      & 6 \\\hline
n \neq 3\ell  & 2 & 1 & 2(n-1) & (n-1)(n-2)/6 \\
n = 3\ell     & 2 & 4 & 2(n-1) & n(n-3)/6 
\end{array}
\end{equation}
Just as for $\Delta(3n^2)$, one has to
distinguish between three being a divisor of $n$ or not.

Finally, we discuss the six-dimensional irreps of $\Delta(6n^2)$.
The general form of such irreps of the groups of
type~D in appendix~\ref{irrepsD} applies also here:
\begin{equation}
a \mapsto \left( \begin{array}{cc} E & 0 \\ 0 & E^2 
\end{array} \right),
\quad
b \mapsto \left( \begin{array}{cc} 0 & \bone_3 \\ \bone_3 & 0 
\end{array} \right),
\quad
c \mapsto \left( \begin{array}{cc} c_1 & 0 \\ 0 & c_2 
\end{array} \right),
\quad
d \mapsto \left( \begin{array}{cc} d_1 & 0 \\ 0 & d_2 
\end{array} \right).
\end{equation}
Then from the presentation~(\ref{presD6n2}) one infers that there is a
series of six-dimensional irreps
\begin{equation}
\six^{(k,l)}: \quad
c_1 = d_2^{-1} = \diag( \eta^l,\, \eta^k, \, \eta^{-k-l}), \quad
c_2 = d_1^{-1} = \diag( \eta^{k+l},\, \eta^{-l}, \eta^{-k})
\end{equation}
with $k,l = 1, \ldots, n-1$.
For some pairs $(k,l)$ one obtains equivalent irreps, and for some pairs 
the representation $\six^{(k,l)}$ is not irreducible. Moreover, one
has to distinguish between $n$ being a multiple of three or not. 
A detailed discussion of these issues can be found in~\cite{luhn6n2}. 

We complete this section with a discussion of 
\begin{equation}\label{z3z3s3}
\Delta(54) \cong (\zz_3 \times \zz_3) \rtimes S_3.
\end{equation}
First we compute the conjugacy classes. For simplicity we regard 
$\Delta(54)$ as a matrix group, in the same way as we treated 
$\Delta(27)$---see section~\ref{D3}.
Thus, a minimal set of generators is
\begin{equation}\label{minimalD54}
\Delta(54): \quad 
E = \left( \begin{array}{ccc} 
0 & 1 & 0 \\ 0 & 0 & 1 \\ 1 & 0 & 0 
\end{array} \right), \quad
C = \left( \begin{array}{ccc} 
1 & 0 & 0 \\ 0 & \omega & 0 \\ 0 & 0 & \omega^2 
\end{array} \right), \quad 
B = \left( \begin{array}{rrr} 
-1 & 0 & 0 \\ 0 & 0 & -1 \\ 0 & -1 & 0
\end{array} \right).
\end{equation}
We note that $\Delta(27)$,
generated by $E$ and $C$, is a normal subgroup of $\Delta(54)$ with
\begin{equation}
\Delta(54)/\Delta(27) \cong \zz_2. 
\end{equation}
This allows us to
use theorem~\ref{cc2} for constructing the conjugacy classes of
$\Delta(54)$ from those of $\Delta(27)$.  
The matrix $B$ corresponds to the element $b$ occurring
in theorem~\ref{cc2}. We denote the conjugacy classes of $\Delta(54)$
by $C'_k$. We apply the relations 
\begin{equation}\label{conjB}
B C B^{-1} = C^2, \quad B E B^{-1} = E^2
\end{equation}
to the conjugacy classes of $\Delta(27)$ in equation~(\ref{cDelta27})
and readily find 
\begin{equation}
\begin{array}{llll}
C'_1 = C_1, & C'_2 = C_2, & C'_3 = C_3, & \\
C'_4 = C_4 \cup C_5, &
C'_5 = C_6 \cup C_7, &
C'_6 = C_8 \cup C_9, &
C'_7 = C_{10} \cup C_{11}.
\end{array}
\end{equation}
This exhausts the elements of $\Delta(27)$.
Then we first search for the conjugacy class associated with $B$. The
relations we need are 
\begin{equation}
C B C^{-1} = C^2B, \quad E B E^{-1} = E^2B.
\end{equation}
With some reordering, e.g. $EC = \omega CE$, we obtain the result 
\begin{equation}
C'_8 = \{ B,\, EB,\, E^2B,\, CB,\, C^2B,\, \omega C^2EB,\,
\omega CE^2B,\, \omega^2 CEB,\, \omega^2 C^2E^2B \}.
\end{equation}
Obviously, the missing conjugacy classes are then given by
\begin{equation}
C'_9 = \omega C'_8, \quad C'_{10} = \omega^2 C'_8.
\end{equation}
For more details see~\cite{GL10}. 

There are ten classes of $\Delta(54)$ and, therefore, 
ten inequivalent irreps.
According to the list presented in equation~(\ref{dims}),
$\Delta(54)$ which corresponds to
$n=3$ has no six-dimensional irreps, but six-dimensional irreps do occur for 
$n \geq 4$. 
The two one-dimensional irreps are given 
by equation~(\ref{1'}). Denoting the
defining irrep given in equation~(\ref{minimalD54}) by $\three$, 
we immediately find three other three-dimensional irreps
by the procedures explained in section~\ref{remarks}, namely by complex
conjugation and multiplication with $\one'$. In summary, the 
three-dimensional irreps we have found so far are
\begin{equation}\label{3333}
\three, \quad \three^*, \quad 
\three' := \one' \otimes \three, \quad (\three')^* := \one'
\otimes \three^*.
\end{equation}
There are four remaining irreps. Denoting their dimensions by 
$d_7,\ldots,d_{10}$, it turns out that 
theorem~\ref{n2} completely fixes these dimensions:
\begin{equation}
2 \times 1^2 + 4 \times 3^2 + d_7^2 + d_8^2 + d_9^2 + d_{10}^2 = 54
\quad \Rightarrow \quad d_7 = d_8 = d_9 = d_{10} = 2.
\end{equation}
Therefore, equation~(\ref{3333}) contains all three-dimensional irreps
and the four two-dimensio\-nal irreps are given by equations~(\ref{21})
and~(\ref{23})--(\ref{22}). Clearly, the result obtained here is in
agreement with the list~(\ref{dims}). 
The character table of $\Delta(54)$ is presented in table~\ref{d54ct}.

\begin{table}
\begin{center}
\begin{tabular}{|c|cccccccccc|}
\hline 
$\Delta(54)$ & $C'_1$ & $C'_2$ & $C'_3$ & $C'_4$ & $C'_5$ &
$C'_6$ & $C'_7$ & $C'_8$ & $C'_9$ & $C'_{10}$ \\
(\# $C$) & (1) & (1) & (1) & (6) & (6) & (6) & (6) & (9) & (9) & (9) \\
$\mathrm{ord}(C)$ & 1 & 3 & 3 & 3 & 3 & 3 & 3 & 2 & 6 & 6 \\
\hline 
$\one$
& $1$ & $1$ & $1$ & $1$ & $1$ & $1$ & $1$ & $1$ & $1$ & $1$ \\
$\one'$
& $1$ & $1$ & $1$ & $1$ & $1$ & $1$ 
& $1$ & $-1$ & $-1$ & $-1$ \\ 
$\two$
& $2$ & $2$ & $2$ & $2$ & $-1$ 
& $-1$ & $-1$ & $0$ & $0$ & $0$ \\ 
$\two'$
& $2$ & $2$ & $2$ & $-1$ & $-1$ 
& $2$ & $-1$ & $0$ & $0$ & $0$ \\ 
$\two''$
& $2$ & $2$ & $2$ & $-1$ & $-1$ 
& $-1$ & $2$ & $0$ & $0$ & $0$ \\ 
$\two'''$
& $2$ & $2$ & $2$ & $-1$ & $2$ 
& $-1$ & $-1$ & $0$ & $0$ & $0$ \\ 
$\three$ 
& $3$ & $3\omega$ & $3\omega^2$ 
& $0$ & $0$ & $0$ & $0$ & $-1$ & $-\omega$ & $-\omega^2$ \\
$\three'$ 
& $3$ & $3\omega$ & $3\omega^2$ 
& $0$ & $0$ & $0$ & $0$ & $1$ & $\omega$ & $\omega^2$ \\
$\three^*$ 
& $3$ & $3\omega^2$ & $3\omega$ 
& $0$ & $0$ & $0$ & $0$ & $-1$ & $-\omega^2$ & $-\omega$ \\
$\left( \three' \right)^*$ 
& $3$ & $3\omega^2$ & $3\omega$ 
& $0$ & $0$ & $0$ & $0$ & $1$ & $\omega^2$ & $\omega$ \\
\hline 
\end{tabular}
\caption{Character table of $\Delta(54)$.} \label{d54ct}
\end{center}
\end{table}
Before we discuss tensor products of irreps of $\Delta(54)$, we make a
digression and sketch how its principal series could be used to find
the one and two-dimensional irreps. This is meant to be an illustration
of the usage of principal series outlined at the end of
section~\ref{remarks}. The principal series of $\Delta(54)$ is given
by~\cite{GL10}
\begin{equation}
\{e\} \llhd \zz_3 \llhd \zz_3\times\zz_3 \llhd \Delta(27) \llhd \Delta(54),
\end{equation}
where $\zz_3$ is the center of $\Delta(54)$ generated by 
$\omega \bone$ and $\zz_3\times\zz_3$ is the normal subgroup of
equation~(\ref{z3z3s3}) on which $S_3$ acts. 
According to the analysis presented at the end of section~\ref{remarks}
the series of factor groups
\begin{equation}\label{fg}
\Delta(54)/\Delta(27) \cong \zz_2,\enspace
\Delta(54)/(\zz_3\times\zz_3) \cong S_3, \enspace 
\Delta(54)/\zz_3, \enspace
\Delta(54)/\{e\} \cong \Delta(54)
\end{equation}
has the property that the irreps of any of its members are also
irreps of all groups to the right of it. Starting with the smallest
factor group in equation~(\ref{fg}), it is obvious that its
representations correspond to the one-dimensional irreps of
$\Delta(54)$. The next factor group gives the two-dimensional
irrep of $S_3$, i.e.\ the irrep of equation~(\ref{21}). Then we come
to the group $H \equiv \Delta(54)/\zz_3$, which is a group with 18
elements of which we 
have already found two one-dimensional and one two-dimensional
irrep. Its remaining three two-dimensional irreps correspond to
equations~(\ref{23})--(\ref{22}). 
Note that because in $H$ the center is factored out and $EC = \omega CE$
differs from $CE$ by $\omega \bone$, the two-dimensional irreps of
$H$ correspond to irreps of $\Delta(54)$ where $E$ and $C$ commute,
in agreement with the general observations concerning the
two-dimensional irreps of $\Delta(6n^2)$.

Finally, we want to discuss some tensor products of $\Delta(54)$. We confine
ourselves to two three-dimensional examples, namely
$\three \otimes \three$ and $\three^* \otimes \three$. Using the
relation~(\ref{ma}) and the character table~\ref{d54ct},
we obtain
\begin{equation}\label{t33}
\three \otimes \three = \three^* \oplus (\three')^* \oplus
(\three')^* 
\quad \mbox{and} \quad
\three^* \otimes \three = \one \oplus \two \oplus \two' 
\oplus \two'' \oplus \two'''.
\end{equation}
The first tensor product can be understood in the following way. The
$\three^*$ resides in the antisymmetric basis~(\ref{b3}); the complex
conjugation is just the consequence of equation~(\ref{3''}). The two
$(\three')^*$ can be thought of as corresponding to the symmetric
basis and the ``diagonal'' basis $\{ e_i \otimes e_i \,|\, i= 1, 2,3 \}$.
The second tensor product of equation~(\ref{t33}) can be understood
with the basis~(\ref{B}):
\begin{eqnarray}
&& \one: \;\; \mathcal{B}_{00}, \\
&& \label{222}
\two: \;\; \mathcal{B}_{10}, \; \mathcal{B}_{20}, \quad
\two': \;\; \mathcal{B}_{12}, \; \mathcal{B}_{21}, \quad
\two'': \;\; \mathcal{B}_{11}, \; \mathcal{B}_{22}, \quad
\two''': \;\; \mathcal{B}_{01}, \; \mathcal{B}_{02}.
\end{eqnarray}
That these basis vectors reproduce correctly $E$ and $C$ of
equations~(\ref{21})--(\ref{24}) follows from the construction of the
basis~(\ref{B})---see equation~(\ref{ECB}). That $B \otimes B$
exchanges the basis vectors in each of the four instances in
equation~(\ref{222}) can easily be verified.

\section{Unitary versus special unitary groups}
\label{unitary-vs-su}

\subsection{General considerations}

We already know that every finite group $G$
has a faithful representation. Since every representation of a finite
group is equivalent to a unitary representation, $G$ is isomorphic to
a group of unitary matrices. Thus every finite group is isomorphic to
a finite subgroup of $U(n)$ for a suitable $n$. We can choose
the minimal $n$ such that $G$ has a faithful $n$-dimensional representation, not
necessarily irreducible---see section~\ref{flavour symmetries}.

Since there are three generations of fermions the finite subgroups of
$U(3)$ are particularly interesting, because they offer the possibility to
group the three members of a fermionic gauge multiplet into triplets
transforming under faithful representations of the family group. 
While the finite subgroups of $SU(3)$ have been studied intensively in
the past, there is, to our knowledge, no such systematic classification of
the finite subgroups of $U(3)$. 

As a first step 
it is expedient to investigate the relation between
the finite subgroups of $SU(n)$ and $U(n)$. The following theorem, which
can be found in~\cite{Speiser}, theorem 157 (p.177), tells us at
least a partial solution to this problem. 
\begin{theorem}\label{U3SU3theorem}
If $G$ is a finite subgroup of $U(n)$, then 
	\be
	N:=\{a\in G\,|\,\det a = 1 \}
	\ee
is a normal subgroup of $G$ and the
factor group $G/N$ is cyclic. In other words: \textit{Every finite
  subgroup of $U(n)$ is a cyclic extension of a finite subgroup
  of $SU(n)$}. 
\\
\textit{Proof:} Clearly $N$ is normal in $G$, because for $a \in N$ we
have $\det (bab^{-1}) = \det a = 1$ $\forall\, b \in G$.
In order to show that $G/N$ is cyclic we define the mapping
\begin{equation}
\phi: 
\begin{array}{ccc} 
G/N &\rightarrow& \mathbbm{C}, \\
aN  &\mapsto& \det a.
\end{array}
\end{equation}
Via the property of the determinant the mapping $\phi$ is a
homomorphism with values on the unit circle and, consequently, the set of
complex numbers $\phi(aN)$ with $a \in G$ is a finite subgroup of $U(1)$. 
Thus $G/N$ is isomorphic to
a finite subgroup of $U(1)$ and is, therefore, cyclic. Q.E.D. 
\end{theorem}

Theorem~\ref{U3SU3theorem} reduces the problem of finding all finite subgroups
of $U(n)$ to an extension problem of the finite subgroups of $SU(n)$. Though we
cannot solve it in general, we can list the 
necessary and sufficient requirements for the existence of a
solution---see also~\cite{Hall}, chapter 15.3 (p.224). 
\begin{theorem}
Let $N$ be a finite subgroup of $SU(n)$, then there exists a finite subgroup $G$
of $U(n)$ with $G/N \cong \zz_m$ if and only if there exists a unitary 
$n\times n$-matrix $x$ such that 
	\begin{enumerate}
	 \item $x^i\not\in N\enspace\forall\, i\in \{1,\ldots,m-1\}$,
	 \item $x^{m}:= \alpha\in N$ and
	 \item $xnx^{-1} \in N\quad\forall n\in N$.
	\end{enumerate}
Thus we find all finite subgroups $G$ of $U(n)$ by taking all subgroups of
$SU(n)$ and forming all matrix groups 
	\be\label{xN}
	G=N \cup xN \cup x^2 N \cup \ldots \cup x^{m-1}N,
	\ee
where $x$ fulfills the above conditions.
\end{theorem}

Conversely, if we have a finite subgroup $G$ of $U(n)$, the
subset $N$ of matrices $a$ with $\det a = 1$ is an $SU(n)$ subgroup,
and there always exists an element $x$ such that the above conditions
are fulfilled. Note that if we find a matrix $x$ such that 
$x^m = \alpha = \bone$, then we even have $G \cong N \rtimes \zz_m$, where
$\zz_m$ acts on $N$ via the usual matrix multiplication:
\begin{equation}
\left( n_1 x^{k_1} \right) \left( n_2 x^{k_2} \right) = 
\left( n_1 x^{k_1} n_2 x^{-k_1} \right) x^{k_1+k_2}
\quad \mbox{with} \quad k_1, k_2 \in \{0,1,\ldots, m-1 \}
\end{equation}
and $n_1, n_2 \in N$.

\subsection{The group $\Sigma(81)$}

This group is defined as a subgroup of $U(3)$, generated by the
matrix $E$ of equation~(\ref{F}) and the three $\zz_3$
generators~\cite{ma81} 
\begin{equation}\label{S123}
S_1 = \diag(\omega,\,1,\,1), \quad
S_2 = \diag(1,\,\omega,\,\,1), \quad
S_3 = \diag(\,1,\,1,\,\omega) \quad \mbox{with} \quad \omega = e^{2\pi i/3}.
\end{equation}
By replacing $\omega$ with 
$\eta = \exp(2\pi i/n)$ 
where $n=2,3,4,\ldots$, one
finds a whole series, $\Sigma(3n^3)$, of groups of this
type~\cite{tanimoto}. For $n=3$, a set of generators equivalent to $E$
and the matrices of equation~(\ref{S123}) is
\begin{equation}
\label{SEC1}
\Sigma(81): \quad E, \;\; C,\;\; 
S:= \diag( 1,\, 1,\, \omega^2),
\end{equation}
where $C$ is defined in equation~(\ref{minimalD54}).
The generator $S$ fulfills
\begin{equation}\label{SEC}
SES^{-1} = CE, \quad SCS^{-1} = C, \quad S^3 = \bone.
\end{equation}
Therefore, we find the structure
\begin{equation}
\Sigma(81) \cong \Delta(27) \rtimes \zz_3.
\end{equation}
Note that this structure can easily be generalized.
Replacing $\omega$ and $\omega^2$ in $C$ and $S$ 
by $\eta = \exp(2\pi i/n)$ and $\eta^{-1}$, respectively, 
it is easy to show that equation~(\ref{SEC}) holds with $S^n = \bone$.
Therefore, one finds 
$\Sigma(3n^3) \cong \Delta(3n^2) \rtimes \zz_n$. 

In order to obtain the presentation of $\Sigma(81)$, we 
add a generator $u$ to the presentation of $\Delta(27)$---see
equation~(\ref{presD3n2}) with $n=3$. The generator $u$ 
corresponds to $S$. Using equation~(\ref{SEC}), one arrives at the following
presentation of $\Sigma(81)$:
\begin{equation}\label{presS81}
\begin{array}{l}
a^3 = c^3 = d^3 = u^3 = e, \\
cd = dc, \quad cu = uc, \quad du = ud, \\
aca^{-1} = c^{-1}d^{-1}, \quad ada^{-1} = c, \quad uau^{-1} = cda.
\end{array}
\end{equation}

Now we sketch how to find all irreps of $\Sigma(81)$. For details and
tensor products of irreps we
refer the reader to the references~\cite{tanimoto,smirnov,ma81,BenTov81}. The
character table is given in~\cite{ma81}.
In order to find the one-dimensional irreps, we follow the line of 
arguments used in the case of $\Delta(3n^2)$. We obtain that $c$ and
$d$ are mapped into the same number and $c^3 \mapsto 1$. Moreover, in
the case of $\Sigma(81)$ we have the requirement $cd \mapsto
1$. Therefore, we conclude that $c \mapsto 1$ and $a^3 \mapsto 1$,
$u^3 \mapsto 1$ and arrive at nine one-dimensional irreps:
\begin{equation}
\one^{(p,q)}: \quad a \mapsto \omega^p, \quad c \mapsto 1,  
\quad d \mapsto 1, \quad u \mapsto \omega^q \quad (p,q = 0,1,2).
\end{equation}
Denoting the defining irrep by $\three$, 
we resort to equation~(\ref{1xd}) to discover further three-dimensional
irreps. In this context we note that replacing $E$ by $\omega^p E$
leads to equivalent irreps, as discussed in
equation~(\ref{equiv1x3}). This leaves us with the six irreps 
\begin{equation}
\one^{(0,q)} \otimes \three \quad \mbox{and} \quad 
\one^{(0,q)} \otimes \three^*,
\end{equation}
which are indeed inequivalent and which are simply the extensions of
the irreps $\three$ and $\three^*$ of $\Delta(27)$. 

Using the information from~\cite{ma81} that $\Sigma(81)$ has 17
conjugacy classes, theorem~\ref{n2} tells us that there are two
remaining three-dimensional irreps. Interestingly, these two irreps
are not extensions of irreps of $\Delta(27)$, but irreps where $c$ and
$d$ of the presentation~(\ref{presS81}) commute with $a$. Therefore,
both $c$ and $d$ are mapped onto $\omega^p \bone$ with $p=1,2$. Since
the product $cd$ corresponds to $C$, cf.\ equation~(\ref{acd}), 
the presentation~(\ref{presS81}) yields the irreps 
\begin{equation}
\three^{(p)}: \quad E \mapsto E, \quad C \mapsto \omega^{2p} \bone,
\quad S \mapsto \diag(1,\, \omega^p,\, \omega^{2p}) 
\quad \mbox{with} \quad p=1,2.
\end{equation}

\section{Concluding remarks}
\label{conclusions}
Finite symmetry groups in flavour physics are a fascinating subject.
Unfortunately, no consensus has yet been reached as to which is the most
promising class of finite groups, 
let alone the most promising group. Even if we
postulate that the existence of three families necessitates a group with one
or more three-dimensional irreps and if, in addition, we confine ourselves to
groups $G$ with $\oo G \leq 100$, we are faced with the disenchanting number
of 90 groups---see~\cite{wingerter} where these groups are listed. 
Moreover, for each of these groups there are several possibilities for its
breaking to a subgroup. So the number of possible cases is a multiple
of the number of groups. 
Which breaking is realized will in general be model-dependent and in
models which show promise with respect to explaining features of
fermion masses and mixing matrices one usually faces the problem of
vacuum alignment, i.e.\ the problem of achieving the desired symmetry
breaking. 
This is exactly the problem from which many models based on $A_4$
suffer. Much emphasis has been put in recent years on this group
which, with its 12 elements, is the smallest group with a
three-dimensional irrep. The motivation for $A_4$ derived
mainly from near tri-bimaximal lepton mixing~\cite{HPS}. However, on the one
hand it was noticed that $A_4$ is not the only group which can generate
tri-bimaximal mixing, on the other hand recent neutrino
data~\cite{nudata}, which indicate that the so-called reactor angle is
not very small, have weakened the case for tri-bimaximal mixing. 

It is instructive to consider the groups of low order which possess a
three-dimensional irrep. There are seven such groups with $\oo G \leq
30$~\cite{wingerter}:  
\begin{equation}
A_4, \quad T_7, \quad \widetilde T, \quad S_4, \quad \zz_2 \times A_4, \quad
\Delta(27), \quad \zz_9 \rtimes \zz_3.
\end{equation}
All these groups have been treated in this review with the exception of 
$\zz_9 \rtimes \zz_3$. This group has the following structure:
Denoting the generators of $\zz_9$ and $\zz_3$ by $a$ and $b$, respectively,
then the semidirect product is defined via $b a b^{-1} = a^3$---see
section~\ref{GroupstructureI}. It is a genuine $U(3)$ subgroup, its generators
and irreps can be found in~\cite{ludl10}. In~\cite{maT7} a comparison
has been made for the groups $A_4$, $T_7$ and $\Delta(27)$ with
respect to their properties as flavour groups. 

Recently, various arguments in favour of four Standard Model
fermion generations have been put forward, which makes it worthwhile
to consider groups with four-dimensional 
irreps~\cite{kephart,schmidt}. 
Among the groups discussed in the
review $A_5$ has a four-dimensional irrep, on which the model
of~\cite{kephart} is based. In~\cite{schmidt} the group 
$\zz_5 \rtimes_\phi \zz_4$ was utilized, the smallest group with a
four-dimensional irrep.

Last but not least we want to mention that for the investigation of finite
groups the computer algebra system GAP~\cite{GAP} with its associated
libraries, in particular, the SmallGroups library~\cite{SGL} is an
indispensable tool. Without it the systematic search for 
groups with specific properties would be very hard, if not impossible. 

\newpage
\appendix

\setcounter{section}{0}
\renewcommand{\theequation}{\Alph{section}\arabic{equation}}

\section{A counterintuitive example of a flavour group for three families} 
\label{example}
\setcounter{equation}{0}

In section~\ref{flavour symmetries} we have addressed the question
whether, in the case of models with three fermion families, one can confine 
oneself to the consideration of finite subgroups of $U(3)$. 
Here we will show that in general this is not true. This might be
counterintuitive, but we will corroborate this statement by a concrete  
example of a model proposed in~\cite{GL07}. In that model, the flavour
symmetry group is defined in the following way. 
We define a set $N$ of diagonal $3 \times 3$ matrices whose diagonal
elements are $\pm 1$ and a set $\hat S_3$ of $3 \times 3$ permutation
matrices. The set $N$ is a group under matrix multiplication
such that $N$ is isomorphic to 
$\mathbbm{Z}_2 \times \mathbbm{Z}_2 \times \mathbbm{Z}_2$.
Now we introduce the set of triples $\left( m,n,s \right)$
with $m,n \in N$ and $s \in \hat S_3$ and endow it with 
the multiplication law 
\be
\label{law}
\left( m_1, n_1, s_1 \right) \left( m_2, n_2, s_2 \right)
= \left( m_1 s_1 m_2 s_1^{-1}, n_1 s_1 n_2 s_1^{-1}, s_1 s_2 \right).
\ee
It is tedious but straightforward to check that the group axioms are
fulfilled. Actually, the group is the semidirect product 
$(N \times N) \rtimes S_3$---see section~\ref{GroupstructureI}.
Thus we have obtained a group with 
$2^3 \times 2^3 \times 3! = 384$ elements, and the mappings 
\begin{equation}\label{D123}
\begin{array}{cccc}
D_1: & \left( m,n,s \right) &\to& ms, \\
D_2: & \left( m,n,s \right) &\to& ns, \\
D_3: & \left( m,n,s \right) &\to& mns 
\end{array}
\end{equation}
define three-dimensional irreps. If we assume that, for instance, 
$\ell_L$ and $e_R$ transform according to $D_1$ and $D_2$,
respectively, and that we have a triplet of Higgs doublets, 
i.e.\ $n_H =3$, 
transforming according to $D_3$, then the Yukawa Lagrangian of
equation~(\ref{Delta}) is given by 
\begin{equation}
\mathcal{L}_Y^{(e)} = -y \sum_{i=1}^3 \bar \ell_{iL} \phi_i e_{iR} 
+ \mbox{H.c.}
\end{equation}
with a single Yukawa coupling constant $y$. Here, the index $i$ can be
conceived as the flavour index. None of the three irreps of
equation~(\ref{D123}) is faithful because each irrep has a non-trivial
kernel with eight elements, i.e., eight elements are mapped onto
the unit matrix. Moreover, all the three kernels are 
different. Therefore, it is not possible to replace the group by one
of the three irreps, say by $D_1$ and conceive $D_2$ and $D_3$ as
representations of $D_1$. 
Furthermore, $(N \times N) \rtimes S_3$ has no faithful three-dimensional
(or lower-dimensional) irreps. 
Thus it is not possible to consider the above
flavour group as a subgroup of $U(3)$.

For the group $(N \times N) \rtimes S_3$ one can show that the minimal
dimension in which a faithful representation exists is six. Such a
representation is for instance given by $D_1 \oplus D_2$. Therefore,
this group can be considered as a subgroup of $U(6)$. 
One can also demonstrate that this group has no
faithful irrep~\cite{GL07} at all.

\section{The generators of $S_3$ and their representation as rotation matrices} 
\label{XS3}
\setcounter{equation}{0}

In section~\ref{SSS} the generators $s = (123)$ and $t = (12)$ of $S_3$ are 
represented in two different but equivalent ways as $3 \times 3$ rotation
matrices. A matrix $X$ which procures the similarity transformation between
the two representations has to fulfill
\begin{equation}
X
\renewcommand{\arraystretch}{1.3}
\left( \begin{array}{rrr}
-\frac{1}{2} & -\frac{\sqrt{3}}{2} & 0 \\
\frac{\sqrt{3}}{2} & -\frac{1}{2} & 0 \\
0 & 0 & \hphantom{-}1 
\end{array}\right)
X^\dagger =
 \left( \begin{array}{rrr}
0 & 1 & 0 \\ 0 & 0 & 1 \\ 1 & 0 & 0
\end{array}\right)
\end{equation}
and
\begin{equation}
X
\renewcommand{\arraystretch}{1.3}
\left( \begin{array}{rrr}
0 & \hphantom{-}1 & 0 \\
1 & 0 & 0 \\
0 & 0 & -1 
\end{array}\right)
X^\dagger =
 \left( \begin{array}{rrr}
-1 & 0 & 0 \\ 0 & 0 & -1 \\ 0 & -1 & 0
\end{array}\right)
\end{equation}
for $s$ and $t$, respectively.
By explicit computation it is straightforward to check that 
a possible $X$ is given by
\begin{equation}
X=\left( \begin{array}{ccc}
\frac{1}{\sqrt{3}} & 
\frac{-1}{\sqrt{3}} & 
\frac{1}{\sqrt{3}} \\ 
\frac{-3-\sqrt{3}}{6} & 
\frac{-3+\sqrt{3}}{6} &
\frac{1}{\sqrt{3}} \\ 
\frac{3-\sqrt{3}}{6} & 
\frac{3+\sqrt{3}}{6} & 
\frac{1}{\sqrt{3}} 
\end{array} \right).
\end{equation}

\section{$A_5$ and the rotation matrix $W$}
\label{A5W}
\setcounter{equation}{0}
In this appendix we compute the possible solutions $W$ of equation~(\ref{W}). 
By a reformulation the third relation of equation~(\ref{W}) can easily be
solved for $W$:
\begin{equation}
EWE = W \quad \Rightarrow \quad
W = \left( 
\begin{array}{ccc}
\alpha & \beta & \gamma \\ 
\beta & \gamma & \alpha \\ 
\gamma & \alpha & \beta
\end{array} \right).
\end{equation}
With this form of $W$ we find 
\begin{equation}\label{W1}
W^2 = \bone \quad \Rightarrow \quad 
\alpha^2 + \beta^2 + \gamma^2 = 1, \quad
\alpha \beta + \beta \gamma + \gamma \alpha = 0
\end{equation}
and 
\begin{equation}\label{W2}
WAW = AWA \quad \Rightarrow \quad 
\begin{array}{cc@{\,=}c}
\alpha^2 - \beta^2 - \gamma^2 = \alpha, &
\beta \gamma - \alpha (\beta + \gamma) & \alpha, \\
\beta^2 - \gamma^2 - \alpha^2 = \gamma, & 
\alpha \beta - \gamma (\alpha + \beta) & -\beta, \\
\gamma^2 - \alpha^2 - \beta^2 = \beta, & 
\alpha \gamma - \beta (\alpha + \gamma) & -\gamma.
\end{array}
\end{equation}
Together, equations~(\ref{W1}) and~(\ref{W2}) give us eight
relations for three real parameters $\alpha, \beta, \gamma$.
To solve this system of equations we begin with 
$\alpha^2 + \beta^2 + \gamma^2 = 1$ and 
$\alpha^2 - \beta^2 - \gamma^2 = \alpha$. Adding these two relations
we obtain $2\alpha^2 - \alpha -1 = 0$. Therefore, $\alpha = 1$ or
$-1/2$. However, for $\alpha = 1$ it follows that 
$\beta = \gamma = 0$, which is in contradiction to 
$\beta \gamma - \alpha (\beta + \gamma) = \alpha$. 
Thus we conclude
\begin{equation}\label{alpha}
\alpha = -1/2. 
\end{equation}
Next we take the sum of 
$\beta^2 - \gamma^2 - \alpha^2 = \beta$ and 
$\gamma^2 - \alpha^2 - \beta^2 = \gamma$, and the sum of 
$\alpha \beta - \gamma (\alpha + \beta) = -\beta$ and 
$\alpha \gamma - \beta (\alpha + \gamma) = -\gamma$. This gives 
two equations in $\beta$ and $\gamma$, namely
$\beta + \gamma = -1/2$ and 
$\beta + \gamma = 2\beta\gamma$, respectively. 
This system of equations for $\beta$ and $\gamma$ has two
solutions:
\begin{equation}\label{beta-gamma}
\beta  =  \frac{1}{2}\, \mu_2, \quad
\gamma =  \frac{1}{2}\, \mu_1 
\quad \mbox{or} \quad
\beta  =  \frac{1}{2}\, \mu_1, \quad
\gamma =  \frac{1}{2}\, \mu_2 
\quad \mbox{with} \quad \mu_{1,2} = \frac{-1 \pm \sqrt{5}}{2}.
\end{equation}
It is straightforward to check that 
with the results~(\ref{alpha}) and~(\ref{beta-gamma}) 
all eight relations above are fulfilled.

\section{$A_5$ as the symmetry group of the icosahedron}
\label{A5I}
\setcounter{equation}{0}

Here we prove that the rotation representation of $A_5$ is isomorphic
to the group $I$ of rotation symmetries of the regular icosahedron. 
This representation is given 
by the matrices $A$ and $E$
of equation~(\ref{AE}) and $W$ of equation~(\ref{WW'}). 
As shown in section~\ref{SSS}, the rotation $AW\!E$ has a
five-fold rotation axis with rotation angle $\alpha = 72^\circ$.  
The corresponding axis is readily computed by using equation~(\ref{SO3calc}). 
In this way we find 
	\be\label{Icosahedronvertex}
	AW\!E=R(2\pi/5,\,\vec{v}) 
        \quad \mbox{with} \quad
        \vec{v}=\frac{1}{\sqrt{1+\mu_2^2}}\bmat 0\\ \mu_2\\ 1\emat. 
	\ee
The constants $\mu_1$ and $\mu_2$, which we will frequently use in the
following, are defined in equation~(\ref{mu12}).
Now we perform a successive conjugation of $AW\!E$ with the elements of the
tetrahedral group $T$ generated by $A$ and $E$. According to
equation~(\ref{Icosahedronvertex}), we obtain eleven further rotations through
an angle of $2\pi/5$. In summary, the twelve rotation axes are given by 
        \be\label{verI}
	\begin{array}{ccc}
	\pm\displaystyle{\frac{1}{N}}\bmat 0\\ \mu_2\\ 1\emat, & 
	\pm\displaystyle{\frac{1}{N}}\bmat 1 \\0 \\ \mu_2\emat, &
	\pm\displaystyle{\frac{1}{N}}\bmat \mu_2 \\ 1 \\ 0\emat,\\[2mm]
	\pm\displaystyle{\frac{1}{N}}\bmat 0\\ \mu_2\\ -1\emat, &
	\pm\displaystyle{\frac{1}{N}}\bmat -1\\ 0 \\ \mu_2 \emat, &
	\pm\displaystyle{\frac{1}{N}}\bmat \mu_2 \\ -1\\ 0 \emat,
	\end{array}
	\ee
where $N=\sqrt{1+\mu_2^2}$.
Since the conjugacy class of symmetry rotations of the icosahedron through
$72^\circ$ contains twelve elements, it suggests itself to 
interpret the vectors of equation~(\ref{verI}) as coordinates of the vertices
of an icosahedron. Indeed, plotting the endpoints of the vectors~(\ref{verI})
and connecting the nearest neighbors with lines, 
one obtains fi\-gure~\ref{icosahedron}, which is the picture of a regular
icosahedron. By construction, this solid is invariant under the
actions of $A$ and $E$. In order to prove invariance under the full
group generated by $A$, $E$ and $W$, it remains to demonstrate that it is also
invariant under the action of $R(2\pi/5,\,\vec{v})$. This can be seen
by explicit computation. To this end we denote the vectors of
equation~(\ref{verI}) by the symbols $\pm \vec v_{ka}$ , with
$k=1,2,3$ indicating the position of the zero and $a=\pm$ indicating the
occurrence of $+1$ or $-1$. Then the action of $R(2\pi/5,\,\vec{v})$
can be subsumed as
\begin{equation}
R(2\pi/5,\,\vec{v}): \quad
\vec v_{2+} \mapsto -\vec v_{3-} \mapsto -\vec v_{1-} \mapsto 
\vec v_{3+} \mapsto \vec v_{2-} \mapsto \vec v_{2+},
\end{equation}
which completes the proof. 

For consistency we may check that the rotation axis $\vec e_x$ of $A$
points to the center of an edge and that 
the rotation axis $\vec n$, given in equation~(\ref{n}), of
$E$ points to the center of a face of the icosahedron defined by the
vertices~(\ref{verI}). For this purpose we note that 
an edge has the length $2/N$. Therefore, $\vec v_{2+}$ and $- \vec v_{2-}$ are
connected by an edge and the same is true for the vectors 
$\vec v_{k-}$ with $k=1,2,3$. Then the relations 
\begin{equation}
\vec e_x \propto \vec v_{2+} - \vec v_{2-}
\quad \mbox{and} \quad
\vec n \propto \vec v_{1-} + \vec v_{2-} + \vec v_{3-}
\end{equation}
provide the desired consistency check.

\section{Irreps of the groups of type~C}
\label{irrepsC}
\setcounter{equation}{0}

In section~\ref{classification} we have presented a classification of
all subgroups of~$SU(3)$. It turns out that for groups of type~C and~D
one can give a fairly general discussion of their irreps.

We begin with the group $C(n,a,b)$. Denoting a generic element of 
the normal subgroup $N(n,a,b)$ of diagonal matrices of $C(n,a,b)$ by $F$, 
the only properties of the group we will need are 
\begin{equation}\label{EFN}
E F E^{-1} \in N(n,a,b) 
\quad \mbox{and} \quad E^3 = \bone,
\end{equation}
where $E$ is defined in equation~(\ref{F}). 
Suppose we have an irrep $D$ of this
group on the unitary space $\mathcal{V}$. On this space the elements
$g \in C(n,a,b)$ are represented by $D(g)$, which we abbreviate for
simplicity of notation by $\bar g$. We observe that 
there exists a vector $x \in \mathcal{V}$ which is a simultaneous
eigenvector of all $\bar F$ in $N(n,a,b)$. 
But then also the vectors ${\bar E}^k x$ with $k=1,2$ are simultaneous
eigenvectors of all $\bar F$ in $N(n,a,b)$. This follows from 
\begin{equation}
\bar F \left( {\bar E}^k x \right) = {\bar E}^k 
\left( {\bar E}^{-k} \bar F {\bar E}^k \right) x
\end{equation}
and the first property of equation~(\ref{EFN}). At this point we have
to distinguish two cases. \\
\textbf{Case 1:} There is an $F_0$ such that $\bar F_0$ has at least two
different eigenvalues on the vectors ${\bar E}^k x$ with $k=0,1,2$. 
In this case it turns out that the set 
$\mathcal{B} = \{ x,\, \bar E x,\, {\bar E}^2 x \}$
is an orthogonal basis of $\mathcal{V}$. This can be seen in the
following way. First we note that 
no further powers of $E$ have to be considered because $E^3 = \bone$. 
Furthermore, the space spanned by $\mathcal{B}$ is an invariant
subspace of $\mathcal{V}$, but because we assume that we have an irrep
it must be identical with $\mathcal{V}$. Finally we discuss the
orthogonality of the vectors of $\mathcal{B}$. Let us first assume that  
$x$ and $\bar E x$ are eigenvectors of $\bar F_0$ with different
eigenvalues which means that $x$ and $\bar E x$ must be orthogonal to
each other. But then due to 
\begin{equation}
\langle x | \bar E x \rangle = 0 \quad \Rightarrow \quad
\langle x | {\bar E}^2 x \rangle = \langle \bar E x | x \rangle = 0
\quad \mbox{and} \quad 
\langle \bar E x | {\bar E}^2 x \rangle = \langle x | \bar E x \rangle
= 0
\end{equation}
it follows that ${\bar E}^2 x$ is orthogonal to both $x$ and $\bar E
x$ and $\mathcal{B}$ must be an orthogonal basis. 
The other two cases of different eigenvalues of $\bar F_0$ are treated
analogously. \\
\textbf{Case 2:} There is no $F$ such that $\bar F$ has two different
eigenvalues. In this case all elements of $N(n,a,b)$ are represented
as $\bar F \propto \bone$. This means that all representation
operators commute and, since we have an irrep, it must be
one-dimensional. Then the vector $x$ is an
eigenvector of $\bar E$ in a trivial way.

In summary, we have obtained the interesting result that 
all irreps of $C(n,a,b)$ have either dimension one or three.

A group of type~C can have many different one-dimensional irreps, but
from its semidirect-product structure---see
equation~(\ref{Csemi})---it follows that there are at least three,
namely 
\begin{equation}
E \mapsto \omega^p, \quad F \mapsto 1 \; \forall\, F \in N(n,a,b) 
\end{equation}
with $\omega = \exp (2\pi i/3)$ and $p=0,1,2$. 
Furthermore, from the discussion of case~1 we see that 
the basis $\mathcal{B}$, if we order its elements as 
$\{ x,\, {\bar E}^2 x,\, \bar E x \}$,
is very convenient for the three-dimensional irreps,
because in this basis $E$ is simply represented by itself 
and all elements of $N(n,a,b)$ are represented by
diagonal matrices.

\section{A new set of generators for the groups of type~D}
\label{generatorsD}
\setcounter{equation}{0}
To see the structure of $D(n,a,b;d,r,s)$ we define a new
set of generators~\cite{ludl-dt,zwicky}. First we choose the diagonal matrix
\begin{equation}
\label{Fr}
F_r = E {\tilde G}^2 E^{-1} = 
\diag \left( -\delta^{-r},\,-\delta^{-r},\, \delta^{2r} \right).
\end{equation}
Then we define a matrix $B_t$ by
\begin{equation}
B_t = F_r \tilde G = 
\left( \begin{array}{ccc} 
-1 & 0 & 0 \\ 0 & 0 & -\delta^{-t} \\ 0 & -\delta^t & 0 
\end{array} \right) 
\quad \mbox{with} \quad t = r-s.
\end{equation}
We list the properties of this matrix:
\begin{eqnarray}
B_t^2 &=& \bone, \label{bb} \\
B_t E B_t &=& F'_t E^2, \label{beb} \\
B_t\, \diag(\alpha,\beta,\gamma)\, B_t^{-1} &=& \diag(\alpha,\gamma,\beta),
\label{bdb} 
\end{eqnarray}
where the diagonal matrix $F'_t$ is given by
\begin{equation}
\label{F't}
F'_t = \diag( \delta^{-t},\,\delta^{-t},\, \delta^{2t} ).
\end{equation}
According to the discussion we have just accomplished we may take 
\begin{equation}
F(n,a,b),\; F_r,\; E,\; B_t  
\end{equation}
as a new set of generators. 
However, equations~(\ref{bb}) and~(\ref{beb}) suggest to explore
whether an $S_3$ structure is hidden in these equations. To this end we search
for a diagonal matrix $\tilde F$ which resides in the normal
subgroup $N(n,a,b;d,r,s)$ of diagonal matrices of $D(n,a,b;d,r,s)$,
such that
\begin{equation}\label{tFE}
\big( \tilde F E \big)^3 = \bone \quad \mbox{and} \quad
B_t \big( \tilde F E \big) B_t = \big( \tilde F E \big)^2.
\end{equation}
Together with $B_t^2 = \bone$ these relations form a presentation of 
$S_3$---see section~\ref{SSS}.
The first relation of equation~(\ref{tFE}) is solved by 
$\det \tilde F = 1$, therefore, it does not give a genuine restriction on
$\tilde F$. 
The second relation---taking into account our requirements on 
$\tilde F$---has the general solution 
\begin{equation}
\tilde F = \diag(u,\, \delta^{-t},\, \delta^tu^*)
\quad \mbox{with} \; |u| = 1.
\end{equation}
So we may take $u = \delta^{-t}$ and identify $\tilde F$ with $F'_t$ 
and define
\begin{equation}
E_t = F'_tE.
\end{equation}
However, at this point we notice that the phase factors $\delta^t$ in
$E_t$ and $B_t$ are redundant. One can remove these factors by the 
similarity transformation 
\begin{equation}\label{simi}
e^{i \hat \alpha} E_t e^{-i \hat \alpha} = E, \quad
e^{i \hat \alpha} B_t e^{-i \hat \alpha} =: B =
\left( \begin{array}{rrr}
-1 & 0 & 0 \\ 0 & 0 & -1 \\ 0 & -1 & 0
\end{array} \right)
\quad \mbox{with} \quad 
e^{i \hat \alpha} = \diag(1,\, \delta^{-t},\, \delta^{-2t})
\end{equation}
being a diagonal matrix of phase factors.\footnote{In general, the matrix 
  $e^{i \hat \alpha}$ is not in $D(n,a,b;d,r,s)$.}
This similarity transformation leaves the diagonal generators invariant.

Summarizing, a suitable set of generators which clearly shows the structure of
the groups $D(n,a,b;d,r,s)$ is given by 
$F(n,a,b)$, $F_r$, $F'_t$, $E$ and $B$,
where $E$ and $B$ obviously generate a subgroup
isomorphic to $S_3$. 

\section{Irreps of the groups of type~D}
\label{irrepsD}
\setcounter{equation}{0}

Now we discuss the irreps of the groups of type~D. For simplicity of
notation we introduce the symbol 
$\mathcal{N} \equiv N(n,a,b;d,r,s)$ for the
normal subgroup of diagonal matrices. Denoting a generic element
of $\mathcal{N}$ by $F$, the properties relevant for the following
discussion are
\begin{equation}\label{EFBN}
EFE^{-1} \in \mathcal{N}, \quad BFB^{-1} \in \mathcal{N}, \quad
E^3 = B^2 = \bone, \quad B E B^{-1} = E^2.
\end{equation}
Now we assume that we have an irrep of $D(n,a,b;d,r,s)$ and, as before, 
we indicate the representation of the elements of the group by a bar.
Analogously to the case of groups of type~C, we state that in the
representation space $\mathcal{V}$ there is a vector $x$ which is a
simultaneous eigenvector to all elements of $\mathcal{N}$. 
From equation~(\ref{EFBN}) we infer that we obtain further
simultaneous eigenvectors by applying in all possible ways the
operators $\bar E$ and $\bar B$ to $x$. Thus we end up with the six vectors
\begin{equation}\label{bv}
x,\, {\bar E}^2 x,\, \bar E x,\, 
\bar B x, \, \bar B {\bar E}^2 x,\, \bar B \bar E x.
\end{equation}
Therefore, we conclude that any irrep of $D(n,a,b;d,r,s)$ has at most
dimension six. The dimension of the irrep is smaller if the
system~(\ref{bv}) is linearly dependent. We proceed by distinguishing
different cases with respect to the properties of $\mathcal{N}$ which
are related to the number of linear independent vectors in
equation~(\ref{bv}). 

We begin with the following two assumptions:
\begin{enumerate}
\renewcommand{\labelenumi}{\alph{enumi}.}
\item
There is an $F_0 \in \mathcal{N}$ 
such that $\bar F_0$ has at least two
different eigenvalues on the vectors ${\bar E}^k x$ with $k=0,1,2$.
\item
For every pair
$({\bar E}^k x,\, \bar B {\bar E}^l x)$ ($k,l=0,1,2$) there is 
an $F_{kl} \in \mathcal{N}$ such that its eigenvalues are different on
the pair.
\end{enumerate}
From assumption~a) it follows the vectors ${\bar E}^k x$
form an orthogonal system---see the discussion for
the groups of type~C. Obviously, this is then also true for the vectors
$\bar B {\bar E}^k x$. Assumption~b) leads to pairwise orthogonal
vectors. Taking both assumptions together the system~(\ref{bv}) forms
an orthogonal basis in a six-dimensional irrep. 
It is easy to show that, with the basis~(\ref{bv}), the representation
matrices have the form
\begin{equation}
E \mapsto \left( \begin{array}{cc} E & 0 \\ 0 & E^2 
\end{array} \right),
\quad
B \mapsto \left( \begin{array}{cc} 0 & \bone_3 \\ \bone_3 & 0 
\end{array} \right),
\quad
F \mapsto \left( \begin{array}{cc} \bar F_1 & 0 \\ 0 & \bar F_2 
\end{array} \right),
\end{equation}
where $\bar F_1$ and $\bar F_2$ are diagonal phase matrices.

Next we drop assumption~b). This means that there is a pair
$({\bar E}^p x,\, \bar B {\bar E}^q x)$ such that the eigenvalues of any 
$F \in \mathcal{N}$ are the same on both vectors. 
Then, the vectors\footnote{It is possible that one of these vectors is
  zero in a given irrep.} 
\begin{equation}
x_\pm = {\bar E}^p x \pm \bar B {\bar E}^q x
\end{equation}
are simultaneous eigenvectors to all $F \in \mathcal{N}$ 
and so are the orthogonal systems 
\begin{equation}\label{bv1}
x_\epsilon,\, \bar E x_\epsilon,\, {\bar E}^2 x_\epsilon 
\end{equation}
with $\epsilon = \pm 1$. It turns out that
this three-dimensional orthogonal system is already closed under the
action of $B$. This can be seen by direct computation:  
\begin{equation}
\bar B x_\epsilon = \epsilon 
{\bar E}^q x + {\bar E}^{2p+q} \bar B {\bar E}^q x = 
\epsilon {\bar E}^{q-p} \left( {\bar E}^p x + 
\epsilon \bar B {\bar E}^q x \right) = \epsilon {\bar E}^{q-p} x_\epsilon
\end{equation}
because $q-p = (2p+q) \, \mbox{mod}\, 3$.
With a suitable ordering of the basis~(\ref{bv1})
we end up with two types of three-dimensional irreps characterized by
\begin{equation}
E \mapsto E, \quad B \mapsto \mp B,
\end{equation}
where the minus (plus) sign corresponds to $\epsilon = +1$ ($-1$).

Now we relinquish assumption~a) which means that every $F \in
\mathcal{N}$ has one and the same eigenvalue on the vectors 
${\bar E}^k x$ ($k=0,1,2$) and one and the same eigenvalue on 
$\bar B {\bar E}^k x$ ($k=0,1,2$).
Now we consider the vectors 
\begin{equation}\label{yp}
y_p = (\bone + \omega^p \bar E + \omega^{2p} {\bar E}^2) x 
\quad \mbox{with} \quad p=0,1,2,
\end{equation}
which fulfill
\begin{equation}
\bar E y_p = \omega^{2p} y_p.
\end{equation}
It is impossible that all the three vectors of equation~(\ref{yp})
vanish at the same time. Now we distinguish some cases. 

We begin with $y_2 \neq 0$. Then we have $\bar E y_2 = \omega y_2$ and
$\bar E (\bar B y_2) = \omega^2 \bar B y_2$ due to
equation~(\ref{EFBN}). Moreover, we are allowed to represent \emph{all}
elements of $\mathcal{N}$ by the unit matrix. Thus we obtain the
two-dimensional irrep
\begin{equation}\label{2-1}
E \mapsto \left( \begin{array}{cc} \omega & 0 \\ 0 & \omega^2
\end{array} \right),
\quad
B \mapsto \left( \begin{array}{cc} 0 & 1 \\ 1 & 0 
\end{array} \right),
\quad
F \mapsto \bone_2 \; \forall\, F \in \mathcal{N}.
\end{equation}
This irrep is present for all groups of type~D. It simply reflects the
group structure~(\ref{Dsemi}) and is just the two-dimensional irrep of
$S_3$. Starting with $y_1$ instead of $y_2$ leads to an equivalent irrep.

If there is an element $F_0 \in \mathcal{N}$ such that its eigenvalues
$\lambda_1$ and $\lambda_2$ with respect to $x$ and $\bar Bx$, respectively,
are different, then we have two inequivalent two-dimensional irreps,
depending on whether we use $y_1$ or $y_2$. By a basis transformation
we can achieve that the two irreps are distinguished in the
representation of $F_0$: 
\begin{equation}\label{2-2}
E \mapsto \left( \begin{array}{cc} \omega & 0 \\ 0 & \omega^2
\end{array} \right),
\quad
B \mapsto \left( \begin{array}{cc} 0 & 1 \\ 1 & 0 
\end{array} \right),
\quad
F_0 \mapsto \left( \begin{array}{cc} \lambda_1 & 0 \\ 0 & \lambda_2
\end{array} \right) \; \mbox{or} \;
\left( \begin{array}{cc} \lambda_2 & 0 \\ 0 & \lambda_1
\end{array} \right).
\end{equation}

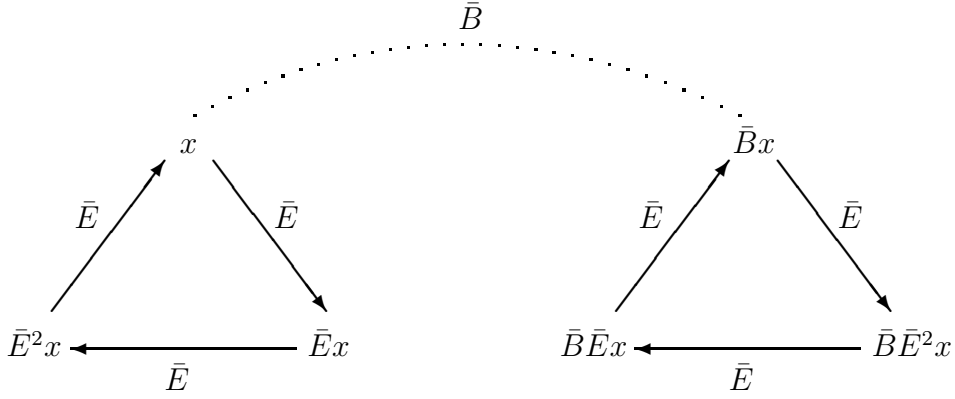
\begin{figure}
\begin{center}
\setlength{\unitlength}{5mm}
\begin{picture}(25,12)\thicklines

\put(5.5,7){\vector(3,-4){3}}
\put(1.2,3){\vector(3,4){3}}
\put(7.7,2){\vector(-1,0){6}}

\put(4.6,7.2){$x$}
\put(0,1.8){$\bar E^2 x$}
\put(8,1.8){$\bar E x$}

\put(7.1,5.2){$\bar E$}
\put(1.8,5.2){$\bar E$}
\put(4.2,0.9){$\bar E$}

\put(20.5,7){\vector(3,-4){3}}
\put(16.2,3){\vector(3,4){3}}
\put(22.7,2){\vector(-1,0){6}}

\put(19.3,7.2){$\bar Bx$}
\put(14.7,1.8){$\bar B \bar E x$}
\put(23,1.8){$\bar B \bar E^2 x$}

\put(22.1,5.2){$\bar E$}
\put(16.8,5.2){$\bar E$}
\put(19.2,0.9){$\bar E$}

\bezier{30}(5,8.2)(12,12)(19.5,8.2)

\put(12,10.5){$\bar B$}

\end{picture}
\end{center}
\vspace{-7mm}
\caption{Graphical representation of the irreps of the groups of
  type~D. At the corners of the two triangles the basis vectors of a 
  six-dimensional irrep are indicated. Furthermore, the vectors show
  the action of $E$, denoted by $\bar E$ in this irrep, the dotted
  line shows the action of $B$, denoted by $\bar B$. For three-dimensional
  irreps the two triangles collapse into one, for two-dimensional irreps each
  triangle shrinks to a point, and for one-dimensional irreps the
  whole figure shrinks to a point. \label{Dtriangle}}
\end{figure}
However, with the $F_0$ above and $\lambda_1 \neq \lambda_2$ we have
one more possibility for a two-dimensional irrep, because in this case
we are allowed to represent $E$ trivially; this corresponds to the
usage of $y_0$. In this case we obtain
\begin{equation}\label{2-3}
E \mapsto \bone_2, 
\quad
B \mapsto \left( \begin{array}{cc} 0 & 1 \\ 1 & 0 
\end{array} \right),
\quad
F_0 \mapsto \left( \begin{array}{cc} \lambda_1 & 0 \\ 0 & \lambda_2
\end{array} \right).
\end{equation}

We stress once more that the irrep~(\ref{2-1}) exists for all groups of
type~D while the irreps~(\ref{2-2}) and~(\ref{2-3}) depend on the
structure of $\mathcal{N}$, which is not specified here; such irreps
need not even exist. 
Moreover, for each type of irrep~(\ref{2-2}) and~(\ref{2-3}) 
there could be several irreps differing in the
representation of~$\mathcal{N}$. For instance it was shown
in~\cite{luhn6n2} that, in the case of $\Delta(6n^2)$, irreps of the
type~(\ref{2-2}) and~(\ref{2-3}) exist only if $n$ is a multiple of
three, in which case there are altogether four inequivalent two-dimensional
irreps; otherwise there is only one two-dimensional irrep, namely that
of equation~(\ref{2-1}). 

We finally discuss the one-dimensional irreps. Clearly, from
equation~(\ref{Dsemi}) it follows that 
\begin{equation}
E \mapsto 1, \quad B \mapsto \pm 1, \quad 
F \mapsto 1 \; \forall\, F \in \mathcal{N}
\end{equation}
are irreps. These two irreps always exist irrespective of the
structure of $\mathcal{N}$. Of course, more one-dimensional irreps,
depending on the properties $\mathcal{N}$ may exist.

In summary, we have found the result that the dimension of an irrep 
of the groups of type~D is either one, two, three or six. 
Figure~\ref{Dtriangle} represents graphically our method for finding
the dimensions of the irreps.

\end{document}